\documentclass[fleqn,10pt]{wlscirep}
\usepackage[utf8]{inputenc}
\usepackage[T1]{fontenc}

\usepackage{subcaption}
\usepackage{xspace}
\usepackage{graphicx}
\usepackage{svg}
\usepackage{amsmath}
\usepackage{makecell}
\usepackage{adjustbox}

\newcommand{\nascondi}[1]{}
\usepackage{pgfplots}
\usepackage{pgfplotstable}
\usepackage{tikz}
\usetikzlibrary{trees}
\usepackage{multirow}	
\usepackage{color, colortbl}
\usepackage{array}

\usepackage{listings}
\lstdefinelanguage{sparql}{
basicstyle=\ttfamily,
morestring=[b][\color{blue}]\",
keywordstyle=\color{red}\bfseries,
morekeywords={SELECT,CONSTRUCT,DESCRIBE,ASK,WHERE,FROM,NAMED,PREFIX,BASE,OPTIONAL,FILTER,GRAPH,LIMIT,OFFSET,SERVICE,UNION,EXISTS,NOT,BINDINGS,MINUS,a},
sensitive=true,
comment=[l]{\#\ },
commentstyle=\scriptsize
}

\pgfplotsset{compat=1.18} 

\newcolumntype{P}[1]{>{\centering\arraybackslash}p{#1}}

\usepackage{changes}
\definechangesauthor[name={EC},color=magenta]{EC}  
\definechangesauthor[name={EM},color=orange]{EM}  
\definechangesauthor[name={GV},color=blue]{GV}
\definechangesauthor[name={MM},color=red]{MM}
\definechangesauthor[name={MC},color=green]{MC}
\definechangesauthor[name={AC},color=blue!70]{AC} 
\definechangesauthor[name={MS},color=green!70]{MS} 
\definechangesauthor[name={PP},color=gray]{PP} 
\definechangesauthor[name={TC},color=brown]{TC} 
\definechangesauthor[name={JR}, color=violet]{JR}  

\newcommand{\rnakg}{RNA-KG\xspace}
\newcommand{\pheno}{PheKnowLator\xspace}

\title{\rnakg: An ontology-based knowledge graph for representing interactions involving RNA molecules}

\author[1]{Emanuele Cavalleri}

\author[1]{Alberto Cabri}

\author[1]{Mauricio Soto-Gomez}

\author[1]{Sara Bonfitto}

\author[1]{Paolo Perlasca}

\author[1]{Jessica Gliozzo}

\author[2]{Tiffany J. Callahan}

\author[3]{Justin Reese}

\author[4]{Peter N Robinson}

\author[1]{Elena Casiraghi}

\author[1]{Giorgio Valentini}

\author[1,*]{Marco Mesiti}

\affil[1]{AnacletoLab, Computer Science Department, University of Milan, 20122, Italy}

\affil[2]{Department of Biomedical Informatics, Columbia University Irving Medical Center, New York, NY 10032, USA}

\affil[3]{Environmental Genomics and Systems Biology Division, Lawrence Berkeley National Laboratory, Berkeley, CA 94720, USA}

\affil[4]{Berlin Institute of Health - Charit\'e, Universit\"atsmedizin, Berlin, 13353, Germany}

\affil[*]{corresponding author(s): Marco Mesiti (marco.mesiti@unimi.it)}

\begin{abstract}
The "RNA world" represents a novel frontier for the study of fundamental biological processes and human diseases and is paving the way for the development of new drugs tailored to the patient's biomolecular characteristics.
Although scientific data about coding and non-coding RNA molecules are continuously produced and available from public repositories, they are scattered across different databases and a centralized, uniform, and semantically consistent representation of the "RNA world" is still lacking.
We propose \rnakg, a knowledge graph encompassing biological knowledge about RNAs gathered from more than 50 public databases, integrating functional relationships with genes, proteins, and chemicals and ontologically grounded biomedical concepts.
To develop \rnakg, we first identified, pre-processed, and characterized each data source; next, we built a meta-graph that provides an ontological description of the KG by representing all the bio-molecular entities and medical concepts of interest in this domain, as well as the types of interactions connecting them.
Finally, we leveraged an instance-based semantically abstracted knowledge model to specify the ontological alignment according to which \rnakg was generated. \rnakg can be downloaded in different formats and also queried by a SPARQL endpoint. 
A thorough topological analysis of the resulting heterogeneous graph provides further insights into the characteristics of the "RNA world". 
\rnakg can be both directly explored and visualized, and/or analyzed by applying computational methods to infer bio-medical knowledge from its heterogeneous nodes and edges.
The resource can be easily updated with new experimental data, and specific views of the overall KG can be extracted according to the bio-medical problem to be studied. 
\end{abstract}

\begin{document}
\maketitle

\section*{Background \& Summary}

The involvement of RNAs in various physiological processes has been ascertained by several studies \cite{Bartel2004,guttman_rinn_2012,Cech2014} that have revealed the pervasive transcription of an unexpected variety of RNA molecules \cite{trnaclover} that can lead to a significant breakthrough in the treatment of cancer, genetic, and neurodegenerative disorders, cardiovascular and infectious diseases~\cite{Damase21}. 
The study of RNA is also one of the most promising 
avenue of research
in therapeutics, as evidenced by the recent success of mRNA-based vaccines for the COVID-19 pandemic~\cite{Barbier22}, for the treatment of melanoma~\cite{Carvalho23}, for the development of new drugs that can target both proteins and mRNA, as well as other non-coding RNA, and for encoding missing or defective proteins, regulating the transcriptome, and mediating DNA or RNA editing~\cite{Winkle21}. 
Thus, RNA technology significantly broadens the set of druggable targets, and is also less expensive than other technologies (e.g., drug synthesis based on recombinant proteins), due to the relatively simple structure of RNA molecules that facilitate their biochemical synthesis and chemical modifications~\cite{Paunovska22}.
Non coding RNAs (ncRNAs) comprise a large range of RNA species, and a large set of scientific data representing different kinds of interactions among them and with other bio-entities (e.g., genes, proteins, chemicals, diseases, and phenotypes) are made publicly available by several genomics laboratories.

The possibility of integrating the interactions that they made available would be of great relevance for knowledge discovery and also for the development of new RNA-based drugs. However, these sources adopt different data models, formats, and conventions for the representation of the bio-entities, and different semantics can be assigned to the proposed interactions. 
The extraction and integration of information from even two data sources for conducting knowledge discovery activity would require a lot of effort from researchers.   
To address these issues, KGs \cite{hogan21} have emerged as a compelling abstraction for organizing interrelated knowledge in different domains and a way for integrating heterogeneous information extracted from multiple data sources with the aim of highlighting complex interdependencies and uncovering hidden relationships. KGs can be represented both with property graphs (e.g., Neo4j~\cite{neo4j}) or according to the Resource Description Framework (RDF~\cite{rdf}) 
with different advantages and disadvantages \cite{Alocci2015PropertyGV}.  When a KG is generated according to an ontology, it contains a schema part (denoted TBOX or terminologies) and a data part (denoted ABOX, facts, or assertions) on top of which different kinds of reasoning activities can be conducted using expressive languages (like OWL~\cite{owl}, DL~\cite{dl}, or SPARQL~\cite{sparql}).
KGs have started to play a central role also in the life sciences \cite{chen2023kg} for the representation of bio-entities and their interactions and for the application of AI approaches for discovering new knowledge and eventually for explaining it. Different ontologies have been proposed for systematizing the corpus of terms used to describe the function and localization of bio-entities and for offering a formal framework to represent biological knowledge. Specific biological KGs (e.g., PrimeKG \cite{Chandak23}, Human Disease benchmark KG \cite{callahan23}, ReproTox-KG \cite{Evangelista2023}, Monarch Knowledge Graph~\cite{shef20}, and Knowledge Base of Biomedicine~\cite{kabob}) have been recently constructed for conducting different kinds of analysis and supporting the research activities.  

In this paper we describe {\em \rnakg}, the first ontology-based knowledge graph for representing coding and non-coding RNA molecules and their interactions with other biomolecular data as well as with pathways, abnormal phenotypes and diseases to support the study and the discovery of the biological role of the ``RNA-world''.
\rnakg contains RDF triples extracted from more than 50 public data sources and also integrates related bio-medical concepts. \rnakg can be exploited for the study of RNA molecules and the development of innovative graph algorithms to support knowledge discovery in data science. A big effort has been dedicated to the characterization of the data sources and to the identification of the bio-medical ontological concepts that better represent the information provided by the considered data sources and the interactions involving RNA molecules.  This work culminated in the construction of a meta-graph that represents all the possible interactions that can be devised from the considered data sources and that can be represented by means of the Relation Ontology (RO \cite{ro}), which ensures common semantics for the different relationships that can be extracted from the sources. Relying on the generated meta-graph and exploiting the Phenotype Knowledge Translator (\pheno~\cite{callahan23}) tool, we extracted 578,384 nodes and 8,768,582 edges of good quality according to the metrics provided in each data source. 
We also experimentally evaluated the main statistical and topological characteristics of the generated KG. \rnakg can be exported according to different knowledge models and can be accessed through a SPARQL endpoint.   

\subsection*{Related Work} 
For a better understanding of the approach that we have followed in the construction of \rnakg, we first outline the methods developed for integrating graph-based biomedical heterogeneous data sources and then summarize the main characteristics of the different types of RNA molecules.
Finally, we outline the bio-ontologies that can be exploited for the characterization of RNA molecules and the bio-entities with which they are related. 

\subsubsection*{Approaches for the construction of bio-medical knowledge graphs} 

The data integration issue is a well-known problem in data management and many approaches have been devised to deal with relational data \cite{Halevy2009}. However, the explosion of data formats (like CSV, JSON, XML) and the variability in the representation of the same types of information \cite{Mesiti,BCM21} has pushed the need to exploit ontologies as global common models both for accessing (OBDA – Ontology-Based Data Access) and integrating (OBDI – Ontology-Based Data Integration) data sources \cite{Poggi2008}. 

In OBDA, queries are expressed in terms of an ontology, and the mappings between the ontology and the data sources' schema are described in the form of declarative rules. Two approaches are usually proposed to enable access and integration of different data sources: {\em materialization}, where data are converted from the local data format according to the ontology concepts and relationships; {\em virtualization}, where the transformation is executed on the fly during the evaluation of queries by exploiting the mapping rules and the ontology. In this case, only the data from the original sources involved in the query are accessed. 
Materialization can provide fast and accurate access to data because already organized in a centralized repository. However, data freshness can be compromised when data sources frequently change.
On the other hand, virtualization allows access to fresh data but requires the application of transformations during query evaluation and can cause delay, and inconsistency when the structures of the local sources change. Several approaches are available for the specification of mapping rules like R2RML \cite{r2rml} (a W3C standard for relational to RDF mapping), and RML \cite{dimouldow2014} that extends the standard for dealing with other formats. Moreover, SPARQL-Generate \cite{sparkGen2017}, YARRRML \cite{yarrrml}, and ShExML \cite{ShExML} were also proposed for dealing with data heterogeneity.

In the biological context, many efforts are nowadays devoted to the construction of KGs by integrating different public sources that exploit the materialization and virtualization approaches previously described. 
For instance, Zhang and colleagues~\cite{Zhang21} applied a Connecting Ontology ($CO$) to integrate all external ontologies that describe the data sources involved. By exploiting algorithms for fusing and integrating annotations, an enriched KG is obtained that spans multiple data sources and is annotated by the integrated biological ontology obtained by gluing together the Gene \cite{go}, Trait \cite{Pan2019}, Disease \cite{do}, and Plant \cite{Cooper2016} ontologies.  
The Precision Medicine KG  (PrimeKG)~\cite{Chandak23} was developed to represent holistic and multimodal views of diseases. PrimeKG integrates more than 20 high-quality resources with more than 4 million relations that capture information like disease-associated perturbations in the proteome, biological processes, and molecular pathways. The considered data were collected and annotated using diverse ontologies such as Disease Gene Network, Mayo Clinical Knowledgebase, Mondo, Bgee, and DrugBank. 
ReproTox-KG \cite{Evangelista2023} combines information about genes, drugs, and preclinical small molecules with knowledge about the association of genes and drugs with birth defects with the aim of predicting the likelihood that preclinical compounds induce specific birth abnormalities, and whether these compounds are likely to cross the placental barrier. The information is extracted from scientific publications by considering several ontologies including HPO \cite{hpo}, CDC birth-defect terms \cite{cdc}, Geneshot \cite{Geneshot19} for connecting genes with birth-defect terms, DrugCentral \cite{drugcentral} for connecting drugs with birth-defect terms, and LINCS L1000 data \cite{LINCS} for drug–gene associations. 
Sima and colleagues \cite{Sima2019} proposed a virtualization approach based on an ontology-based federation of three data sources (Bgee, OMA, and UNIProtKB), i.e., starting from the GenEx semantic model for gene expression, the authors proposed mapping rules to deal with the different formats of the three sources and faced the issue of joint queries across the sources by leveraging SPARQL endpoints. A preliminary version of the \rnakg meta-graph \cite{iwbbio} was recently presented and here deeply enhanced with the description of the methodology and the generation of \rnakg. 
A fully automated Python 3 library named \pheno was recently proposed for the construction of semantically rich, large-scale biomedical KGs that are Semantic Web compliant and amenable to automatic OWL reasoning, and conform to contemporary property graph standards. The library offers tools to download data, transform and/or pre-processing resources into edge lists, construct knowledge graphs, and generate a wide-range of outputs~\cite{callahan23}. 

All these papers point out the difficulties that arise when trying to integrate different data sources that exploit different data models, formats, and ontologies. Specifically, data redundancies, data duplicates, and lack of common identifier mechanisms must be properly addressed.  
In the case of RNA data integration, we also have to consider the lack of specific ontologies for the description of all possible non-coding RNA sequences, and the presence of ontologies that are not well-recognized by the community because still in their infancy. All these aspects must be properly addressed in the generation of \rnakg.
 
\subsubsection*{RNA molecules}\label{sec:RNA}

\begin{figure}[htbp]
    \centering
\includegraphics[width=0.9\textwidth]{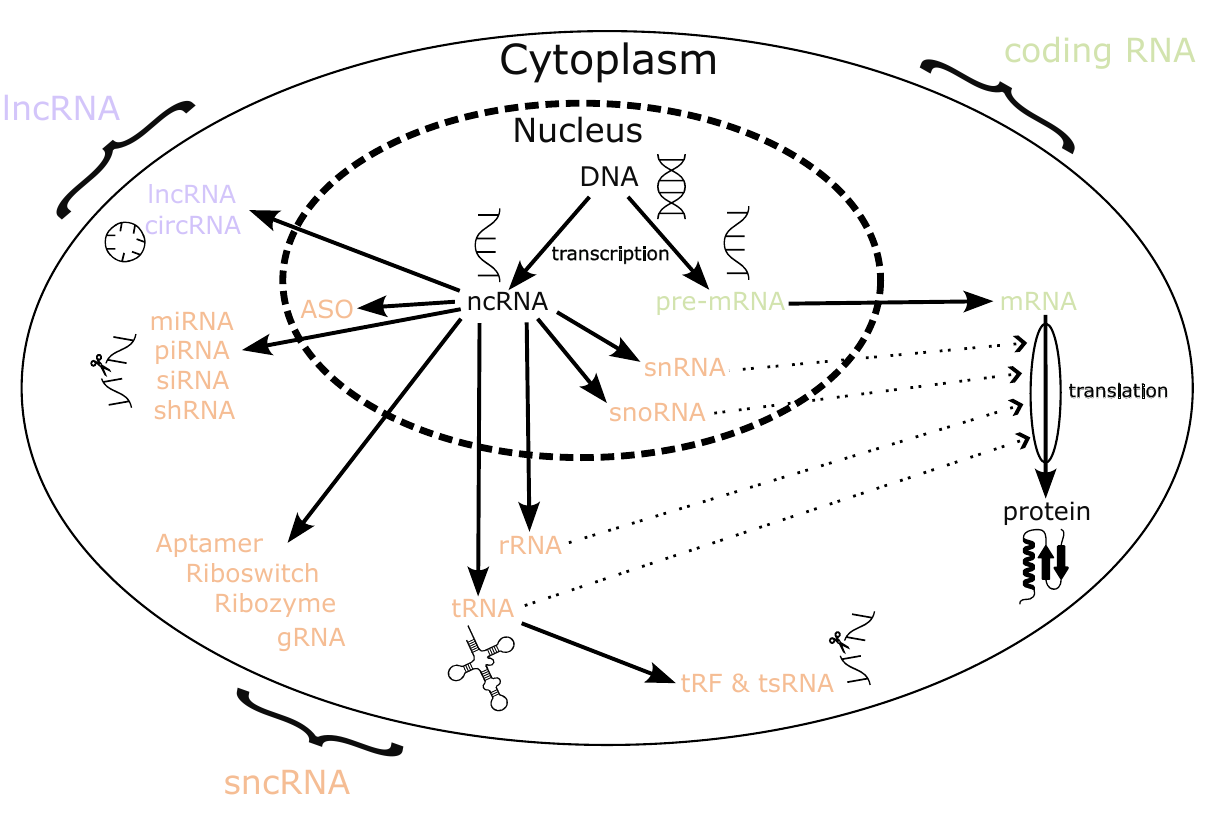}
\caption{Schematic representation of the RNA network within a cell.}
    \label{fig:RNAClassification}
\end{figure}

The wide variety of RNA molecules, which can be classified as sketched in Figure~\ref{fig:RNAClassification}, can be translated into proteins, can regulate gene expression, have enzymatic activity, and can modify or regulate other RNAs. 

\noindent {\bf Coding RNA.}
In Eukaryotes, messenger RNA (mRNA) primary transcripts follow a cascade of biological processes to transform them into mature functional mRNA molecules that are read by ribosomes, translated into amino acid chains and finally assembled in proteins through peptide bonds~\cite{vorlaender22}.

\noindent {\bf Non-coding RNA.}
Transcripts that are not translated into proteins are named non-coding RNAs (ncRNAs). They can be further classified into long non-coding RNAs (lncRNAs -- with more than 200 nucleotides) and small non-coding RNAs (sncRNAs -- with less than 200 nucleotides)~\cite{Hombach16}. lncRNAs are the majority of transcription products and play a pivotal role in disease development and progression \cite{Mattick23}. Circular RNAs (circRNAs) are lncRNAs produced from alternative splicing events, and may play a role as splicing event regulators. circRNAs have been involved in many human diseases, including cancer and neurodegenerative disorders such as Alzheimer’s and Parkinson’s disease, due to their aberrant expression in different pathological conditions \cite{CircRNAs2021}.

\noindent {\bf Small non-coding RNA (sncRNA).}
sncRNAs are involved in several cellular biological processes, including: translation processes; RNA interference (RNAi) pathways; splicing and self-cleavage processes; catalysis of biochemical reactions, and targeted gene editing.

\noindent {\bf sncRNAs involved in the translation process.}
Several sncRNAs are involved in this process, including ribosomal RNAs (rRNAs), transfer RNAs (tRNAs), small nuclear RNAs (snRNAs), small nucleolar RNAs (snoRNAs), and Small Cajal body-specific RNAs (scaRNAs, snoRNAs specific to the Cajal body). While rRNA constitutes the core structural and enzymatic framework of the ribosome, tRNAs are characterized by a structure consisting of an acceptor stem that links to a particular amino acid and of a specific anticodon sequence of 3 bases complementary to the corresponding mRNA codon, thus assuring the translation from the mRNA codon triplets to the corresponding sequence of amino acids.
snRNAs and snoRNAs primarily guide chemical modifications of other RNAs, mainly rRNAs and tRNAs, and control chromatin compaction and accessibility.

\noindent {\bf sncRNAs associated with RNA interference pathways.}
RNA interference pathways play a central role in gene expression and their misregulation is associated with several diseases~\cite{Hannon02}.
sncRNAs associated with RNAi pathways include: microRNAs (miRNAs), short interfering RNAs (siRNAs), short hairpin RNAs (shRNAs), antisense oligonucleotides (ASOs), piwi-interacting RNAs (piRNAs), tRNA-derived fragments (tRFs), and tRNA-derived small RNAs (tsRNAs). 
Mature miRNAs, siRNAs, shRNAs, and ASOs regulate the mRNA expression  by blocking translation or promoting degradation of the target mRNA (complementary base pairings).  Unlike siRNAs, each miRNA can simultaneously regulate the expression and the activity of hundreds of protein-coding genes and Transcription Factors (TFs). 
miRNAs from various exogenous sources, which are present in human circulation, are named  xeno-miRNAs \cite{Xeno-miRNA2020}. 
By contrast, ASOs are more effective to knock down nuclear targets, whereas siRNAs are superior at suppressing mRNA cytoplasmic targets by recruiting, via Watson-Crick base paring, the RNA-induced Silencing Complex (RISC) that catalyzes the mRNA cleavage. Similarly, throughout RNAi, piRNAs and tRFs promote genome integrity, avoiding potential threats to cellular homeostasis, by silencing transposons, retrotransposons and repeat sequences~\cite{yu20}.
 
\noindent {\bf Aptamers, riboswitches, ribozymes and guide-RNA.}
The tertiary structure of RNA sequences can also be investigated to identify interferences. Aptamers are short single-stranded nucleic acids that can bind to a variety of targets (e.g., proteins, peptides, carbohydrates, DNA, and RNA) thanks to their 3D conformation.
Riboswitches are small non-coding RNAs involved in alternative splicing and self-cleavage processes that cause gene expression control and mRNA degradation, critical for survival of the cell 
\cite{Machtel2016}.
Some RNAs, such as ribozymes, even possess enzymatic activity therefore catalyzing biochemical reactions (e.g., mRNA and protein cleavage). Synthetic ribozymes can and have already been designed to target viral RNA.
Synthetic guide RNAs (gRNAs) are usually involved in the application of CRISPR-Cas9 technique, used for gene editing and gene therapy \cite{CRISPR16}.

\subsubsection*{Biomedical ontologies for the semantic characterization of \rnakg}
Several standard biomedical ontologies can be used to set up common semantics in the considered data sources. Table~\ref{tab:ontologies} shows those considered during \rnakg construction (their specifications are made available in the web portals {\tt ebi.ac.uk/ols4} and {\tt bioportal.bioontology.org}).
We selected these ontologies because their terms and hierarchical structures are commonly accepted by the scientific community to unequivocally describe biological classes and entities such as diseases, phenotypes, chemicals, biological processes, proteins, and relations between them. In the case of \rnakg, we have also taken into account the lack of specific ontologies for the description of all possible RNA sequences (especially non-coding ones), and the presence of bio-ontologies that are yet not well-recognized by the community.

\subsection*{Results}

\rnakg is open-source and available through a SPARQL endpoint (hosted at \url{http://fievel.anacleto.di.unimi.it:9999}). The code for generating and maintaining \rnakg is available on GitHub (\url{https://github.com/AnacletoLAB/RNA-KG}).
\rnakg comprises a comprehensive collection of relationships involving RNA molecules from various recognized RNA sources. The considered data sources have been characterized from different perspectives and compared (see Tables \ref{tab:database1}-\ref{tab:database2}). These tables also provide the different kinds of relationships that can be established and their respective occurrences are summarized in Figure \ref{fig:RNAMoleculesRelations}. Data sources might contain similar relationships and their possible overlapping has been visually highlighted in Figure \ref{fig:venn}. Relationships have been represented according to the Relation Ontology (RO) and the resulting meta-graph (Figure \ref{fig:metagraph}) has been used for the generation of an ontological description of \rnakg. Different analyses have been conducted to characterize the types of nodes and interactions that are represented in \rnakg (Figure \ref{fig:pie+inter}), their distribution (Figures \ref{fig:nodes_count}-\ref{fig:edges_count}), and the topological structure analysis (Table \ref{tab:basic_props}). Details on the obtained results are discussed in the {\em Data Records} section.

The methodology we employed to construct RNA-KG enabled us to generate a high-quality knowledge graph that includes reliable interactions, validated through experimental methods and/or strongly endorsed by data providers, and whose meaning was meticulously verified to ensure a consistent representation of domain knowledge.

In the supplementary section, we have also included further figures and tables that better describe \rnakg. Specifically, Supplementary Table \ref{tab:databasesont2} delineates the bio-ontologies that have been exploited for representing concepts in each RNA source. Supplementary Tables \ref{tab:stats1}-\ref{tab:stats4} report the descriptive statistics of the triples that have been extracted from the different data sources. Moreover, Supplementary Figures~\ref{fig:RNADBxMolecule}-\ref{fig:RNADBxRO} show bio-entities present in RNA sources and their mapping to RO terms we used to represent relationships within the sources. Supplementary Table \ref{tab:nodeTypes} highlights the primary node types and their corresponding identifiers with an instance sample.
Finally, Supplementary Listings~\ref{lst:sparql1}-~\ref{lst:sparql3} provide a few examples of SPARQL queries, illustrating the kind of information that can be extracted from \rnakg.

\subsection*{Discussion}
The structure of the obtained KG has been analyzed 
by computing various graph metrics that provide a macroscopic description of the network topology derived from the entity relations (see Table \ref{tab:basic_props}). By means of a t-SNE representation of an embedding of the nodes/edges in \rnakg (Figure~\ref{fig:grape}) we have highlighted that the embedding of the node type is able to effectively identify the similarities among the nodes of the same type, thus capturing their function in the network. Moreover, the edge embedding is able to capture, in most cases, the similarity between edges. 
In our analysis we have also identified the nodes having the highest out-degree and observed that the (undirected) degree distribution follows a heavy-tailed distribution (see Figure~\ref{fig:degree_distribution}).
Moreover, we computed bounds for the treewidth and the closeness centrality distribution, which suggest a sparse yet well-connected structure.
All of these characteristics are commonly found in networks with a scale-free structure, which is often seen in networks representing complex interactions.
Further investigation will be conducted in this direction to confirm such a structure. This property would yield benefits in the network analysis both in terms of structure understanding and algorithmic design. 
Finally, we have investigated the presence of isomorphic node groups, that is nodes that are topologically indistinguishable because they present exactly the same neighbors. These nodes deserve further investigation to check whether the involved molecules are duplicates and should thus be collapsed to improve the information quality of RNA-KG.  A detailed discussion of the results is reported in Section {\em Technical Validation}.  

\rnakg can generate heterogeneous biomedical graphs in different formats that can be processed by graph-based computational tools to infer biomedical knowledge, provide insights into biomolecular mechanisms and biological processes underlying diseases, support the discovery of new drugs, especially those based on RNA, and evaluate biomedical hypotheses in silico.
In particular, \rnakg is specifically designed to deal with computational tasks involving RNAs, by e.g. exploiting the information about ncRNA interactions for gene and protein expression regulation, collected from tens of publicly available databases. By leveraging the biomedical concepts represented in the biomedical ontologies embedded in the KG, \rnakg can be also analyzed to predict associations and causal relationships of the "RNA world" with diseases and abnormal phenotypes. We also observe that the rich information embedded in the \rnakg can be leveraged for classical biomedical prediction tasks, including e.g. gene-disease prioritization, drug-target prediction, and drug repurposing.

Most of these biomedical tasks can be modeled as link or node-label prediction problems in heterogeneous graphs.
Even if, in principle we could apply methods developed for homogeneous graphs~\cite{Hamilton17-rev}, to leverage the rich information scattered across the different types of modes and edges of the \rnakg, we suggest applying methods specifically designed for heterogeneous graphs~\cite{Yang22}. To this end, several AI graph-based methods have been recently proposed to deal with heterogeneous graphs, also in the context of biomedical Knowledge Graphs~\cite{johnson23}.
In particular we foresee that Graph Representation Learning methods, by leveraging the topology of the complex bio-medical heterogeneous graphs to embed them into compact vectorial spaces, could be the most promising choice to properly analyze the complex heterogeneous structure of \rnakg~\cite{Li22}.

We are currently working on enhancing the proposed KG in different directions. First, we are identifying key properties associated with RNA molecules and their interactions to be included in \rnakg. This is a tough problem because we need to integrate and make uniform the information about the same bio-entity obtained from different sources.
However, we have already identified the main properties characterizing the same molecule. Moreover, we have established systematic rules for aligning the representation and fusing records representing the same bio-entities. In this line of research, we are also considering the possibility of proposing an RNA Ontology for describing RNAs with a particular emphasis on non-coding RNA molecules. We aim to craft this ontology on the data that we have identified in the sources considered in this study. This could be a relevant contribution toward the definition of a standard representation of RNA molecules and the interactions that might be determined among them.
Another research direction is the use of Transformers~\cite{transformer} in combination with \rnakg for the extraction of new triples from textual documents. In this direction, we are currently leveraging the SPIRES engine~\cite{spires} which enables the user to guide OntoGPT on the kinds of interactions that we are interested in, by specifying our meta-graph in terms of LinkML~\cite{linkml}. The identified triples can be validated by using RNA-KG as a gold standard in this domain. This approach can limit the hallucination issue that is typical of large language models, ensuring the reliability and accuracy of the extracted information. 
We would like also to analyze \rnakg with cutting-edge AI graph representation learning algorithms~\cite{Xia21} to support the discovery of novel RNA drugs. 
Finally, we are developing graphical facilities for supporting the user in the data acquisition and maintenance processes and thus reducing the manual effort required for mapping the data available in the different data sources into \rnakg~\cite{dataplat} and also for showing the graph at different granularities~\cite{MPP23,PFB20}.

\section*{Methods}

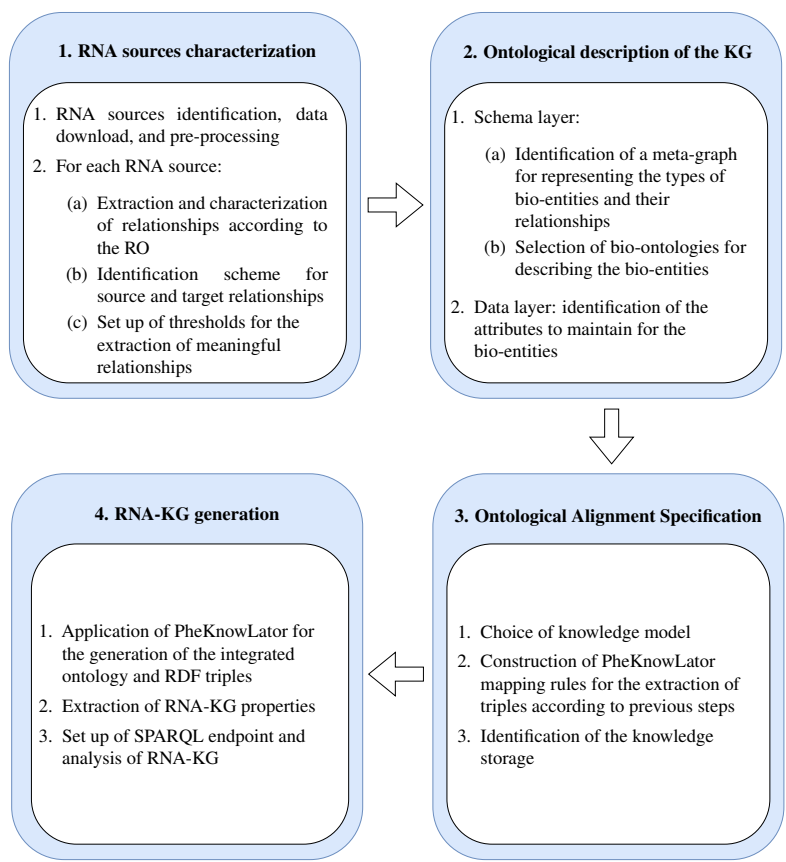
\begin{figure}[htbp]
    \centering
    \begin{adjustbox}{max width=0.6\textwidth}{\definecolor{c6c8ebf}{RGB}{108,142,191}
\definecolor{cdae8fc}{RGB}{218,232,252}

\def \globalscale {1.000000}
\begin{tikzpicture}[y=1cm, x=1cm, yscale=\globalscale,xscale=\globalscale, inner sep=0pt, outer sep=0pt]
  \path[draw=c6c8ebf,fill=cdae8fc,rounded corners=0.9922cm] (0.0, 7.329) rectangle (6.6146, 0.0529);

  \begin{scope}[]
    \node[text=black,anchor=center] (text7552) at (3.35, 6.6){\textbf{1. RNA sources characterization}};

  \end{scope}
  \path[draw=black,fill=white,rounded corners=0.8493cm] (0.2646, 6.006) rectangle (6.35, 0.344);

\begin{scope}[]
  \node[text=black, anchor=center] (text4252) at (3, 3.0692) {
        \begin{minipage}{6cm}
            \begin{enumerate}
            \setlength\itemsep{0em}
                \item RNA sources identification, data download, and pre-processing
                \item For each RNA source:
                \begin{enumerate}
                \setlength\itemsep{0em}
                    \item Extraction and characterization of relationships according to the RO
                    \item Identification scheme for source and target relationships
                    \item \begin{flushleft} Set up of thresholds for the extraction of meaningful relationships \end{flushleft}
                \end{enumerate}
            \end{enumerate}
        \end{minipage}
    };
\end{scope}
  \path[draw=c6c8ebf,fill=cdae8fc,rounded corners=0.9922cm] (7.9375, 7.329) rectangle (14.5521, 0.0529);

  \begin{scope}[]
    \node[text=black,anchor=south west] (text7784) at (8.56, 6.4){\textbf{2. Ontological description of the KG}};

  \end{scope}
  \path[draw=black,miter limit=10.0] (6.7601, 3.5666) -- (6.7601, 3.8312) -- (7.2893, 3.8312) -- (7.2893, 4.109) -- (7.792, 3.6989) -- (7.2893, 3.2888) -- (7.2893, 3.5666) -- cycle;

  \path[draw=black,fill=white,rounded corners=0.8493cm] (8.2021, 6.006) rectangle (14.2875, 0.344);

  \begin{scope}[]
    \node[text=black,anchor=center] (text4692) at (10.88, 3.225) {
        \begin{minipage}{6cm}
            \begin{enumerate}
            \setlength\itemsep{0em}
                \item Schema layer:
                \begin{enumerate}
                \setlength\itemsep{0em}
                    \item \begin{flushleft} Identification of a meta-graph for representing the types of bio-entities and their relationships
                    \end{flushleft}
                    \item Selection of bio-ontologies for describing the bio-entities
                \end{enumerate}
                \item \begin{flushleft} Data layer: identification of the attributes to maintain for the bio-entities \end{flushleft}
            \end{enumerate}
        \end{minipage}
    };
    \end{scope}

\begin{scope}[xshift=7.65cm,yshift=14.6cm]
\path[draw=black, miter limit=10.0] (3.5666, -14.7505) -- (3.8312, -14.7505) -- (3.8312, -15.2797) -- (4.109, -15.2797) -- (3.6989, -15.7824) -- (3.2888, -15.2797) -- (3.5666, -15.2797) -- cycle;
\end{scope}

\begin{scope}[xshift=-15.8852cm,yshift=-8.7cm]

  \path[draw=c6c8ebf,fill=cdae8fc,rounded corners=0.9922cm] (15.9279, 7.329) rectangle (22.5425, 0.0529);

  \begin{scope}[]
    \node[text=black,anchor=center] (text8586) at (19.2352, 6.55){\textbf{4. RNA-KG generation}};
  \end{scope}

  \path[draw=black,fill=white,rounded corners=0.8493cm] (16.3, 6.006) rectangle (22.2779, 0.344);

  \begin{scope}[]
    \node[text=black,anchor=center] (text9550) at (19, 3.2) {
        \begin{minipage}{6cm}
            \begin{enumerate}
            \setlength\itemsep{0em}
                    \item \begin{flushleft} Application of PheKnowLator for the generation of the integrated ontology and RDF triples
                    \end{flushleft}
                    \item Extraction of RNA-KG properties
                \item \begin{flushleft} Set up of SPARQL endpoint and analysis of RNA-KG \end{flushleft}
            \end{enumerate}
        \end{minipage}
    };

  \end{scope}
  \path[draw=c6c8ebf,fill=cdae8fc,rounded corners=0.9922cm] (23.8654, 7.329) rectangle (30.48, 0.0529);

  \begin{scope}[]
    \node[text=black,anchor=center] (text6013) at (27.1727, 6.5){\textbf{3. Ontological Alignment Specification}};
  \end{scope}
  
  \begin{scope}[xshift=46.375cm,yshift=7.25cm]
  \path[draw=black,miter limit=10.0, rotate=180] (22.688, 3.5666) -- (22.688, 3.8312) -- (23.2172, 3.8312) -- (23.2172, 4.109) -- (23.7199, 3.6989) -- (23.2172, 3.2888) -- (23.2172, 3.5666) -- cycle;
  \end{scope}

  \path[draw=black,fill=white,rounded corners=0.8493cm] (24.13, 6.006) rectangle (30.2154, 0.344);

  \begin{scope}[shift={(-0.0132, 0.5132)}]
    \node[text=black,anchor=south west] (text6395) at (23.9, 1.26){
        \begin{minipage}{6cm}
            \begin{enumerate}
            \setlength\itemsep{0em}
                \item Choice of knowledge model
                \item \begin{flushleft} Construction of PheKnowLator mapping rules for the extraction of triples according to previous steps
                \end{flushleft}
                \item \begin{flushleft} Identification of the knowledge storage
                \end{flushleft}
            \end{enumerate}
        \end{minipage}
    };

  \end{scope}
  
\end{scope}
\end{tikzpicture}}
    \end{adjustbox}
    \caption{Workflow for the construction of \rnakg.}
    \label{fig:workflow}
\end{figure}

The creation of a KG is a complex task that requires facing several phases that can be organized in a workflow such as the one reported in Figure~\ref{fig:workflow}.
In the remainder of the section, we provide a detailed description of the different issues and adopted solutions for each phase of the creation of our RNA-based KG.

\subsection*{RNA sources characterization}
In this phase, we have identified and analyzed the characteristics of relevant data sources from which the information for feeding the KG has been extracted. This is a well recognized critical initial step in constructing a KG \cite{wang21}.

To this aim, an extensive literature review was carried out to identify repositories dealing with RNA sequences and annotations developed by well-reputed organizations, published in top journals, periodically updated, and containing significant amounts of RNA molecules and relevant relationships with other types of molecules and bio-entities.
Furthermore, RNA sources that are included in other bigger repositories have been considered, as well as those that are used as collectors of other repositories.
We have also identified the presence of controlled vocabularies, thesauri, reference ontologies that formally describe the repository content, and the presence of well-recognized identification schemes. 

Sources provide data in different formats (e.g., CSV, TSV, gaf, hpoa, reactome, xlsx, JSON, and HTML) or by issuing queries on content management systems. 
Once the data were downloaded, Pandas \cite{pandas} DataFrames were used to transform the data into a common format (TSV files) and remove syntactic inconsistencies. The data obtained can be then processed through \pheno to extract the relationships among bio-entities. 

For the characterization of the relationships that can be extracted from the different data sources, we applied the Relation Ontology. Moreover, the hierarchical organization of concepts in RO allows the expression of different kinds of relationships at different granularities (e.g., the general term {\tt interacts with}  can be substituted with more specific terms such as {\tt molecularly interacts with} or {\tt genetically interacts with}).
Moreover, in case of a lack of specific terms for describing relationships identified in a data source, two strategies have been devised. 
The first one is to approximate the concept/relationship type with a term already present in RO. 
This implies a better coherence with RO semantics but misses details that the concept/relationship can carry. 
The second strategy is to extend the RO ontology with new terms specifically tailored for the representation of the concept/relationship. In this case, non-standard, more precise terms can be introduced for the representation of the concept/relationship. In the construction of \rnakg, we adopted the first strategy for a 
larger agreement on the meaning of the used terms
and uilized the {\tt interacts with} relationship for representing any connection among bio-entities when a more specific one is missing.

\begin{figure}[htbp]
    \centering 
    \begin{adjustbox}{max width=0.8\textwidth}{\input{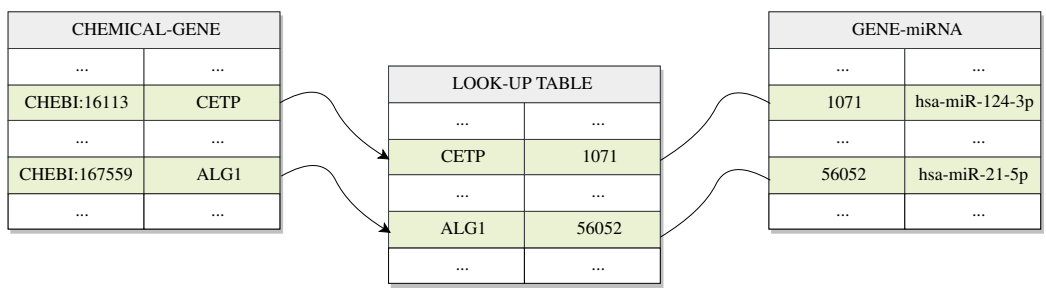}}
    \end{adjustbox}
    \caption{The relationship between chemicals and miRNA cannot be decoded directly because of the use of different identification schemes. However, by means of a look-up table the relationship can be highlighted. 
    }
    \label{fig:lutable}
\end{figure}

The adoption of different types of molecular identifiers represents another issue.
Indeed, the identification scheme encountered in the considered data may vary from the source and target of the relation and could be characterized by different accuracy levels.
Four levels have been detected: {\em Well-Reputed} (denoted {\tt WR}), when the identifiers are widely accepted by the scientific community (e.g., NCBI Entrez Gene identifiers); 
{\em Ontology-based} (denoted {\tt O}), when the identifiers are directly represented with ontological terms; {\em Mapping-based} (denoted {\tt M}), when the identifiers can be obtained by exploiting look-up tables; and {\em Proprietary} (denoted {\tt P}), when all the previous techniques cannot be applied. 
Once the identification scheme adopted in a source has been classified, appropriate {\em look-up} tables for their mapping into the reference ontology have been realized by analyzing synonyms in the reference ontology or by examining the ones provided by the sources themselves to facilitate interoperability with other sources dealing with the same entities.
For instance, we employed \textit{NCBI Gene Entrez} identifiers to represent genes in \rnakg, but many sources provide the correspondent \textit{Gene Symbol}. 
In this case, a look-up table has been used to map gene identifiers into the chosen representation (Figure~\ref{fig:lutable}).

To guarantee a high level of homogeneity in the KG, a few tuples have been omitted when the mapping to the reference ontology was not possible. For some types of RNA molecules (especially ncRNA sequences), the look-up tables cannot be adopted because of the lack of a reference ontology with which these molecules can be represented.
In these cases, the {\em NCBI Entrez Gene} identifiers of the gene from which the specific RNA is transcribed have been extended with a suffix that corresponds to the type of non-coding RNA (e.g., in case of small nucleolar RNA molecules the suffix is {\tt ?snoRNA}). We remark that the lack of a common ontological representation among heterogeneous sources can cause the duplication of molecules. At the current stage of development, we decided to admit the presence of duplicates for this kind of molecule, but we will consider de-duplication techniques~\cite{Altschul1990,Pearson1988} that rely on the use of similarity measures on the molecule sequences in future releases.

To guarantee a high-level of reliability of the relationships to be included in \rnakg, only {\em meaningful} relationships have been considered, that is those satisfying constraints that take into account p-values or FDR -- False Discovery Rate (e.g., $p_{val}<0.01$), experimental validation of results, or scores (denoted with $\sigma$) defined as reliable in the considered data source.

All of these activities led to the identification of the reference ontologies for each data source and the relationships that can be extracted and represented according to the RO ontology. 
Moreover, by studying the organization and format of the data sources, the extraction patterns and the look-up tables to apply have been realized. 
Finally, pruning strategies have been devised specifically tailored to the characteristics of each data source. 

\subsection*{Ontological description of the KG}
In this phase, we identified the classes of bio-entities that need to be managed and of the kinds of relationships that can exist among them ({\em schema layer}). 
Moreover, specific instances and the properties that need to be maintained have been identified ({\em data layer}). 
This design activity plays a fundamental role in the hierarchy, structure and content filling of the knowledge graph, and it is the basis for determining the kind of reasoning that can be supported.

Starting from the knowledge gained from the characterization of RNA sources, we moved toward the construction of the ontological schema underlying \rnakg. A {\em meta-graph} was built to include all the kinds of bio-entities and relationships between them outlined in the previous phase. 
The meta-graph provides both direct and inverse relationships that are considered to guarantee bi-directional navigation of the generated KG.

Once classes of bio-entities and their relationships have been identified, we determined the properties that should be kept for them. At the current stage, only fundamental properties of bio-entities have been collected (identifiers, node types, and source provenance). This choice has the advantage of avoiding the explosion of the KG size. However, in future implementations, we wish to enhance the properties that can be stored within \rnakg.

\subsection*{Ontological alignment specification}
In this phase, we identified the KG representation and the kind of storage system to adopt.
RDF triples have turned out to be suitable because of their common, flexible, and uniform data model.
These properties result in an ontologically-grounded knowledge graph for conducting different kinds of analysis and reasoning.

\begin{figure}[htbp]
    \centering
    \includegraphics[width=0.5\textwidth]{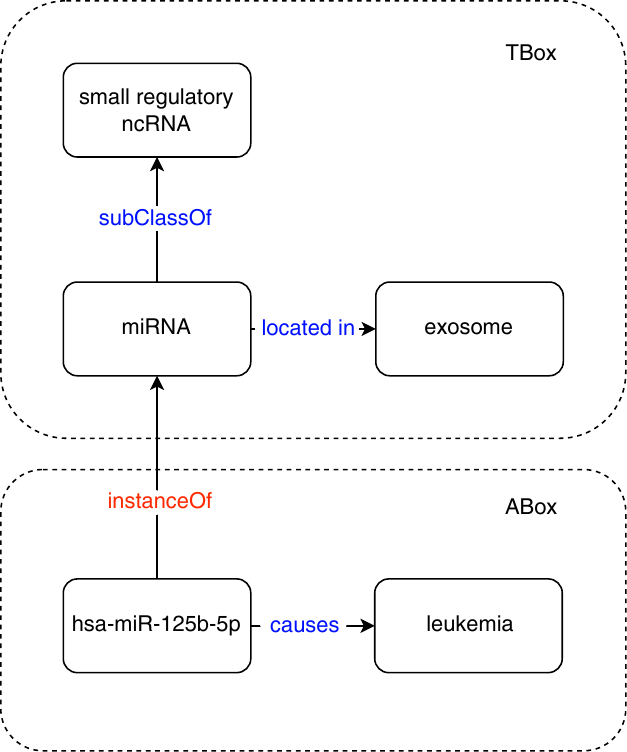}\caption{An example of the use of Description Logic (DL) for knowledge modeling. The TBox includes classes (i.e., {\tt miRNA}, {\tt small regulatory ncRNA}, and {\tt exosome}), and the assertions between classes (i.e., ``{\tt miRNA} {\tt subClassOf} {\tt small regulatory ncRNA}'' and ``{\tt miRNA} is {\tt located in} {\tt exosome}''). The ABox includes instances of classes (i.e., {\em hsa-miR-125b-5p}) represented in the TBox and assertions about those instances (i.e., ``{\em hsa-miR-125b-5p} {\tt instanceOf} miRNA'' and ``{\em hsa-miR-125b-5p} {\tt causes} leukemia'').
    }
        \label{fig:tabox}
\end{figure}

Since a standardized formal definition for the concept of a KG is still lacking, we considered the one
adopted by Callahan and colleagues~\cite{callahan23} where a KG is a pair $<T, A>$, where T is the TBox and A the ABox. The TBox represents the taxonomy of a particular domain including classes, properties/relationships, and assertions that are assumed to generally hold within a domain (e.g., a miRNA is a small regulatory ncRNA located in an exosome as depicted in Figure \ref{fig:tabox}). The ABox describes attributes and roles of class instances (i.e., individuals) and assertions about their membership in classes within the TBox (e.g., {\em hsa-miR-125b-5p} is a type of miRNA that may cause {\em leukemia}). Non-ontological entities (i.e., entities from a data source that are not compliant to a given set of ontologies such as RNA molecules) can be integrated with ontologies using either a TBox (i.e., class-based) or ABox (i.e., instance-based) knowledge model. For the class-based approach, each database entity is represented as {\tt subClassOf} an existing ontology class, while for the instance-based approach it is represented as {\tt instanceOf} an existing ontology class.

For the construction of the KG we have employed the \pheno ecosystem~\cite{callahan23} because it offers both approaches for the representation of bio-entities and their relationships, and also because of its simplicity in the identification of the columns containing the molecules' identifiers and for the specification of their relationships in terms of the RO ontology. 
\pheno also provides tools to easily generate the ontology that better describes the content of the KG that, besides the terms and relationships of the meta-graph, also includes other ontological terms for supporting the reasoning.

\begin{figure}[htbp]
    \centering
    \includegraphics[width=0.9\textwidth]{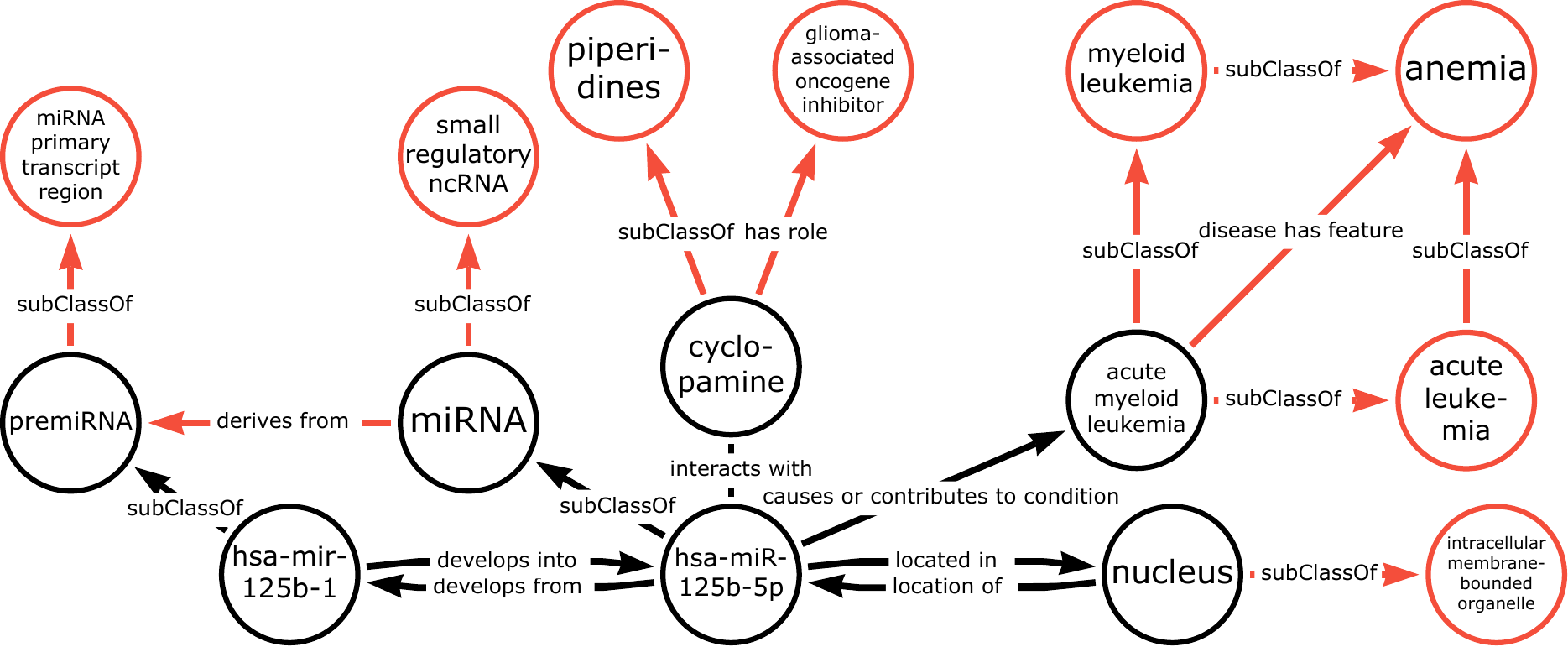}\caption{Example of a \rnakg subgraph realized according to the instance-based, inverse relations, semantically abstracted (OWL-NETS without harmonization) parameters.
    Black nodes and edges come from the integration of RNA  sources, whereas red ones from the integration of selected ontologies.}
        \label{fig:instancebased}
\end{figure}

KGs can be easily exported according to different kinds of models offered by \pheno depending on the analyses to be conducted. Even if \rnakg is made available in all the supported knowledge models, we think that the instance-based, inverse relation, semantically abstracted (OWL-NETS~\cite{owlnets} without harmonization) configuration is the most suitable to be processed by different kinds of ML algorithms for node and link prediction. 
This solution ensures that RNA molecules (which lack semantic characterizations in bio-ontologies) and other non-ontological data can be specified as {\tt subClassOf} specific ontological classes. Moreover, this approach enables the automatic specification of inverse relations among the involved bio-entities.
Lastly, OWL-NETS reversibly abstracts ontological biomedical knowledge into a network representation containing only biologically meaningful concepts and relations.
Figure~\ref{fig:instancebased} shows a small toy-example subgraph extracted from \rnakg according to the proposed set-up. We can notice the presence of inverse relationships ({\tt located in} and its inverse {\tt location of}), and the relation RDF {\tt subClassOf} connected to entities that do not have a corresponding term in a reference ontology (miRNA molecules are specified as {\tt subClassOf} the SO term {\tt miRNA}).

By studying the characteristics of the data sources, specific mapping rules have been devised through \pheno to extract triples compliant with the adopted ontologies.
Mapping rules contain the position of the source and object in the TSV file, the two human-readable labels for subject and object (e.g., mRNA and disease), the type of relationship that holds/exists between them according to RO (e.g., {\tt RO\_0003302} corresponds to {\tt causes or contributes to condition} relation), and further detailed options (e.g., thresholds for considering the tuple, row filtering options, transformation options according to the look-up table). These rules will be exploited for the extraction of the triples according to the reference ontology. 

Since many ontologies are used in our context, we adopted the \pheno tools to clean ontology files (i.e., remove and normalize errors, eliminate obsolete and/or deprecated entities, remove duplicate classes and class concepts) and merge cleaned ontology files into a single ontology file. The so-obtained merged ontology describes entirely the structure of \rnakg and is compliant with our meta-graph.

\subsection*{\rnakg generation}
In this final phase, the \pheno mapping rules have been issued on the pre-processed data with the aim of generating a KG compliant with the meta-graph identified in Phase 2 (ontological description of the KG).

In order to evaluate the characteristics of the generated KG, we used the GRAPE library that we recently developed for fast and efficient graph processing and embedding~\cite{grape}.
By importing \rnakg into the GRAPE environment, we were able to retrieve relevant topological information and topological oddities that can be useful in identifying biological and biomedical KG inconsistencies.
Moreover, GRAPE can be exploited to implement different types of graph embedding techniques that cannot be realized by means of other tools because of the size of the generated KG. 

Finally, a Blazegraph endpoint \cite{blazegraph} has been realized to make \rnakg freely available and accessible. Using SPARQL, it is possible to extract portions of the graph and use it for different kinds of analysis (see the examples reported in the Supplementary Listings~\ref{lst:sparql1}-~\ref{lst:sparql3}). Moreover, the entire \rnakg can be downloaded from our lab website.

\section*{Data Records}
\subsection*{Identification and characterization of RNA entities and relationships}
\label{sec:DB}

By applying the method discussed in the previous section, we identified more than 50 public repositories. 
The papers describing these data sources were published in top bioinformatics, bio-medical, and database focused journals 
between 2008 and 2023 but the majority were published in the last 5 years. 
The main characteristics of the identified repositories are reported in Tables~\ref{tab:database1}-\ref{tab:database2}, whose entries are organized according to the main type of RNA molecules made available by the source.  
Sources with {\tt miRNA} entities can contain hairpin miRNA, xeno-miRNA, and mature miRNA molecules (last ones, in turn, can be classified in {\tt -3p} and {\tt -5p} transcripts). {\tt Inter RNA} sources are those that do not focus on a single RNA type but propose multiple relationships among different types of RNA molecules and bio-entities (e.g., disease in the case of RNADisease or cellular component in the case of RNALocate). 
Note that no species is present for aptamers because they are synthetic and none of the databases are specific for piRNA molecules related to \textit{Homo sapiens} (although relationships involving piRNA sequences are stored in {\tt Inter RNA} sources).
Regarding the format, the majority of the considered data sources ($\geq$ $80$\%) export data in a flat-file format ({\em CSV}). Only a small fraction of them (around $20$\%) provide an API for accessing data stored in a relational database. Only DrugBank offers a RDF data representation coupled with a SPARQL endpoint.

\begin{figure}[htbp]
    \centering
    \definecolor{darkgray176}{RGB}{176,176,176}
\definecolor{gray}{RGB}{128,128,128}

\begin{tikzpicture}[font=\small]
\begin{axis}[
x=0.35cm,
y=0.35cm,
colorbar,
colorbar style={
    ylabel={Count},
    ytick={0, 0.1, 1, 2, 3, 4, 5, 6, 7, 8},
    yticklabels={$10^-1$,$10^0$, $10^1$, $10^2$, $10^3$,$10^4$, $10^5$, $10^6$, $10^7$,$10^8$},
},
colormap={mymap}{[1pt]
  rgb(0pt)=(1,0.96078431372549,0.941176470588235);
  rgb(1pt)=(0.996078431372549,0.87843137254902,0.823529411764706);
  rgb(2pt)=(0.988235294117647,0.733333333333333,0.631372549019608);
  rgb(3pt)=(0.988235294117647,0.572549019607843,0.447058823529412);
  rgb(4pt)=(0.984313725490196,0.415686274509804,0.290196078431373);
  rgb(5pt)=(0.937254901960784,0.231372549019608,0.172549019607843);
  rgb(6pt)=(0.796078431372549,0.0941176470588235,0.113725490196078);
  rgb(7pt)=(0.647058823529412,0.0588235294117647,0.0823529411764706);
  rgb(8pt)=(0.403921568627451,0,0.0509803921568627)
},
point meta max=8.16509841072963,
point meta min=0.301029995663981,
tick align=outside,
tick pos=left,
x grid style={darkgray176},
xmin=0, xmax=23,
xtick style={color=black},
xtick={0.5,1.5,2.5,3.5,4.5,5.5,6.5,7.5,8.5,9.5,10.5,11.5,12.5,13.5,14.5,15.5,16.5,17.5,18.5,19.5,20.5,21.5,22.5},
xticklabel style={rotate=90.0},
xticklabels={
  miRNA,
  mRNA,
  disease,
  chemical,
  lncRNA,
  circRNA,
  siRNA,
  shRNA,
  protein,
  aptamer,
  ASO,
  gRNA,
  gene,
  ribozyme,
  viral RNA,
  riboswitch,
  tRF \& tsRNA,
  snoRNA,
  pseudogene,
  rRNA,
  snRNA,
  tRNA,
  scaRNA
},
y dir=reverse,
y grid style={darkgray176},
ymin=0, ymax=23,
ytick style={color=black},
ytick={0.5,1.5,2.5,3.5,4.5,5.5,6.5,7.5,8.5,9.5,10.5,11.5,12.5,13.5,14.5,15.5,16.5,17.5,18.5,19.5,20.5,21.5,22.5},
yticklabels={
  miRNA,
  mRNA,
  disease,
  chemical,
  lncRNA,
  circRNA,
  siRNA,
  shRNA,
  protein,
  aptamer,
  ASO,
  gRNA,
  gene,
  ribozyme,
  viral RNA,
  riboswitch,
  tRF \& tsRNA,
  snoRNA,
  pseudogene,
  rRNA,
  snRNA,
  tRNA,
  scaRNA
}
]
\addplot graphics [includegraphics cmd=\pgfimage,xmin=0, xmax=23, ymin=23, ymax=0] {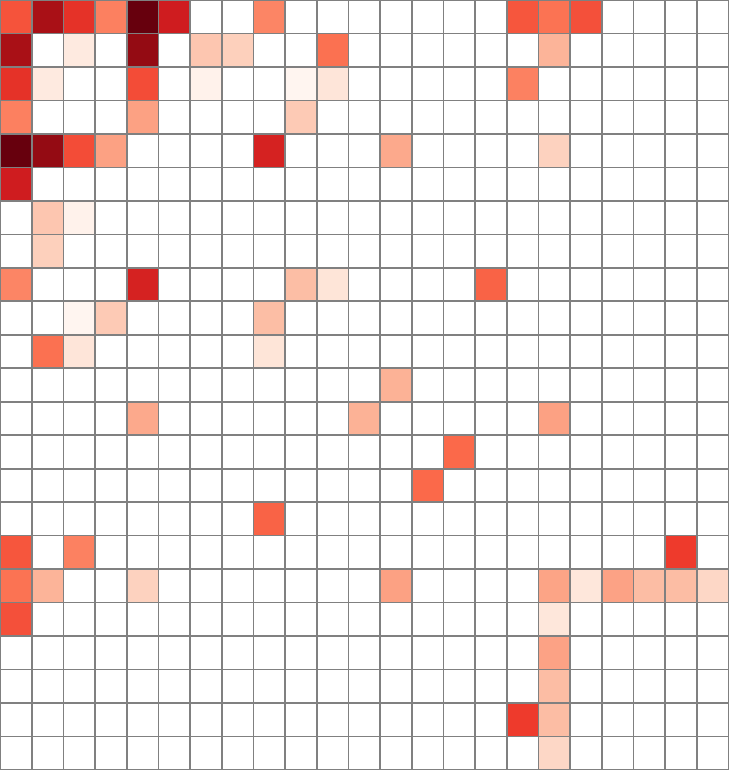};
\addplot [very thick, gray]
table {%
-4.44089209850063e-16 23
-4.44089209850063e-16 0
};
\addplot [very thick, gray]
table {%
23 23
23 0
};
\end{axis}
\end{tikzpicture}
    \caption{Number of relationships involving RNA molecules and relevant bio-entities (gene, protein, chemical, and disease) within the considered RNA sources. Colors represent the ranges of relationships in log scale, as reported in the legend.
\label{fig:RNAMoleculesRelations}}
\end{figure}

Figure~\ref{fig:RNAMoleculesRelations} summarizes the available relations involving RNA molecules and bio-entities (i.e., gene, protein, chemical, and disease) that we have identified in the different data sources. miRNA-lncRNA interactions are the most numerous. We can retrieve around 150 million distinct relationships of this type from public RNA-based data sources.
In terms of cardinality, they are followed by lncRNA-mRNA interactions ($\sim$28 million) and miRNA-mRNA interactions ($\sim$12 million).
Around 800 thousand distinct relationships can also be identified for protein-lncRNA interactions. These categories of molecules often interact with each other in specific diseases.
RNA aptamer-disease is the less represented one because at the current stage only two approved (or under-investigation) RNA aptamer drugs are present in DrugBank and, in general, RNA drugs are less represented than others because they are synthetic (DrugBank siRNA and mRNA vaccine categories contain only 4 approved or under-investigation drugs, ASO drugs are only 12, and RNA aptamer drugs only 2).
In addition to the so far discussed data sources, RNAcentral~\cite{rnacentral} is a collector coordinated by the European Bioinformatics In\-sti\-tu\-te~(EBI~\cite{ebi}), which imports non-coding RNA sequences from multiple databases and enables integrated text search, sequence similarity search, bulk downloads, and programmatic data access through a reliable API.

Starting from the need to understand whether the content of the different data sources overlap, we examined the entities and relationships made available in the considered data sources and identified containment (or overlapping) data sources. 
The result of our study is reported in Figure~\ref{fig:venn} where bubbles represent the relationships made available by the data sources. We can note the presence of two prominent clusters (miRNet and RNAcentral) that properly include or overlap the relationships made available by other data sources. The identification of these containments has been exploited to reduce the issue of semantic compatibility.  
Furthermore, many miRNA and lncRNA sources contain relations that either overlap or are properly included within other sources. For the sake of readability, we have included some of these RNA sources in a legend.
We remark that the {\tt Inter RNA} sources RNAInter, RNALocate, RNAdisease, ncRDeathDB, cncRNAdb, and ViRBase are nicknamed ``Sister Projects'' because they are updated and maintained by the same research team. Common semantics in ``Sister Projects'' result useful for data handling because they share a practically identical structure.

\subsection*{Construction of the RNA meta-graph and ontological description of the KG}\label{sec:metagraph}

Besides the sequences, these data sources also contain different kinds of relationships that can be exploited for the KG construction.
Table~\ref{tab:relations} reports the main relationships that have been identified in the considered data sources according to the RO ontology. 
For each relation, Table~\ref{tab:relations} reports the RO identifier, the corresponding meaning, and, whenever feasible, we have introduced the inverse relationships in case only a unidirectional relationship is available in the data source (e.g., {\tt develops from} and {\tt develops into}).
The general relationships {\tt interacts with} available in RO with the meaning ``A relationship that holds between two entities in which the processes executed by the two entities are causally connected" have been specified in the most specific relationships {\tt molecularly interacts with} in our classification to represent the situation in which the two partners are molecular entities that directly physically interact with each other (e.g., via a stable binding interaction or a brief interaction during which one modifies the other). We use this relationship to represent a specific interaction process at the molecular level (e.g., aptamer-protein binding interaction or tRNA molecule charged with a specific amino acid). We remark that some authors~\cite{Guo2012,LAI2004} suggest that miRNA molecules are {\tt involved in negative regulation of} complementary miRNA molecules by forming base-pairing interactions. However, this kind of relationship is not present in the considered data sources.

\begin{figure}[htbp]
    \centering
    \includegraphics[width=\textwidth]{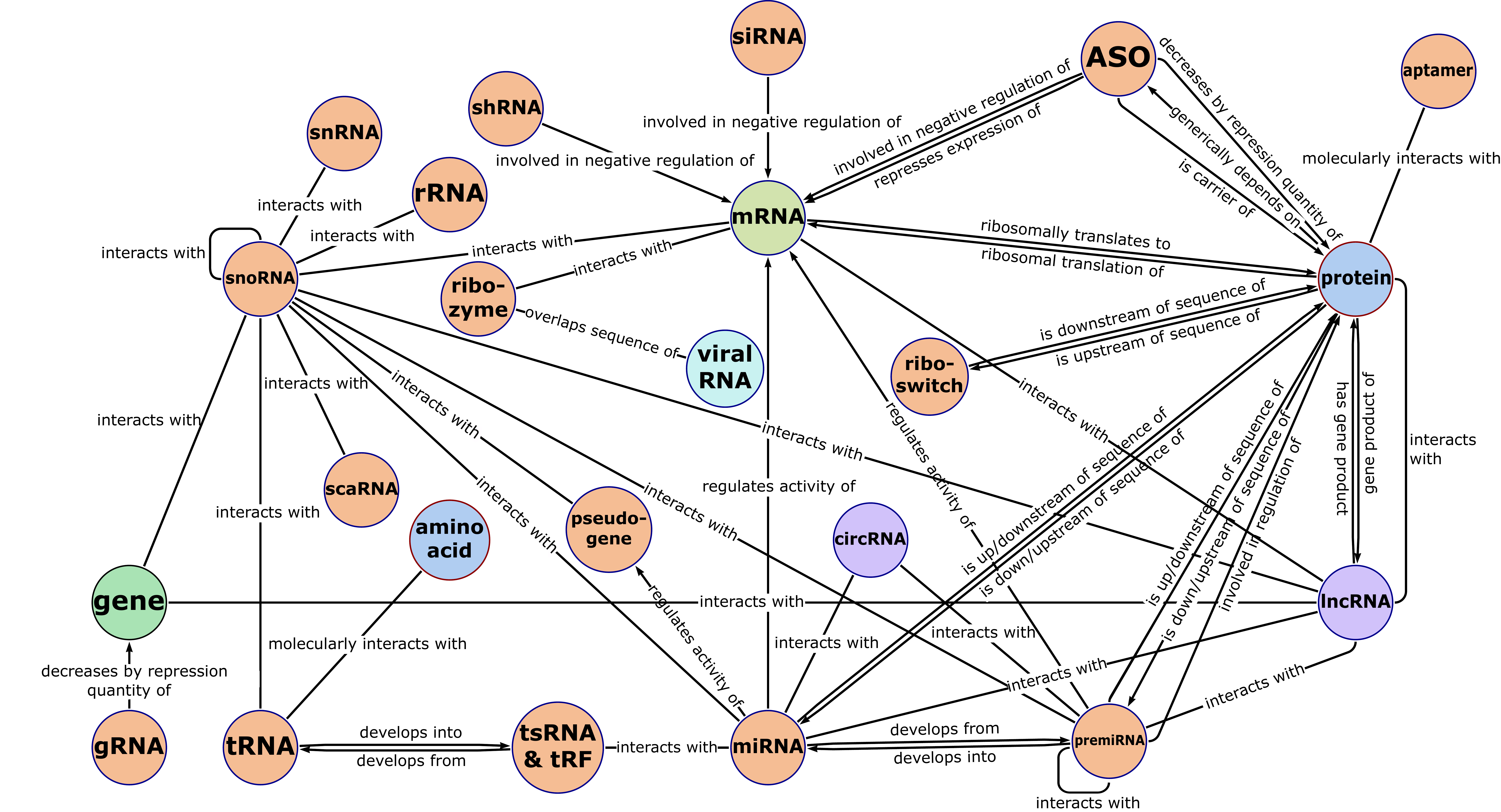}
     \caption{\rnakg meta-graph. Most non-RNA entities are not represented to simplify the visualization.}
    \label{fig:metagraph}
\end{figure}

\begin{figure}[htbp]
    \centering
    \includegraphics[width=0.85\textwidth]{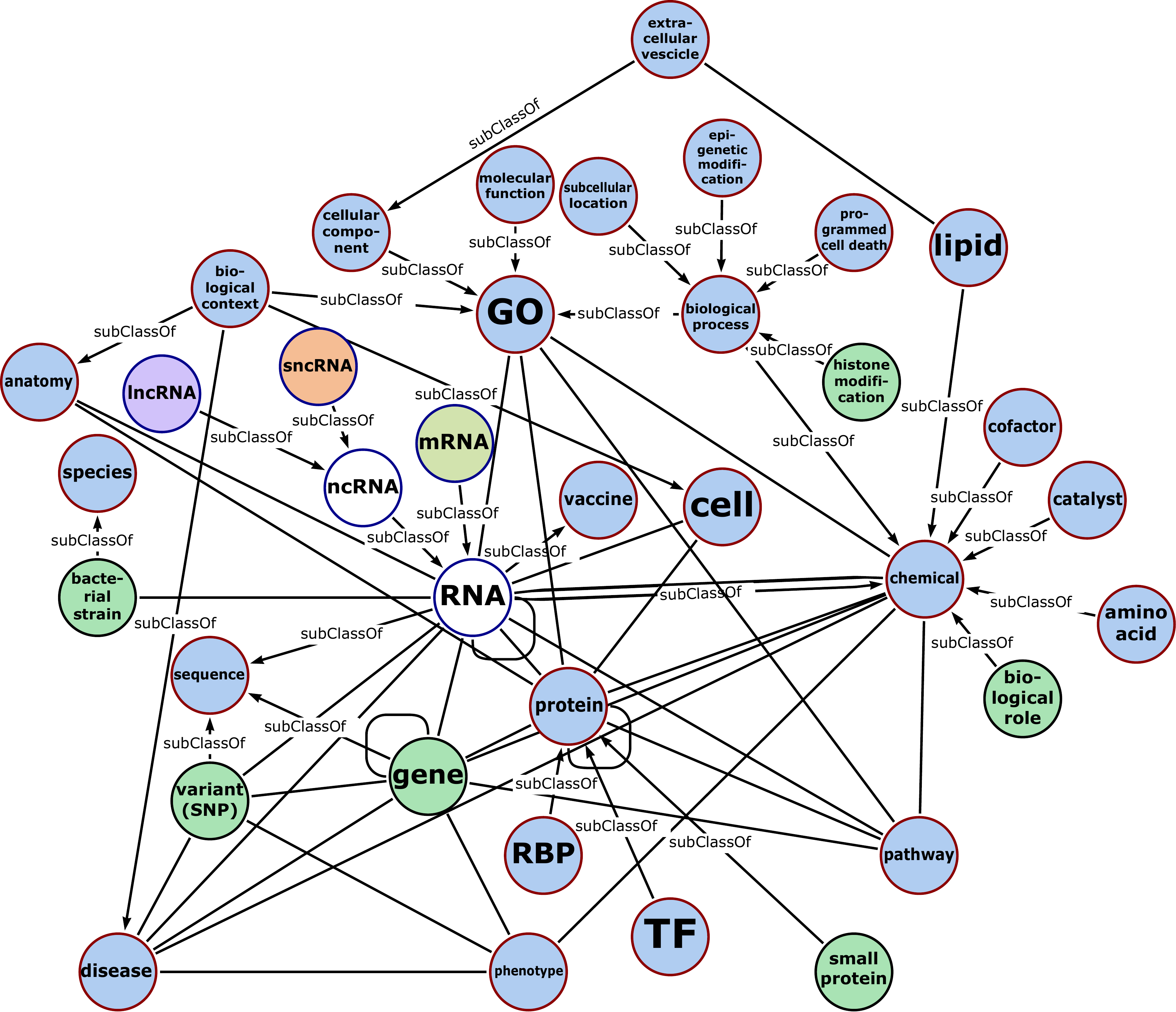}
    \caption{The complete conceptual \rnakg meta-graph.}
\label{fig:completeMetagraph}
\end{figure}

The content of Tables~\ref{tab:database1}--\ref{tab:relations} is the groundwork for the generation of the meta-graph reported in Figure~\ref{fig:metagraph}.
The graphical representation provides a global overview of the richness of information that is currently provided.
To simplify the visualization of the meta-graph, we omitted most of the non-RNA bio-entities that are known to play an important role in studying the biology (and supporting the discovery) of novel RNA drugs. Moreover, we have omitted some of the relationships extracted from the {\tt Inter RNA} data sources (see Table~\ref{tab:database2}) because of the limitation of their occurrences.
The meta-graph in Figure~\ref{fig:metagraph} can be further extended to include other nodes representing other bio-entities (e.g., diseases, epigenetic modifications, small molecules, tissues, biological pathways, and cellular components) and relationships relevant to the analysis.
This "enlarged" meta-graph is quite complex and difficult to be graphically represented. Figure \ref{fig:completeMetagraph} shows a very abstract representation by clustering in a single RNA node all the kinds of RNA molecules described in Figure~\ref{fig:metagraph}. Then, this node is connected with various bio-entities based on insights extracted from RNA sources and literature. It is worth noting that \rnakg has the potential for expansion by integrating additional KGs, and to set the basis for an RNA ontology thanks to the hierarchical structure introduced by the {\tt subClassOf} relationship.
 
\subsection*{\rnakg statistical analysis}

The current version of \rnakg has a single connected component containing 578,384 nodes and 8,768,582 edges. The number of nodes and edges has been deeply reduced by considering only the relationships with a high reliability.
The construction process of the graph is designed to be periodically updated, including data from other public RNA and related biomedical sources. Moreover, thresholds can be tuned for enlarging or reducing the KG size.
Table~\ref{tab:basic_props} depicts the main macroscopic topological and structural properties of the current \rnakg. 

\begin{figure}[htbp]
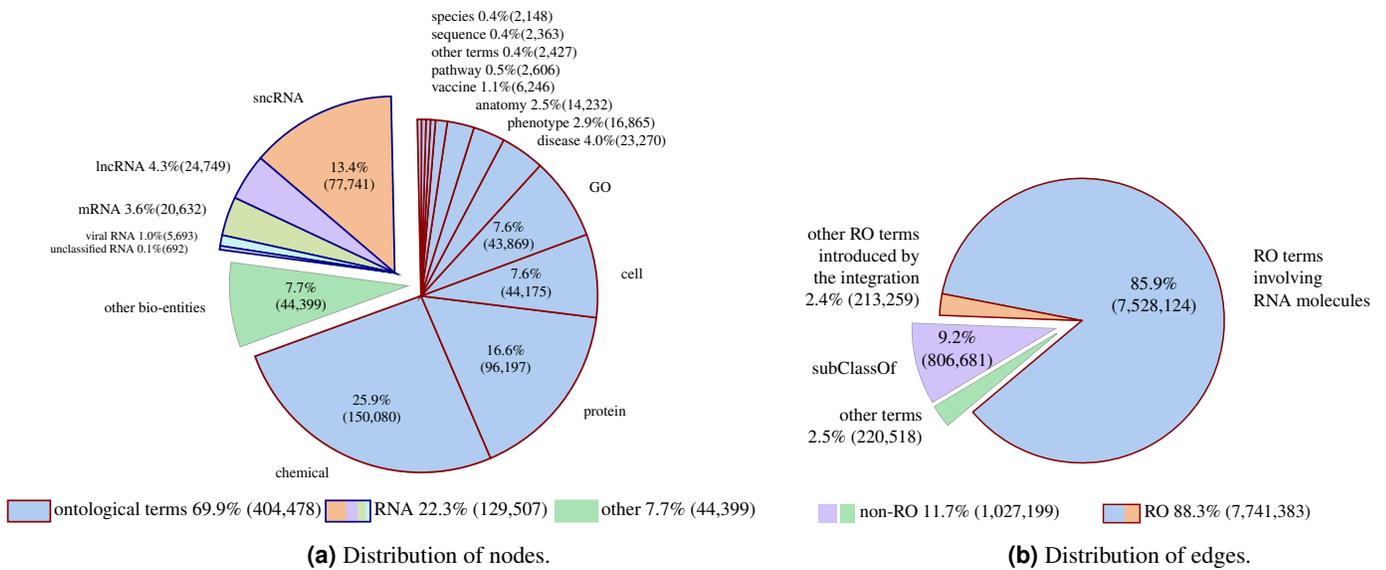

\begin{subfigure}[b]{0.65\linewidth}
\resizebox{.9\linewidth}{!}{\input{images/nodes_piechart_rna3}}
\caption{Distribution of nodes.}
    \label{fig:pie_ont}
\end{subfigure} 
\ \hspace*{-1cm}\
\begin{subfigure}[b]{0.5\linewidth}
\resizebox{.9\linewidth}{!}{\input{images/edges_piechart_rna3}
}
    \caption{Distribution of edges.}
    \label{fig:pie_edges}
\end{subfigure}
    \caption{Pie-chart of: (a) node distribution according to node types.
    (b) edge distribution according to edge types.}
    \label{fig:pie+inter}
\end{figure}

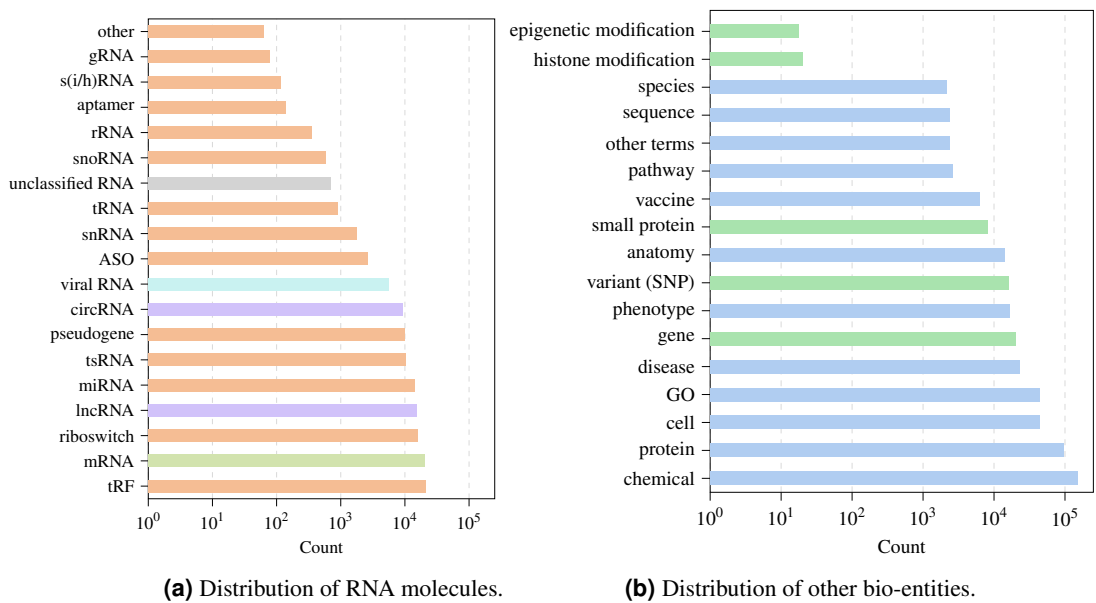
\begin{figure}[htbp]
\centering
 \begin{subfigure}[b]{0.5\linewidth}
\resizebox{.75\linewidth}{!}{
\begin{tikzpicture}

\definecolor{burlywood245189148}{RGB}{245,189,148}
\definecolor{darkgray176}{RGB}{176,176,176}
\definecolor{grayEma}{RGB}{211,211,211}
\definecolor{gray139126102}{RGB}{139,126,102}
\definecolor{palegoldenrod210227174}{RGB}{210,227,174}
\definecolor{paleturquoise201242241}{RGB}{201,242,241}
\definecolor{thistle209194250}{RGB}{209,194,250}

\begin{axis}[
log basis x={10},
tick align=outside,
tick pos=left,
x grid style={gray!30, dashed},
xlabel={Count},
xmajorgrids,
xmin=1, xmax=251188.643150958,
xmode=log,
xtick style={color=black},
xtick={0.1,1,10,100,1000,10000,100000,1000000,10000000},
xticklabels={
  \(\displaystyle {10^{-1}}\),
  \(\displaystyle {10^{0}}\),
  \(\displaystyle {10^{1}}\),
  \(\displaystyle {10^{2}}\),
  \(\displaystyle {10^{3}}\),
  \(\displaystyle {10^{4}}\),
  \(\displaystyle {10^{5}}\),
  \(\displaystyle {10^{6}}\),
  \(\displaystyle {10^{7}}\)
},
y grid style={darkgray176},
ymin=-0.5, ymax=18.5,
ytick style={color=black},
ytick={0,1,2,3,4,5,6,7,8,9,10,11,12,13,14,15,16,17,18},
yticklabels={
  tRF,
  mRNA,
  riboswitch,
  lncRNA,
  miRNA,
  tsRNA,
  pseudogene,
  circRNA,
  viral RNA,
  ASO,
  snRNA,
  tRNA,
  unclassified RNA,
  snoRNA,
  rRNA,
  aptamer,
  s(i/h)RNA,
  gRNA,
  other
},
y=1cm/2
]
\draw[draw=none,fill=burlywood245189148] (axis cs:1e-10,-0.25) rectangle (axis cs:20897,0.25);
\draw[draw=none,fill=palegoldenrod210227174] (axis cs:1e-10,0.75) rectangle (axis cs:20632,1.25);
\draw[draw=none,fill=burlywood245189148] (axis cs:1e-10,1.75) rectangle (axis cs:15805,2.25);
\draw[draw=none,fill=thistle209194250] (axis cs:1e-10,2.75) rectangle (axis cs:15533,3.25);
\draw[draw=none,fill=burlywood245189148] (axis cs:1e-10,3.75) rectangle (axis cs:14295,4.25);
\draw[draw=none,fill=burlywood245189148] (axis cs:1e-10,4.75) rectangle (axis cs:10243,5.25);
\draw[draw=none,fill=burlywood245189148] (axis cs:1e-10,5.75) rectangle (axis cs:9837,6.25);
\draw[draw=none,fill=thistle209194250] (axis cs:1e-10,6.75) rectangle (axis cs:9216,7.25);
\draw[draw=none,fill=paleturquoise201242241] (axis cs:1e-10,7.75) rectangle (axis cs:5693,8.25);
\draw[draw=none,fill=burlywood245189148] (axis cs:1e-10,8.75) rectangle (axis cs:2635,9.25);
\draw[draw=none,fill=burlywood245189148] (axis cs:1e-10,9.75) rectangle (axis cs:1803,10.25);
\draw[draw=none,fill=burlywood245189148] (axis cs:1e-10,10.75) rectangle (axis cs:894,11.25);
\draw[draw=none,fill=grayEma] (axis cs:1e-10,11.75) rectangle (axis cs:692,12.25);
\draw[draw=none,fill=burlywood245189148] (axis cs:1e-10,12.75) rectangle (axis cs:583,13.25);
\draw[draw=none,fill=burlywood245189148] (axis cs:1e-10,13.75) rectangle (axis cs:350,14.25);
\draw[draw=none,fill=burlywood245189148] (axis cs:1e-10,14.75) rectangle (axis cs:140,15.25);
\draw[draw=none,fill=burlywood245189148] (axis cs:1e-10,15.75) rectangle (axis cs:118,16.25);
\draw[draw=none,fill=burlywood245189148] (axis cs:1e-10,16.75) rectangle (axis cs:77,17.25);
\draw[draw=none,fill=burlywood245189148] (axis cs:1e-10,17.75) rectangle (axis cs:64,18.25);
\end{axis}

\end{tikzpicture}}
\caption{Distribution of RNA molecules.}
\label{fig:nodes_countA}
\end{subfigure} 
\ \hspace*{-2.5cm} \ 
\begin{subfigure}[b]{0.45\linewidth}
 \resizebox{\linewidth}{!}{
\begin{tikzpicture}

\definecolor{darkgray176}{RGB}{176,176,176}
\definecolor{lightblue176205241}{RGB}{176,205,241}
\definecolor{palegreen169227174}{RGB}{169,227,174}

\begin{axis}[
log basis x={10},
tick align=outside,
tick pos=left,
x grid style={gray!30, dashed},
xlabel={Count},
xmajorgrids,
xmin=1, xmax=251188.643150958,
xmode=log,
xtick style={color=black},
xtick={0.1,1,10,100,1000,10000,100000,1000000,10000000},
xticklabels={
  \(\displaystyle {10^{-1}}\),
  \(\displaystyle {10^{0}}\),
  \(\displaystyle {10^{1}}\),
  \(\displaystyle {10^{2}}\),
  \(\displaystyle {10^{3}}\),
  \(\displaystyle {10^{4}}\),
  \(\displaystyle {10^{5}}\),
  \(\displaystyle {10^{6}}\),
  \(\displaystyle {10^{7}}\)
},
y grid style={darkgray176},
ymin=-0.5, ymax=16.75,
ytick style={color=black},
ytick={0,1,2,3,4,5,6,7,8,9,10,11,12,13,14,15,16
},
yticklabels={
  chemical,
  protein,
  cell,
  GO,
  disease,
  gene,
  phenotype,
  variant (SNP),
  anatomy,
  small protein,
  vaccine,
  pathway,
  other terms,
  sequence,
  species,
  histone modification,
  epigenetic modification,
},
y=1cm/2
]
\draw[draw=none,fill=lightblue176205241] (axis cs:1e-10,-0.25) rectangle (axis cs:150080,0.25);
\draw[draw=none,fill=lightblue176205241] (axis cs:1e-10,0.75) rectangle (axis cs:96197,1.25);
\draw[draw=none,fill=lightblue176205241] (axis cs:1e-10,1.75) rectangle (axis cs:44175,2.25);
\draw[draw=none,fill=lightblue176205241] (axis cs:1e-10,2.75) rectangle (axis cs:43869,3.25);
\draw[draw=none,fill=lightblue176205241] (axis cs:1e-10,3.75) rectangle (axis cs:23270,4.25);
\draw[draw=none,fill=palegreen169227174] (axis cs:1e-10,4.75) rectangle (axis cs:20081,5.25);
\draw[draw=none,fill=lightblue176205241] (axis cs:1e-10,5.75) rectangle (axis cs:16865,6.25);
\draw[draw=none,fill=palegreen169227174] (axis cs:1e-10,6.75) rectangle (axis cs:16091,7.25);
\draw[draw=none,fill=lightblue176205241] (axis cs:1e-10,7.75) rectangle (axis cs:14232,8.25);
\draw[draw=none,fill=palegreen169227174] (axis cs:1e-10,8.75) rectangle (axis cs:8178,9.25);
\draw[draw=none,fill=lightblue176205241] (axis cs:1e-10,9.75) rectangle (axis cs:6246,10.25);
\draw[draw=none,fill=lightblue176205241] (axis cs:1e-10,10.75) rectangle (axis cs:2606,11.25);
\draw[draw=none,fill=lightblue176205241] (axis cs:1e-10,11.75) rectangle (axis cs:2427,12.25);
\draw[draw=none,fill=lightblue176205241] (axis cs:1e-10,12.75) rectangle (axis cs:2363,13.25);
\draw[draw=none,fill=lightblue176205241] (axis cs:1e-10,13.75) rectangle (axis cs:2148,14.25);
\draw[draw=none,fill=palegreen169227174] (axis cs:1e-10,14.75) rectangle (axis cs:20,15.25);
\draw[draw=none,fill=palegreen169227174] (axis cs:1e-10,15.75) rectangle (axis cs:18,16.25);
\end{axis}

\end{tikzpicture}}
 \caption{Distribution of other bio-entities.}
 \label{fig:nodes_countB}
\end{subfigure} 
\caption{Node distribution according to node types.}\label{fig:nodes_count}
\end{figure}

Figure~\ref{fig:pie_ont} shows the distribution of nodes contained in \rnakg. Nodes can be classified into nodes representing bio-entities
and those that represent the ontological terms. Bio-entities have been further subdivided into RNA nodes (gathering together sncRNA, mRNA, lncRNA, viral RNA, and unclassified RNA nodes), and non-RNA nodes (named {\tt other bio-entities}) that contain, for instance, gene and variant (SNP)-typed nodes.
Furthermore, Figure~\ref{fig:nodes_countA} presents the distribution of nodes according to the main type of RNA molecules, detailing the different categories of sncRNA. mRNA, lncRNA, and miRNA available in \rnakg. These RNA molecules are the most represented in \rnakg because they are well-studied (many RNA sources have been categorized/typed as lncRNA and miRNA, and mRNA are in relationships with many other ncRNAs as already discussed in the characterization of the data sources). 
Also tRF molecules (that are classified in the sncRNA category) are very numerous because they are ``fragments'' of tRNAs (one tRNA can generate more than one tRF or tsRNA). The {\tt unclassified RNA} category includes 692 RNA nodes for which a better semantic characterization cannot be assigned because, in the original sources, they are specified as ``other RNA'', ``miscellaneous RNA'', ``unknown RNA'', ``ncRNA'', or ``RNA molecules to be experimentally confirmed". Finally, the {\tt other} category includes sncRNA molecules whose distribution is negligible in \rnakg (64 sncRNA molecules among ribozymes, piRNAs, enhancer RNAs, vault RNAs, Y RNAs, retained introns, mitochondrial RNAs, small conditional RNAs, and scaRNAs).
The total number of  mRNA, and in general, RNA, is consistent with experimental studies regarding the number of genes in human ($\sim22$-$25K$ protein coding genes and more than 100K total genes \cite{Salzberg2018}).
Ontological terms shown in Figure~\ref{fig:pie_ont} are introduced in the generation of the KG for supporting reasoning activities and can be further classified according to the specific bio-ontology from which they are extracted (e.g., ChEBI for chemicals and HPO for phenotypes).
Among them, chemical and protein nodes cover around 42.5\% of the total amount of nodes in \rnakg. This is due to the fact that ChEBI and PRO both contain many terms representing chemical entities and proteins for \textit{Homo sapiens}. Figure~\ref{fig:nodes_countB} further details the distribution of ontological terms.  
Since the considered ontologies contain also terms that do not follow the usual pattern for their identification (e.g., terms representing glycans belong to PRO but their identifier starts with the prefix {\tt GNO} which differs from the usual one adopted for identifying proteins), we have introduced the category {\tt species}, with the terms representing the species (all species start with the prefix {\tt NCBITaxon}), and the category {\tt other terms}, generally containing all the others.

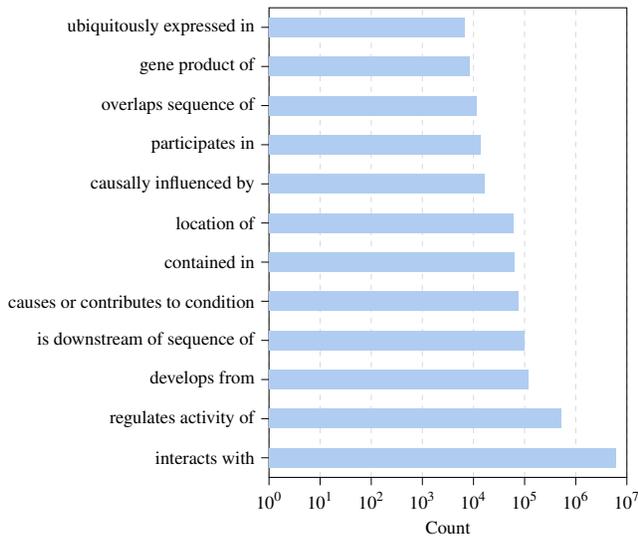
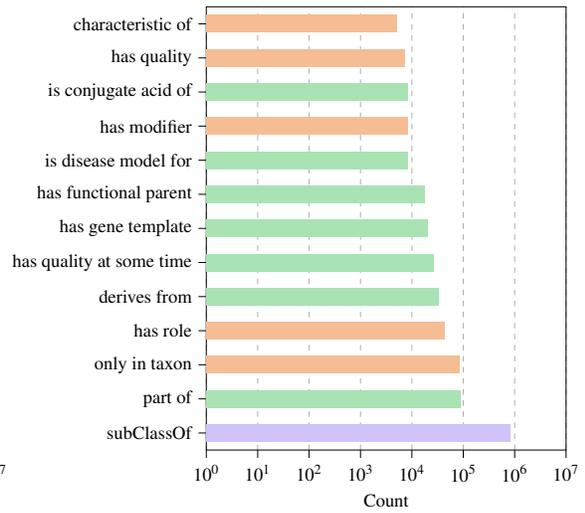
\begin{figure}[htbp]
\centering
 \begin{subfigure}[b]{0.49\linewidth}
\resizebox{\linewidth}{!}{
\begin{tikzpicture}

\definecolor{darkgray176}{RGB}{176,176,176}
\definecolor{lightblue176205241}{RGB}{176,205,241}

\begin{axis}[
log basis x={10},
tick align=outside,
tick pos=left,
x grid style={gray!30, dashed},
xlabel={Count},
xmajorgrids,
xmin=1, xmax=10000000,
xmode=log,
xtick style={color=black},
xtick={0.1,1,10,100,1000,10000,100000,1000000,10000000,100000000},
xticklabels={
  \(\displaystyle {10^{-1}}\),
  \(\displaystyle {10^{0}}\),
  \(\displaystyle {10^{1}}\),
  \(\displaystyle {10^{2}}\),
  \(\displaystyle {10^{3}}\),
  \(\displaystyle {10^{4}}\),
  \(\displaystyle {10^{5}}\),
  \(\displaystyle {10^{6}}\),
  \(\displaystyle {10^{7}}\),
  \(\displaystyle {10^{8}}\)
},
y grid style={darkgray176},
ymin=-0.5, ymax=11.5,
ytick style={color=black},
ytick={0,1,2,3,4,5,6,7,8,9,10,11},
yticklabels={
interacts with,
  regulates activity of,
  develops from,
  is downstream of sequence of,
  causes or contributes to condition,
  contained in,
  location of,
  causally influenced by,
  participates in,
  overlaps sequence of,
  gene product of,
  ubiquitously expressed in
},
y=1.5cm/2,
]
\draw[draw=none,fill=lightblue176205241] (axis cs:1e-10,-0.25) rectangle (axis cs:6130892,0.25);
\draw[draw=none,fill=lightblue176205241] (axis cs:1e-10,0.75) rectangle (axis cs:519228,1.25);
\draw[draw=none,fill=lightblue176205241] (axis cs:1e-10,1.75) rectangle (axis cs:119639,2.25);
\draw[draw=none,fill=lightblue176205241] (axis cs:1e-10,2.75) rectangle (axis cs:100289,3.25);
\draw[draw=none,fill=lightblue176205241] (axis cs:1e-10,3.75) rectangle (axis cs:76574,4.25);
\draw[draw=none,fill=lightblue176205241] (axis cs:1e-10,4.75) rectangle (axis cs:63064,5.25);
\draw[draw=none,fill=lightblue176205241] (axis cs:1e-10,5.75) rectangle (axis cs:60729,6.25);
\draw[draw=none,fill=lightblue176205241] (axis cs:1e-10,6.75) rectangle (axis cs:16459,7.25);
\draw[draw=none,fill=lightblue176205241] (axis cs:1e-10,7.75) rectangle (axis cs:14037,8.25);
\draw[draw=none,fill=lightblue176205241] (axis cs:1e-10,8.75) rectangle (axis cs:11792,9.25);
\draw[draw=none,fill=lightblue176205241] (axis cs:1e-10,9.75) rectangle (axis cs:8417,10.25);
\draw[draw=none,fill=lightblue176205241] (axis cs:1e-10,10.75) rectangle (axis cs:6610,11.25);
\end{axis}

\end{tikzpicture}}
\caption{Distribution of edges involving RNA molecules.}\label{fig:edges_countA}
\end{subfigure} 
\ \hspace*{-.5cm} \ 
\begin{subfigure}[b]{0.49\linewidth}
 \resizebox{.9\linewidth}{!}{\begin{tikzpicture}

\definecolor{burlywood245189148}{RGB}{245,189,148}
\definecolor{darkgray176}{RGB}{176,176,176}
\definecolor{palegreen169227180}{RGB}{169,227,180}
\definecolor{thistle209194250}{RGB}{209,194,250}

\begin{axis}[
log basis x={10},
tick align=outside,
tick pos=left,
x grid style={darkgray176, dashed},
xlabel={Count},
xmajorgrids,
xmin=1, xmax=10000000,
xmode=log,
xtick style={color=black},
xtick={0.1,1,10,100,1000,10000,100000,1000000,10000000,100000000},
xticklabels={
  \(\displaystyle {10^{-1}}\),
  \(\displaystyle {10^{0}}\),
  \(\displaystyle {10^{1}}\),
  \(\displaystyle {10^{2}}\),
  \(\displaystyle {10^{3}}\),
  \(\displaystyle {10^{4}}\),
  \(\displaystyle {10^{5}}\),
  \(\displaystyle {10^{6}}\),
  \(\displaystyle {10^{7}}\),
  \(\displaystyle {10^{8}}\)
},
y grid style={darkgray176},
ymin=-0.5, ymax=12.5,
ytick style={color=black},
ytick={0,1,2,3,4,5,6,7,8,9,10,11,12},
yticklabels={
  subClassOf,
  part of,
  only in taxon,
  has role,
  derives from,
  has quality at some time,
  has gene template,
  has functional parent,
  is disease model for,
  has modifier,
  is conjugate acid of,
  has quality,
  characteristic of
},
y=1.3cm/2,
]
\draw[draw=none,fill=thistle209194250] (axis cs:1e-10,-0.25) rectangle (axis cs:806682,0.25);
\draw[draw=none,fill=palegreen169227180] (axis cs:1e-10,0.75) rectangle (axis cs:89431,1.25);
\draw[draw=none,fill=burlywood245189148] (axis cs:1e-10,1.75) rectangle (axis cs:86370,2.25);
\draw[draw=none,fill=burlywood245189148] (axis cs:1e-10,2.75) rectangle (axis cs:43438,3.25);
\draw[draw=none,fill=palegreen169227180] (axis cs:1e-10,3.75) rectangle (axis cs:33678,4.25);
\draw[draw=none,fill=palegreen169227180] (axis cs:1e-10,4.75) rectangle (axis cs:26090,5.25);
\draw[draw=none,fill=palegreen169227180] (axis cs:1e-10,5.75) rectangle (axis cs:20030,6.25);
\draw[draw=none,fill=palegreen169227180] (axis cs:1e-10,6.75) rectangle (axis cs:17871,7.25);
\draw[draw=none,fill=palegreen169227180] (axis cs:1e-10,7.75) rectangle (axis cs:8429,8.25);
\draw[draw=none,fill=burlywood245189148] (axis cs:1e-10,8.75) rectangle (axis cs:8226,9.25);
\draw[draw=none,fill=palegreen169227180] (axis cs:1e-10,9.75) rectangle (axis cs:8184,10.25);
\draw[draw=none,fill=burlywood245189148] (axis cs:1e-10,10.75) rectangle (axis cs:7276,11.25);
\draw[draw=none,fill=burlywood245189148] (axis cs:1e-10,11.75) rectangle (axis cs:5147,12.25);
\end{axis}

\end{tikzpicture}}
 \caption{Distribution of other edges in \rnakg.}
 \label{fig:edges_countB}
\end{subfigure} 
\caption{Edges distribution. Only direct edges with more than 5,000 occurrences are reported.}\label{fig:edges_count}
\end{figure}

Figure~\ref{fig:pie_edges} shows the distribution of edges in \rnakg. Edges have been subdivided into three categories: $i)$ edges representing RO terms that have been further classified in those that describe the interactions among RNA molecules and RO terms introduced by the integrations of the bio-ontologies; $ii)$ edges representing the {\tt subClassOf} relationships; and $iii)$ edges representing other kinds of relationships not included in RO (e.g., \texttt{has gene template} belonging to PRO).  
Figure~\ref{fig:edges_countA} details the distribution of the types of edges involving RNA molecules. As reflected by the organization of the meta-graph, {\tt interacts with} is the most represented edge type because it is symmetric, whereas the presence of many {\tt regulates activity of} edges is justified by the vast majority of miRNA molecules within \rnakg that indeed regulate the activity of, for example, pseudogene and mRNA molecules. 
Moreover, Figure~\ref{fig:edges_countB} shows the distribution of the remaining edges included in \rnakg. We can notice the prevalence of many {\tt subClassOf} due to the integration of the bio-ontologies, and because we specified each RNA molecule as {\tt subClassOf} an appropriate class within SO (e.g., {\tt SO\_0000276} for miRNA molecules).

\section*{Technical Validation}
In order to evaluate the quality of the generated KG, several analyses have been performed whose results are reported in the following paragraphs. 

\begin{figure}[htbp]
\centering
 \begin{subfigure}[b]{0.53\linewidth}
 \includegraphics[width=\linewidth, keepaspectratio]{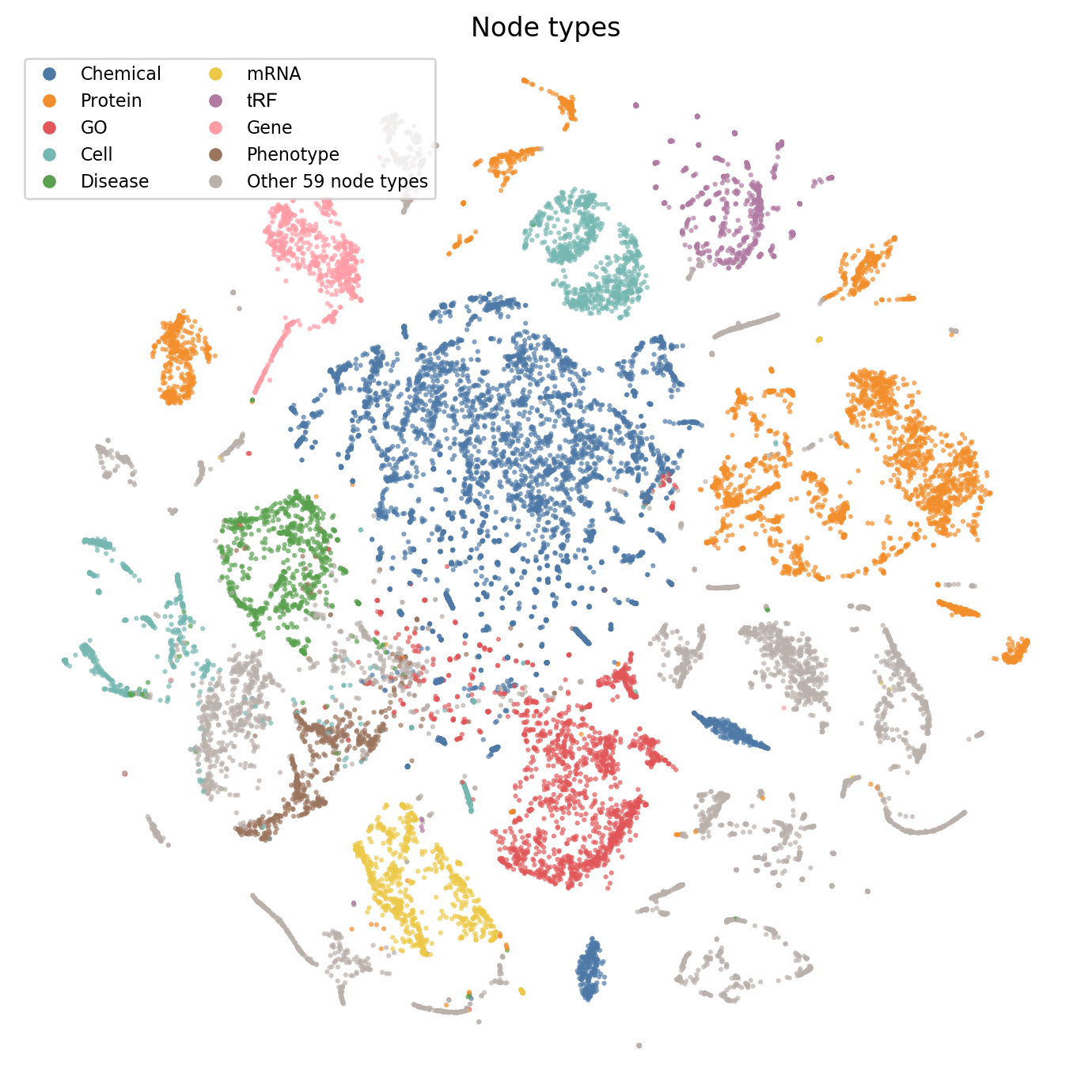}
\caption{Node types embedding.}\label{fig:grapeNode}
\end{subfigure} 
\ \hspace*{-1.5cm} \ 
\begin{subfigure}[b]{0.53\linewidth}
\includegraphics[width=\linewidth, keepaspectratio]{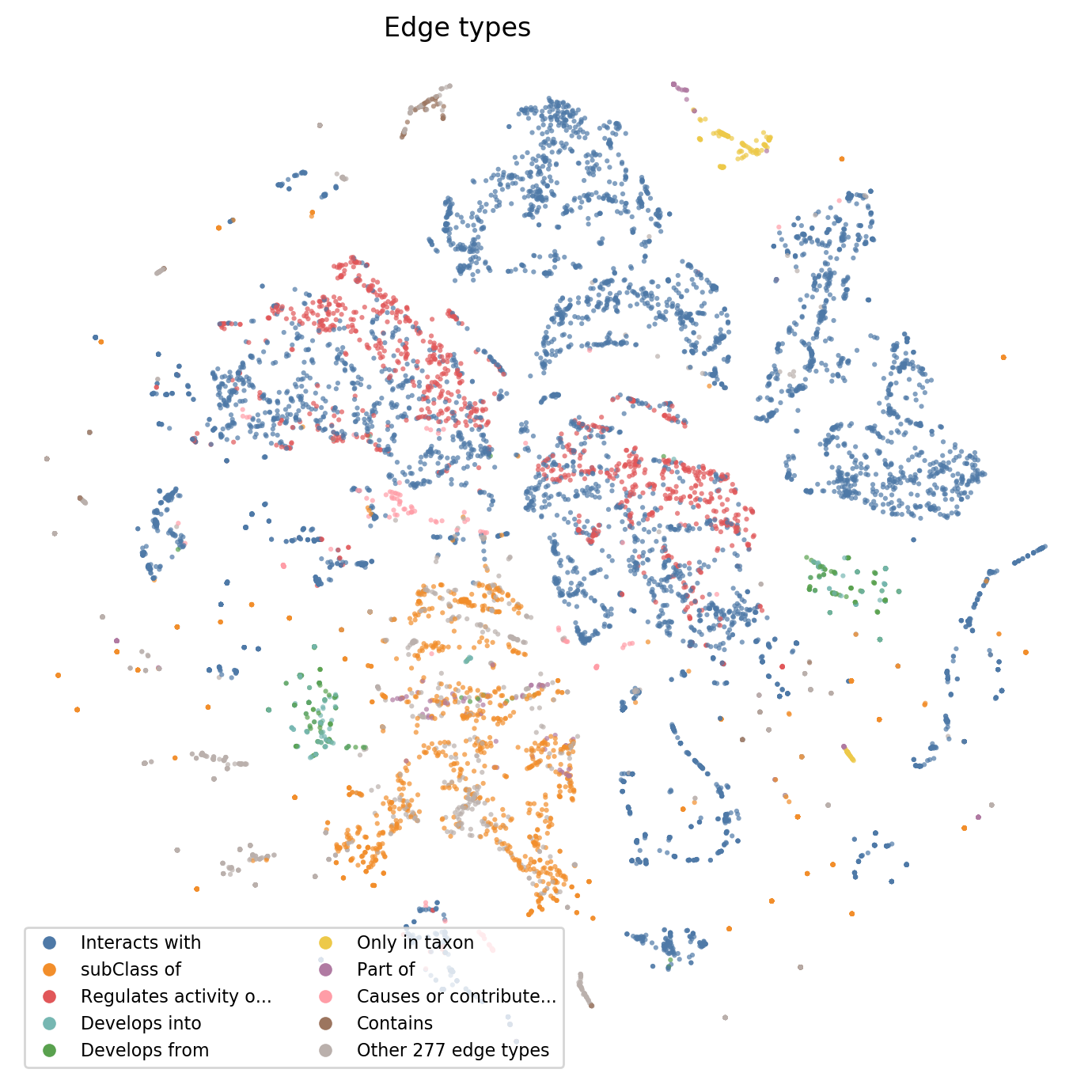}
 \caption{Edge types embedding.
 }\label{fig:grapeEdge}
\end{subfigure} 
\caption{Bidimensional view of \rnakg embeddings.}
\label{fig:grape}
\end{figure}

\paragraph{t-SNE representation.} 
Figure~\ref{fig:grape} shows the t-SNE representation of an embedding of the nodes/edges in  \rnakg by using the GRAPE implementation of Node2Vec with CBOW, a random walk-based second-order embedding algorithm \cite{node2vec}, with walk length equal to 5. 
Figure~\ref{fig:grapeNode} shows how the embedding of the node type is able to effectively identify the similarities among the nodes of the same type, thus capturing their function in the network.
On the other hand, Figure~\ref{fig:grapeEdge} depicts the edge embedding for \rnakg. Also in this case, the embedding is able to capture the similarity between edges with the only exception of the \texttt{interacts with} relation which seems to overlap several other edge types. This fact is not so surprising considering that this relation is also used to denote a generic relation between nodes.

\paragraph{Topological analysis.}  
The topological analysis led to the identification of top-5 nodes with the highest degree centrality: 
{\em microvesicle} ({\tt GO\_1990742}) with degree 27.11K (whose type is {\tt GO}); 
{\em nucleus} ({\tt GO\_0005634}, degree 20.15K), 
{\em hcmv-miR-US25-1-5p} ({\tt human cytomegalovirus hcmv-miR-US25-1-5p mature miRNA}, degree 18.18K and node type {\tt miRNA}), 
{\em hFOXA1} ({\tt PR\_P55317}, degree 31.80K and node type {\tt protein}), and 
{\em cytosol} ({\tt GO\_0005829}, degree 17.26K).
We remark that the nodes cytosol, nucleus, and microvescicle represent cellular components used for aggregating different bio-entities existing in the context of a cell and this is the main reason for the high node degree within \rnakg.
Moreover, the RNA relationships with these kinds of cellular components are enhanced by the semantics {\tt contained in} and {\tt location of} together with their respective RO inverse {\tt contains} and {\tt located in}.
On the other hand, {\em hcmv-miR-US25-1-5p} is a human cytomegalovirus (HCMV)-encoded -5p miRNA transcript, whose diagnostic and prognostic value has been proved valid for several human diseases and their clinical implications \cite{FernndezMoreno2021}. Finally, hFOXA1 is a forkhead TF known to be the main target of insulin signaling, to regulate metabolic homeostasis in response to oxidative stress, and to interact with chromatin. The central role assumed by hFOXA1 in \rnakg is quite interesting since this TF is implicated in various human malignancies characterized by altered expression of ncRNAs \cite{Peng2020}.

\begin{figure}[htbp]
\centering
\begin{subfigure}[b]{0.32\linewidth}
 \resizebox{.9\linewidth}{!}{
\begin{tikzpicture}

\definecolor{darkgray176}{RGB}{176,176,176}
\definecolor{steelblue31119180}{RGB}{180,180,180}

\begin{axis}[
log basis y={10},
tick align=outside,
tick pos=left,
title={Degree distribution},
x grid style={darkgray176},
xlabel={Degree},
xmin=-6657.075375, xmax=143586.075375,
xtick style={color=black},
y grid style={darkgray176},
ylabel={Count},
ymin=0.514257498980282, ymax=1162001.52877676,
ymode=log,
ytick style={color=black},
ytick={0.01,0.1,1,10,100,1000,10000,100000,1000000,10000000,100000000},
yticklabels={
  \(\displaystyle {10^{-2}}\),
  \(\displaystyle {10^{-1}}\),
  \(\displaystyle {10^{0}}\),
  \(\displaystyle {10^{1}}\),
  \(\displaystyle {10^{2}}\),
  \(\displaystyle {10^{3}}\),
  \(\displaystyle {10^{4}}\),
  \(\displaystyle {10^{5}}\),
  \(\displaystyle {10^{6}}\),
  \(\displaystyle {10^{7}}\),
  \(\displaystyle {10^{8}}\)
}
]
\draw[draw=none,fill=steelblue31119180] (axis cs:147.42625,1e-10) rectangle (axis cs:1318.83625,577388);
\draw[draw=none,fill=steelblue31119180] (axis cs:1611.68875,1e-10) rectangle (axis cs:2783.09875,792);
\draw[draw=none,fill=steelblue31119180] (axis cs:3075.95125,1e-10) rectangle (axis cs:4247.36125,103);
\draw[draw=none,fill=steelblue31119180] (axis cs:4540.21375,1e-10) rectangle (axis cs:5711.62375,25);
\draw[draw=none,fill=steelblue31119180] (axis cs:6004.47625,1e-10) rectangle (axis cs:7175.88625,21);
\draw[draw=none,fill=steelblue31119180] (axis cs:7468.73875,1e-10) rectangle (axis cs:8640.14875,12);
\draw[draw=none,fill=steelblue31119180] (axis cs:8933.00125,1e-10) rectangle (axis cs:10104.41125,7);
\draw[draw=none,fill=steelblue31119180] (axis cs:10397.26375,1e-10) rectangle (axis cs:11568.67375,3);
\draw[draw=none,fill=steelblue31119180] (axis cs:11861.52625,1e-10) rectangle (axis cs:13032.93625,5);
\draw[draw=none,fill=steelblue31119180] (axis cs:13325.78875,1e-10) rectangle (axis cs:14497.19875,4);
\draw[draw=none,fill=steelblue31119180] (axis cs:14790.05125,1e-10) rectangle (axis cs:15961.46125,2);
\draw[draw=none,fill=steelblue31119180] (axis cs:16254.31375,1e-10) rectangle (axis cs:17425.72375,4);
\draw[draw=none,fill=steelblue31119180] (axis cs:17718.57625,1e-10) rectangle (axis cs:18889.98625,2);
\draw[draw=none,fill=steelblue31119180] (axis cs:19182.83875,1e-10) rectangle (axis cs:20354.24875,3);
\draw[draw=none,fill=steelblue31119180] (axis cs:20647.10125,1e-10) rectangle (axis cs:21818.51125,5);
\draw[draw=none,fill=steelblue31119180] (axis cs:22111.36375,1e-10) rectangle (axis cs:23282.77375,1e-10);
\draw[draw=none,fill=steelblue31119180] (axis cs:23575.62625,1e-10) rectangle (axis cs:24747.03625,1e-10);
\draw[draw=none,fill=steelblue31119180] (axis cs:25039.88875,1e-10) rectangle (axis cs:26211.29875,3);
\draw[draw=none,fill=steelblue31119180] (axis cs:26504.15125,1e-10) rectangle (axis cs:27675.56125,1e-10);
\draw[draw=none,fill=steelblue31119180] (axis cs:27968.41375,1e-10) rectangle (axis cs:29139.82375,1e-10);
\draw[draw=none,fill=steelblue31119180] (axis cs:29432.67625,1e-10) rectangle (axis cs:30604.08625,1e-10);
\draw[draw=none,fill=steelblue31119180] (axis cs:30896.93875,1e-10) rectangle (axis cs:32068.34875,1);
\draw[draw=none,fill=steelblue31119180] (axis cs:32361.20125,1e-10) rectangle (axis cs:33532.61125,1e-10);
\draw[draw=none,fill=steelblue31119180] (axis cs:33825.46375,1e-10) rectangle (axis cs:34996.87375,1);
\draw[draw=none,fill=steelblue31119180] (axis cs:35289.72625,1e-10) rectangle (axis cs:36461.13625,1e-10);
\draw[draw=none,fill=steelblue31119180] (axis cs:36753.98875,1e-10) rectangle (axis cs:37925.39875,1e-10);
\draw[draw=none,fill=steelblue31119180] (axis cs:38218.25125,1e-10) rectangle (axis cs:39389.66125,1e-10);
\draw[draw=none,fill=steelblue31119180] (axis cs:39682.51375,1e-10) rectangle (axis cs:40853.92375,1);
\draw[draw=none,fill=steelblue31119180] (axis cs:41146.77625,1e-10) rectangle (axis cs:42318.18625,1e-10);
\draw[draw=none,fill=steelblue31119180] (axis cs:42611.03875,1e-10) rectangle (axis cs:43782.44875,1e-10);
\draw[draw=none,fill=steelblue31119180] (axis cs:44075.30125,1e-10) rectangle (axis cs:45246.71125,1e-10);
\draw[draw=none,fill=steelblue31119180] (axis cs:45539.56375,1e-10) rectangle (axis cs:46710.97375,1e-10);
\draw[draw=none,fill=steelblue31119180] (axis cs:47003.82625,1e-10) rectangle (axis cs:48175.23625,1e-10);
\draw[draw=none,fill=steelblue31119180] (axis cs:48468.08875,1e-10) rectangle (axis cs:49639.49875,1e-10);
\draw[draw=none,fill=steelblue31119180] (axis cs:49932.35125,1e-10) rectangle (axis cs:51103.76125,1e-10);
\draw[draw=none,fill=steelblue31119180] (axis cs:51396.61375,1e-10) rectangle (axis cs:52568.02375,1e-10);
\draw[draw=none,fill=steelblue31119180] (axis cs:52860.87625,1e-10) rectangle (axis cs:54032.28625,1e-10);
\draw[draw=none,fill=steelblue31119180] (axis cs:54325.13875,1e-10) rectangle (axis cs:55496.54875,1);
\draw[draw=none,fill=steelblue31119180] (axis cs:55789.40125,1e-10) rectangle (axis cs:56960.81125,1e-10);
\draw[draw=none,fill=steelblue31119180] (axis cs:57253.66375,1e-10) rectangle (axis cs:58425.07375,1e-10);
\draw[draw=none,fill=steelblue31119180] (axis cs:58717.92625,1e-10) rectangle (axis cs:59889.33625,1e-10);
\draw[draw=none,fill=steelblue31119180] (axis cs:60182.18875,1e-10) rectangle (axis cs:61353.59875,1e-10);
\draw[draw=none,fill=steelblue31119180] (axis cs:61646.45125,1e-10) rectangle (axis cs:62817.86125,1e-10);
\draw[draw=none,fill=steelblue31119180] (axis cs:63110.71375,1e-10) rectangle (axis cs:64282.12375,1e-10);
\draw[draw=none,fill=steelblue31119180] (axis cs:64574.97625,1e-10) rectangle (axis cs:65746.38625,1e-10);
\draw[draw=none,fill=steelblue31119180] (axis cs:66039.23875,1e-10) rectangle (axis cs:67210.64875,1e-10);
\draw[draw=none,fill=steelblue31119180] (axis cs:67503.50125,1e-10) rectangle (axis cs:68674.91125,1e-10);
\draw[draw=none,fill=steelblue31119180] (axis cs:68967.76375,1e-10) rectangle (axis cs:70139.17375,1e-10);
\draw[draw=none,fill=steelblue31119180] (axis cs:70432.02625,1e-10) rectangle (axis cs:71603.43625,1e-10);
\draw[draw=none,fill=steelblue31119180] (axis cs:71896.28875,1e-10) rectangle (axis cs:73067.69875,1e-10);
\draw[draw=none,fill=steelblue31119180] (axis cs:73360.55125,1e-10) rectangle (axis cs:74531.96125,1e-10);
\draw[draw=none,fill=steelblue31119180] (axis cs:74824.81375,1e-10) rectangle (axis cs:75996.22375,1e-10);
\draw[draw=none,fill=steelblue31119180] (axis cs:76289.07625,1e-10) rectangle (axis cs:77460.48625,1e-10);
\draw[draw=none,fill=steelblue31119180] (axis cs:77753.33875,1e-10) rectangle (axis cs:78924.74875,1e-10);
\draw[draw=none,fill=steelblue31119180] (axis cs:79217.60125,1e-10) rectangle (axis cs:80389.01125,1e-10);
\draw[draw=none,fill=steelblue31119180] (axis cs:80681.86375,1e-10) rectangle (axis cs:81853.27375,1e-10);
\draw[draw=none,fill=steelblue31119180] (axis cs:82146.12625,1e-10) rectangle (axis cs:83317.53625,1e-10);
\draw[draw=none,fill=steelblue31119180] (axis cs:83610.38875,1e-10) rectangle (axis cs:84781.79875,1e-10);
\draw[draw=none,fill=steelblue31119180] (axis cs:85074.65125,1e-10) rectangle (axis cs:86246.06125,1e-10);
\draw[draw=none,fill=steelblue31119180] (axis cs:86538.91375,1e-10) rectangle (axis cs:87710.32375,1e-10);
\draw[draw=none,fill=steelblue31119180] (axis cs:88003.17625,1e-10) rectangle (axis cs:89174.58625,1e-10);
\draw[draw=none,fill=steelblue31119180] (axis cs:89467.43875,1e-10) rectangle (axis cs:90638.84875,1e-10);
\draw[draw=none,fill=steelblue31119180] (axis cs:90931.70125,1e-10) rectangle (axis cs:92103.11125,1e-10);
\draw[draw=none,fill=steelblue31119180] (axis cs:92395.96375,1e-10) rectangle (axis cs:93567.37375,1e-10);
\draw[draw=none,fill=steelblue31119180] (axis cs:93860.22625,1e-10) rectangle (axis cs:95031.63625,1e-10);
\draw[draw=none,fill=steelblue31119180] (axis cs:95324.48875,1e-10) rectangle (axis cs:96495.89875,1e-10);
\draw[draw=none,fill=steelblue31119180] (axis cs:96788.75125,1e-10) rectangle (axis cs:97960.16125,1e-10);
\draw[draw=none,fill=steelblue31119180] (axis cs:98253.01375,1e-10) rectangle (axis cs:99424.42375,1e-10);
\draw[draw=none,fill=steelblue31119180] (axis cs:99717.27625,1e-10) rectangle (axis cs:100888.68625,1e-10);
\draw[draw=none,fill=steelblue31119180] (axis cs:101181.53875,1e-10) rectangle (axis cs:102352.94875,1e-10);
\draw[draw=none,fill=steelblue31119180] (axis cs:102645.80125,1e-10) rectangle (axis cs:103817.21125,1e-10);
\draw[draw=none,fill=steelblue31119180] (axis cs:104110.06375,1e-10) rectangle (axis cs:105281.47375,1e-10);
\draw[draw=none,fill=steelblue31119180] (axis cs:105574.32625,1e-10) rectangle (axis cs:106745.73625,1e-10);
\draw[draw=none,fill=steelblue31119180] (axis cs:107038.58875,1e-10) rectangle (axis cs:108209.99875,1e-10);
\draw[draw=none,fill=steelblue31119180] (axis cs:108502.85125,1e-10) rectangle (axis cs:109674.26125,1e-10);
\draw[draw=none,fill=steelblue31119180] (axis cs:109967.11375,1e-10) rectangle (axis cs:111138.52375,1e-10);
\draw[draw=none,fill=steelblue31119180] (axis cs:111431.37625,1e-10) rectangle (axis cs:112602.78625,1e-10);
\draw[draw=none,fill=steelblue31119180] (axis cs:112895.63875,1e-10) rectangle (axis cs:114067.04875,1e-10);
\draw[draw=none,fill=steelblue31119180] (axis cs:114359.90125,1e-10) rectangle (axis cs:115531.31125,1e-10);
\draw[draw=none,fill=steelblue31119180] (axis cs:115824.16375,1e-10) rectangle (axis cs:116995.57375,1);
\end{axis}

\end{tikzpicture}}
 \caption{}
 \label{fig:degree_distributionB}
\end{subfigure} 
\ \
\begin{subfigure}[b]{0.32\linewidth}
  \resizebox{.9\linewidth}{!}{\input{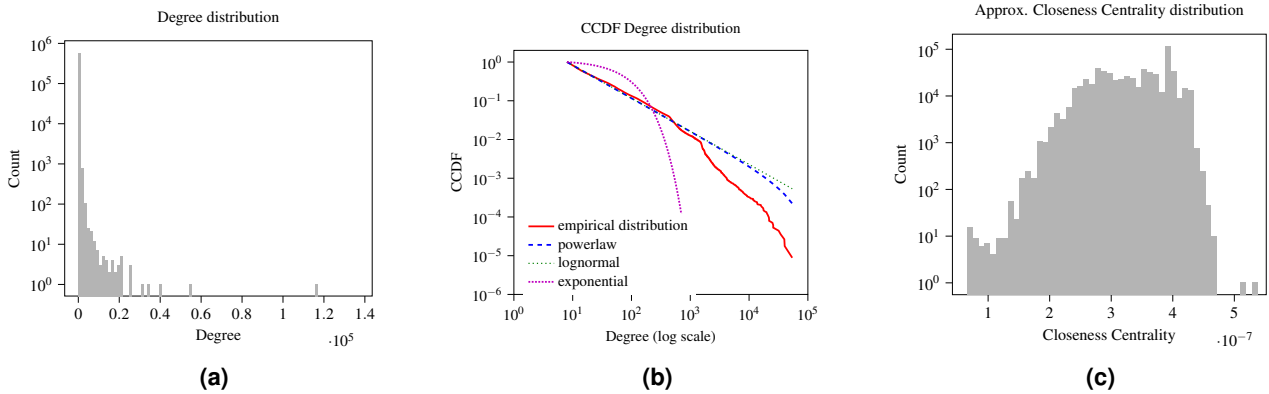}}
  \caption{}
  \label{fig:degree_distributionC}
\end{subfigure} 
\ \
 \begin{subfigure}[b]{0.32\linewidth}
\resizebox{.9\linewidth}{!}{

\begin{tikzpicture}

\definecolor{darkgray176}{RGB}{176,176,176}
\definecolor{steelblue31119180}{RGB}{180,180,180}

\begin{axis}[
log basis y={10},
tick align=outside,
tick pos=left,
title={Approx. Closeness Centrality distribution},
x grid style={darkgray176},
xlabel={Closeness Centrality\hspace{50pt}},
xmin=4.20207733498046e-08, xmax=5.5184404613191e-07,
xtick style={color=black},
xtick={0,1e-07,2e-07,3e-07,4e-07,5e-07,6e-07},
xticklabels={0,1,2,3,4,5,6},
y grid style={darkgray176},
ylabel={Count},
ymin=0.557882914479117, ymax=210182.095484104,
ymode=log,
ytick style={color=black},
ytick={0.01,0.1,1,10,100,1000,10000,100000,1000000,10000000},
yticklabels={
  \(\displaystyle {10^{-2}}\),
  \(\displaystyle {10^{-1}}\),
  \(\displaystyle {10^{0}}\),
  \(\displaystyle {10^{1}}\),
  \(\displaystyle {10^{2}}\),
  \(\displaystyle {10^{3}}\),
  \(\displaystyle {10^{4}}\),
  \(\displaystyle {10^{5}}\),
  \(\displaystyle {10^{6}}\),
  \(\displaystyle {10^{7}}\)
}
]
\draw[draw=none,fill=steelblue31119180] (axis cs:6.56491039308094e-08,1e-10) rectangle (axis cs:7.51004378685138e-08,15);
\draw[draw=none,fill=steelblue31119180] (axis cs:7.51004378685138e-08,1e-10) rectangle (axis cs:8.45517718062183e-08,9);
\draw[draw=none,fill=steelblue31119180] (axis cs:8.45517718062183e-08,1e-10) rectangle (axis cs:9.40030986384954e-08,6);
\draw[draw=none,fill=steelblue31119180] (axis cs:9.40030986384954e-08,1e-10) rectangle (axis cs:1.034544325762e-07,7);
\draw[draw=none,fill=steelblue31119180] (axis cs:1.034544325762e-07,1e-10) rectangle (axis cs:1.12905766513904e-07,4);
\draw[draw=none,fill=steelblue31119180] (axis cs:1.12905766513904e-07,1e-10) rectangle (axis cs:1.22357093346182e-07,9);
\draw[draw=none,fill=steelblue31119180] (axis cs:1.22357093346182e-07,1e-10) rectangle (axis cs:1.31808434389313e-07,9);
\draw[draw=none,fill=steelblue31119180] (axis cs:1.31808434389313e-07,1e-10) rectangle (axis cs:1.4125976122159e-07,56);
\draw[draw=none,fill=steelblue31119180] (axis cs:1.4125976122159e-07,1e-10) rectangle (axis cs:1.50711088053868e-07,23);
\draw[draw=none,fill=steelblue31119180] (axis cs:1.50711088053868e-07,1e-10) rectangle (axis cs:1.60162429096999e-07,175);
\draw[draw=none,fill=steelblue31119180] (axis cs:1.60162429096999e-07,1e-10) rectangle (axis cs:1.69613755929277e-07,248);
\draw[draw=none,fill=steelblue31119180] (axis cs:1.69613755929277e-07,1e-10) rectangle (axis cs:1.79065096972408e-07,175);
\draw[draw=none,fill=steelblue31119180] (axis cs:1.79065096972408e-07,1e-10) rectangle (axis cs:1.88516423804685e-07,1063);
\draw[draw=none,fill=steelblue31119180] (axis cs:1.88516423804685e-07,1e-10) rectangle (axis cs:1.97967750636963e-07,989);
\draw[draw=none,fill=steelblue31119180] (axis cs:1.97967750636963e-07,1e-10) rectangle (axis cs:2.07419091680094e-07,2072);
\draw[draw=none,fill=steelblue31119180] (axis cs:2.07419091680094e-07,1e-10) rectangle (axis cs:2.16870418512372e-07,4231);
\draw[draw=none,fill=steelblue31119180] (axis cs:2.16870418512372e-07,1e-10) rectangle (axis cs:2.26321745344649e-07,3155);
\draw[draw=none,fill=steelblue31119180] (axis cs:2.26321745344649e-07,1e-10) rectangle (axis cs:2.3577308638778e-07,5784);
\draw[draw=none,fill=steelblue31119180] (axis cs:2.3577308638778e-07,1e-10) rectangle (axis cs:2.45224413220058e-07,14169);
\draw[draw=none,fill=steelblue31119180] (axis cs:2.45224413220058e-07,1e-10) rectangle (axis cs:2.54675740052335e-07,16144);
\draw[draw=none,fill=steelblue31119180] (axis cs:2.54675740052335e-07,1e-10) rectangle (axis cs:2.64127066884612e-07,22337);
\draw[draw=none,fill=steelblue31119180] (axis cs:2.64127066884612e-07,1e-10) rectangle (axis cs:2.73578422138598e-07,17724);
\draw[draw=none,fill=steelblue31119180] (axis cs:2.73578422138598e-07,1e-10) rectangle (axis cs:2.83029748970876e-07,38108);
\draw[draw=none,fill=steelblue31119180] (axis cs:2.83029748970876e-07,1e-10) rectangle (axis cs:2.92481075803153e-07,33770);
\draw[draw=none,fill=steelblue31119180] (axis cs:2.92481075803153e-07,1e-10) rectangle (axis cs:3.0193240263543e-07,30594);
\draw[draw=none,fill=steelblue31119180] (axis cs:3.0193240263543e-07,1e-10) rectangle (axis cs:3.11383729467707e-07,21493);
\draw[draw=none,fill=steelblue31119180] (axis cs:3.11383729467707e-07,1e-10) rectangle (axis cs:3.20835084721693e-07,22130);
\draw[draw=none,fill=steelblue31119180] (axis cs:3.20835084721693e-07,1e-10) rectangle (axis cs:3.30286411553971e-07,26708);
\draw[draw=none,fill=steelblue31119180] (axis cs:3.30286411553971e-07,1e-10) rectangle (axis cs:3.39737738386248e-07,24267);
\draw[draw=none,fill=steelblue31119180] (axis cs:3.39737738386248e-07,1e-10) rectangle (axis cs:3.49189065218525e-07,14887);
\draw[draw=none,fill=steelblue31119180] (axis cs:3.49189065218525e-07,1e-10) rectangle (axis cs:3.58640392050802e-07,36687);
\draw[draw=none,fill=steelblue31119180] (axis cs:3.58640392050802e-07,1e-10) rectangle (axis cs:3.68091747304788e-07,31433);
\draw[draw=none,fill=steelblue31119180] (axis cs:3.68091747304788e-07,1e-10) rectangle (axis cs:3.77543074137066e-07,28758);
\draw[draw=none,fill=steelblue31119180] (axis cs:3.77543074137066e-07,1e-10) rectangle (axis cs:3.86994400969343e-07,11809);
\draw[draw=none,fill=steelblue31119180] (axis cs:3.86994400969343e-07,1e-10) rectangle (axis cs:3.9644572780162e-07,117257);
\draw[draw=none,fill=steelblue31119180] (axis cs:3.9644572780162e-07,1e-10) rectangle (axis cs:4.05897054633897e-07,34119);
\draw[draw=none,fill=steelblue31119180] (axis cs:4.05897054633897e-07,1e-10) rectangle (axis cs:4.15348409887883e-07,9046);
\draw[draw=none,fill=steelblue31119180] (axis cs:4.15348409887883e-07,1e-10) rectangle (axis cs:4.24799736720161e-07,14655);
\draw[draw=none,fill=steelblue31119180] (axis cs:4.24799736720161e-07,1e-10) rectangle (axis cs:4.34251063552438e-07,12994);
\draw[draw=none,fill=steelblue31119180] (axis cs:4.34251063552438e-07,1e-10) rectangle (axis cs:4.43702390384715e-07,751);
\draw[draw=none,fill=steelblue31119180] (axis cs:4.43702390384715e-07,1e-10) rectangle (axis cs:4.53153717216992e-07,240);
\draw[draw=none,fill=steelblue31119180] (axis cs:4.53153717216992e-07,1e-10) rectangle (axis cs:4.62605044049269e-07,45);
\draw[draw=none,fill=steelblue31119180] (axis cs:4.62605044049269e-07,1e-10) rectangle (axis cs:4.72056399303256e-07,10);
\draw[draw=none,fill=steelblue31119180] (axis cs:4.72056399303256e-07,1e-10) rectangle (axis cs:4.81507697713823e-07,1e-10);
\draw[draw=none,fill=steelblue31119180] (axis cs:4.81507697713823e-07,1e-10) rectangle (axis cs:4.9095905296781e-07,1e-10);
\draw[draw=none,fill=steelblue31119180] (axis cs:4.9095905296781e-07,1e-10) rectangle (axis cs:5.00410408221796e-07,1e-10);
\draw[draw=none,fill=steelblue31119180] (axis cs:5.00410408221796e-07,1e-10) rectangle (axis cs:5.09861706632364e-07,1e-10);
\draw[draw=none,fill=steelblue31119180] (axis cs:5.09861706632364e-07,1e-10) rectangle (axis cs:5.19313061886351e-07,1);
\draw[draw=none,fill=red] (axis cs:5.19313061886351e-07,1e-10) rectangle (axis cs:5.28764360296918e-07,1e-10);
\draw[draw=none,fill=steelblue31119180] (axis cs:5.28764360296918e-07,1e-10) rectangle (axis cs:5.38215715550905e-07,1);
\end{axis}

\end{tikzpicture}

}
\caption{}
\label{fig:degree_distributionA}
\end{subfigure} 
\caption{(a) Node degree distribution (semi-log). (b) Complementary cumulative distribution (CCDF) for the node degree. (c) Approximated closeness centrality distribution.}\label{fig:degree_distribution}
\end{figure}

\paragraph{Degree distribution.}
As can be seen in Table~\ref{tab:basic_props}, the average degree of the undirected version of RNA-KG is relatively small (9.65).
Despite the graph sparsity, the diameter of the KG is also relatively small (36). 
On the other hand, as shown in Figure~\ref{fig:degree_distributionB}, the degree distribution suggests a heavy-tailed distribution.
All of these properties are usually associated with scale-free graphs, which is a common structure in real-world complex systems.
These properties motivate the computation of the empirical {\em complementary cumulative distribution function} (CCDF) for the degree reported in Figure~\ref{fig:degree_distributionC}.
This curve approximates the probability distribution that a randomly selected node has a degree greater than or equal to $x$.
The linear trend in the plot is usually associated with a powerlaw distribution, where the CCDF is given by a function proportional to $x^{1-\alpha}$.
We estimate the power of the distribution using~\cite{powelaw}, obtaining a value of $\alpha=1.832$. 
The theoretical powerlaw obtained for the degrees is shown in Figure~\ref{fig:degree_distributionC} together with other common heavy-tailed distributions. 
Among these alternatives, we found that the powerlaw distribution fits better according to the log-likelihood ratio criterion~\cite{powerlaw_compare} with $p$-values smaller or equal than $10^{-6}$.
Further exploration should be made to confirm the powerlaw properties of the graph since they are usually associated with a hierarchical modular structure of the network, entailing algorithmic advantages for its analysis.   
For instance, the closeness centrality distribution in Figure~\ref{fig:degree_distributionA} presents a bimodal behavior, which could be explained by the existence of a well-connected core usually present in heavy-tail degree distribution networks. 

\paragraph{Treewidth.} 
Treewidth is a graph parameter measuring the structural similarity between a graph and a tree. It is based on the construction of a tree decomposition which captures how subset of nodes can be grouped to form a tree structure that maintains the global structure of the former graph. For instance, graphs having treewidth equal to one are trees, cycles have treewidth two, and clique graphs have a treewidth equal to the number of nodes minus one. 
The computation of the treewidth is in NP, but several approximation strategies can be used~\cite{treewidth}. 
The upper bound ($8,554$ in Table \ref{tab:basic_props}), computed on the undirected version of the KG, can be considered relatively small because it represents about $1.5\%$ of the KG size. This result is consistent with a tree-like hierarchical structure of \rnakg that has a small and well-connected core.

\paragraph{Isomorphic node groups.}
\rnakg contains 829 {\em isomorphic node groups}, that is nodes with exactly the same neighbours, node and edge types.
Nodes in such groups are topologically indistinguishable, that is, swapping their identifiers would not change the graph topology.
These groups involve a total of 14.30K nodes (2.47\%) and 398.32K edges (4.54\%), with the largest one containing 1.04K nodes and 19.78K edges. 
This particular group has degree 19 and is composed of riboswitches, specifically putative Ile (GAU) T-box riboswitches, and contains sequences that are all located upstream of an isoleucine--tRNA ligase.
Other isomorphic node groups involve other riboswitches, tRNAs, tsRNAs, and tRFs.
The detected isomorphic group components involve sncRNAs all interacting with amino acids at a molecular level or that originate from tRNA with molecular interactions tied to specific amino acids. For example, these groups may include sequences such as {\tt Aspartic acid-GTC} tRNA sequences or {\tt tsRNA-Leucine}s.
All these isomorphic node groups deserve further investigation to check whether the involved molecules correspond to the same tRNA, riboswitch, tsRNA, or tRF and thus pruning them to improve the information quality of \rnakg. Indeed, these groups derive from different RNA sources and contain molecules presenting proprietary identifiers that might collapse. 

\section*{Usage Notes}

While every effort has been made to provide complete, up-to-date and correct information, we make no warranty about the completeness and  accuracy of the \rnakg content. The use of the data contained in this database is limited to research-oriented and non-profit activities only (as requested by the data sources from which the data have been extracted).

\section*{Code Availability}

The \rnakg’s project website is at \url{http://fievel.anacleto.di.unimi.it:9999}. The code to reproduce results, together with documentation and tutorials, is available in \rnakg’s GitHub repository at \url{https://github.com/AnacletoLAB/RNA-KG}. In addition, the repository contains information and Python scripts to build new versions of \rnakg as the underlying primary resources get updated and new data become available. \rnakg data resource is hosted on Zenodo under a persistent identifier \url{https://doi.org/10.5281/zenodo.10078877}. We have deposited the KG and all relevant intermediate files in this repository.

\section*{Acknowledgements}
This research was supported by the National Center for Gene Therapy and Drugs based on RNA Technology, PNRR-NextGenerationEU program (G43C22001320007).

\section*{Author contributions statement}
M.M. and G.V. designed the study.  Em.C. retrieved, processed, and harmonized datasets.  T.J.C., M.M. defined the methodology for the construction of the KG. Em.C. and S.B. worked on the specification of the mapping alignment. Em.C. generated the KG that was analyzed by A.C. and M.S.G. S.B. and P.P. set up the SPARQL endpoint and developed the SPARQL queries on the generated KG.  G.V., E.C., J.G., J.R., P.N.R. identified the possible applications of the KG in conducting knowledge discovery in life science. M.M., Em.C. and G.V. wrote the initial draft of the paper. All authors reviewed the final version of the manuscript and approved it.    

\section*{Competing interests}

The authors declare no competing interests.

\section*{Figures \& Tables}

\clearpage

\begin{figure}[htbp]
    \centering
   \begin{adjustbox}{max width=0.7\textwidth}{\usetikzlibrary{shapes,backgrounds}
\pagestyle{empty}
\def\hmdd{(-25.4,4.2) circle (1.2cm)}
\def\mirnet{(-24.5,8) circle (6.5cm)}
\def\epimir{(-26,13) circle (1.1cm)}
\def\smtwomir{(-28,11.5) circle (1.2cm)}
\def\mirtwodisease{(-25.4,4.6) circle (1.2cm)}
\def\tam{(-26,3.8) circle (1.4cm)}
\def\transmir{(-26.8,3.1) circle (0.8cm)}
\def\putmir{(-28,1.8) circle (1.2cm)}
\def\mirpathdb{(-25.8,2) circle (0.6cm)}
\def\mircancer{(-28.6,5.6) circle (2.5cm)}
\def\noncode{(-20.3,-3.2) circle (3.3cm)}
\def\lnctwocancer{(-22,-2.5) circle (1cm)}
\def\lncrnadisease{(-22,-3) circle (1.1cm)}
\def\lncrnawiki{(-24.8,-2.5) circle (2.7cm)}
\def\mirandola{(-32,9) circle (1.5cm)}
\def\vesciclepedia{(-32,11) circle (2cm)}
\def\rnacentral{(-14,0) circle (10.5cm)}
\def\lncbook{(-21,-1.5) circle (2cm)}
\def\mirbase{(-11,7.5) circle (2cm)}
\def\modomics{(-14.5,-3.8) circle (2.1cm)}
\def\ribocentre{(-8,3) circle (2.5cm)}
\def\snodb{(-11,-7.7) circle (2cm)}
\def\tarbase{(-19.8,6cm) circle (1.4cm)}
\def\mirecords{(-22.3,5.7cm) circle (1.3cm)}
\def\mirtarbase{(-20.9,5.25cm) circle (1.65cm)}
\def\trnadb{(-8,-3) circle (2cm)}
\def\mirdb{(-19.65,9.87) circle (4cm)}
\def\somamir{(-21,2.5) circle (4cm)}
\def\lncexpdb{(-27.8,-2.5) circle (2.5cm)}
\def\dbesslnk{(-32,-9) circle (2.2cm)}
\def\lncatlas{(-26,-9) circle (2.5cm)}
\def\viroiddb{(-37,-7) circle (2cm)}
\def\tbdb{(-33,-4) circle (2cm)}
\def\rswitch{(-33,2) circle (2cm)}
\def\addgene{(-36,5) circle (2cm)}
\def\tsrfun{(-12,13) circle (2cm)}
\def\trfdb{(-10.5,13) circle (2cm)}
\def\mintbase{(-11.3,14.5) circle (2cm)}
\def\drugbank{(-43,-0.5) circle (2.7cm)}
\def\icbpsirna{(-42.75,-3) circle (1.2cm)}
\def\aptaindex{(-40,-1) circle (2cm)}
\def\eskipfinder{(-45.5,-1) circle (2cm)}

\begin{tikzpicture}[thick,scale=10, every node/.style={scale=10},
  hmdd/.style={shape=circle, opacity=0.5, fill=red!60!green},
  mirpathdb/.style={shape=circle, opacity=0.5, fill=red!40!violet},
  tam/.style={shape=circle, opacity=0.5, fill=blue!60!white},
  transmir/.style={shape=circle, opacity=0.5, fill=yellow},
  mirtwodisease/.style={shape=circle, opacity=0.5, fill=red!80!yellow},
  lncrnadisease/.style={shape=circle, opacity=0.5, fill=blue!50!cyan},
  lnctwocancer/.style={shape=circle, opacity=0.5, fill=green!90!violet} 
]

        \fill[red!60!green, opacity=0.5] \hmdd;
        \fill[red!40!violet, opacity=0.5] \mirpathdb;
        \fill[blue!60!white, opacity=0.5] \tam;
        \fill[yellow, opacity=0.5] \transmir;
        \fill[red!80!yellow, opacity=0.5] \mirtwodisease;
        \fill[blue!50!cyan, opacity=0.5] \lncrnadisease;
        \fill[green!90!violet, opacity=0.5] \lnctwocancer;       
        
        \draw[draw = red!60!green] \hmdd node[below] {};
        \draw[draw= blue!60!white] \tam node[left]{};
        \draw[draw = orange]\mirnet node [above left=25pt, font = {\Huge\bfseries\sffamily},text=orange] {$miRNet$};
        \draw[draw=red!80!yellow] \mirtwodisease node [above]{};
        \draw[draw = magenta] \noncode node [below=30pt, font = {\Huge\bfseries\sffamily},text=magenta] {$NONCODE$};    
        \draw [draw = green!90!violet]\lnctwocancer node [above]{};
        \draw[draw = blue!50!cyan] \lncrnadisease node [left]{}; 
        \draw[draw = olive, text width=1cm] \lncrnawiki node [below left=20pt, font = {\huge\bfseries\sffamily},text=olive] {$Lnc$ $RNA$ $Wiki$};
        \draw[draw = blue!50!yellow] \mirandola node [font = {\Huge\bfseries\sffamily}, text= blue!50!yellow] {$miRandola$};
        \draw[draw = red!50!gray] \vesciclepedia node [font = {\Huge\bfseries\sffamily}, text=red!50!gray] {$Vesciclepedia$};
        \draw[draw = blue] \rnacentral node [above, font = {\Huge\bfseries\sffamily},text=blue] {$RNAcentral$};
        \draw \lncbook[draw = orange!40!blue] node [below right, font = {\Large\bfseries\sffamily},text=orange!40!blue] {$LncBook$};
        \draw \mirbase node [font = {\Huge\bfseries\sffamily}] {$miRBase$};
        \draw \modomics node [font = {\Huge\bfseries\sffamily}] {$Modomics$};
        \draw \ribocentre node [font = {\Huge\bfseries\sffamily}] {$Ribocentre$};
        \draw \snodb node [font = {\Huge\bfseries\sffamily}] {$snoDB$};
        \draw[draw = green!70!darkgray] \tarbase node [above right = 20pt, font = {\Large\bfseries\sffamily},text=green!70!darkgray] {$TarBase$};
        \draw[draw = orange!70!green] \mirecords node [above left = 20pt, font = {\Large\bfseries\sffamily},text=orange!70!green] {$miRecords$};
        \draw[draw = red!70] \mirtarbase node [below right= 20pt, font = {\Large\bfseries\sffamily},text=red!70] {$miRTarBase$};
        \draw \trnadb node [font = {\Huge\bfseries\sffamily}] {$tRNAdb$};
        \draw[draw = purple] \mirdb node [above left=23pt, font = {\huge\bfseries\sffamily},text=purple, text width=2cm] {$miRDB \cup TargetScan$};
        \draw \somamir[draw = cyan] node [below=20pt, font = {\huge\bfseries\sffamily},text=cyan] {$SomamiR$};
        \draw \lncexpdb[draw = red] node [left=3pt, font = {\huge\bfseries\sffamily},text=red, text width=2cm] {$Lnc$ $Exp$ $DB$};
        \draw \mircancer[draw = teal] node [left=40pt, font = {\Large\bfseries\sffamily},text=teal, text width=2cm] {$miRCancer \cup miRdSNP \cup PolymiRTS \cup dbDEMC$};
        \draw[draw=yellow] \transmir node [below] {};
        \draw[draw=red!40!violet] \mirpathdb node [below] {};
        \draw \putmir node [below, font = {\Large\bfseries\sffamily}] {$PuTmiR$};
        \draw \epimir node [font = {\Large\bfseries\sffamily}] {$EpimiR$};
        \draw \smtwomir node [font = {\Large\bfseries\sffamily}] {$SM2miR$};
        \draw[draw=red] \tsrfun node [left=15pt, font = {\Huge\bfseries\sffamily},text=red] {$tsRFun$};
        \draw[draw=green!70!darkgray] \trfdb node [right=15pt, font = {\Huge\bfseries\sffamily},text=green!70!darkgray] {$tRFdb$};
        \draw[draw=blue] \mintbase node [above=15pt, font = {\Huge\bfseries\sffamily},text=blue] {$MINTbase$};
        \draw[draw=red] \drugbank node [above=17pt,font = {\Huge\bfseries\sffamily},text=red, text width=2cm] {$Drug$ $Bank$};
        \draw[draw=green!60!darkgray] \icbpsirna node [below=35pt,font = {\Huge\bfseries\sffamily},text=green!60!darkgray, text width=2cm] {$ICBP$ $siRNA$};
        \draw[draw=blue!80!white] \aptaindex node [right=0.0001, font = {\Huge\bfseries\sffamily},text=blue!80!white, text width=2cm] {$Apta$--$Index$};
        \draw[draw=purple!80!darkgray] \eskipfinder node [left=25pt, font = {\Huge\bfseries\sffamily},text=purple!80!darkgray, text width=2cm] {$eSkip$--$Finder$};

        \draw \dbesslnk node [font = {\Huge\bfseries\sffamily}] {$dbEssLnc$};
        \draw \lncatlas node [font = {\Huge\bfseries\sffamily}] {$lncATLAS$};
        \draw \viroiddb node [font = {\Huge\bfseries\sffamily}] {$viroidDB$};
        \draw \tbdb node [font = {\Huge\bfseries\sffamily}] {$TBDB$};
        \draw \rswitch node [font = {\Huge\bfseries\sffamily}] {$RSwitch$};
        \draw \addgene node [font = {\Huge\bfseries\sffamily}] {$Addgene$};

        \matrix [draw, above left, font = {\Large\bfseries\sffamily}] at (-37,9) {
            \node [hmdd, draw=red!60!green, label=right:$HMDD$, minimum size=1.8cm] {}; &
            \node [mirpathdb, draw=red!40!violet, label=right:$miRPathDB$, minimum size=1.8cm] {};\\
            \node [tam, draw=blue!60!white, label=right:$TAM$, minimum size=1.8cm] {}; &
            \node [transmir, draw=yellow, label=right:$TransmiR$, minimum size=1.8cm] {}; \\
            \node [mirtwodisease, draw=red!80!yellow, label=right:$miR2Disease$, minimum size=1.8cm] {}; &
            \node [lncrnadisease, draw=blue!50!cyan, label=right:$LncRNADisease$, minimum size=1.8cm] {}; \\
            \node [lnctwocancer, draw=green!90!violet, label=right:$Lnc2Cancer$, minimum size=1.8cm] {}; \\
        };
        
\end{tikzpicture}}
    \end{adjustbox}
    \caption{Bubbles represent the relationships made available from the considered data sources. Overlapping and inclusions of bubbles show the presence of common relationships among the considered data sources.} 
    \label{fig:venn}
\end{figure}

\clearpage

\clearpage

\newpage
\begin{table}[htbp]
    \centering
    \begin{small}
    \begin{tabular}{|l|l|p{10cm}|}
    \hline\hline
      {\bf Name} & {\bf Abbr.} & {\bf Description}\\ \hline\hline 
      \href{https://hpo.jax.org/app/}{Human Phenotype Ontology} \cite{hpo}
          & HPO & Terms representing medically relevant phenotypes and disease-phenotype annotations.\\
        \hline
         \href{https://geneontology.org}{Gene Ontology} \cite{go}
          & GO & Terms representing attributes of gene products in all organisms. Cellular component, molecular function, and biological process domains are covered.\\
        \hline
\href{http://obofoundry.org/ontology/mondo.html}{Monarch Merged Disease Ontology} \cite{mondo}
          & Mondo & Terms representing human diseases.\\
        \hline
        \href{http://www.violinet.org/vaccineontology}{Vaccine Ontology} \cite{vo}
          & VO & Terms in the domain of vaccine and vaccination. \\
        \hline
        \href{https://www.ebi.ac.uk/chebi/}{Chemical Entities of Biological Interest} \cite{chebi}
          & ChEBI & Structured classification of molecular entities of biological interest focusing on ``small'' chemical compounds. \\
        \hline
        \href{http://obophenotype.github.io/uberon/}{Uber-anatomy Ontology} \cite{uberon}
          & Uberon & Terms representing body parts, organs and tissues in a variety of animal species, with a focus on vertebrates. \\
        \hline
        \href{http://www.clo-ontology.org/}{Cell Line Ontology} \cite{clo}
          & CLO & Terms representing publicly available cell lines. \\
        \hline
        \href{https://proconsortium.org/}{PRotein Ontology} \cite{pro}*
          & PRO & Terms representing protein-related entities (including specific modified forms, orthologous isoforms, and protein complexes). \\
        \hline
        \href{http://www.sequenceontology.org}{Sequence Ontology} \cite{so}
          & SO & Terms representing
          features and properties of nucleic acid used in biological sequence annotation. \\
          \hline
        \href{https://rgd.mcw.edu/rgdweb/ontology/search.html}{Pathway Ontology} \cite{pw}
          & PW & Terms for annotating gene products to pathways. \\
        \hline
        \href{https://github.com/oborel/obo-relations/}{Relation Ontology} \cite{ro}
          & RO & Terms and properties representing relationships used across a wide variety of biological ontologies.\\
        \hline\hline
    \end{tabular} 
    \end{small}
    \caption{Main biomedical ontologies employed for \rnakg construction (* modified to exclude all non-human proteins).}
    \label{tab:ontologies}
\end{table}

\begin{table}[htbp]
    \centering
 \begin{scriptsize}
    \begin{tabular}{|c|l|c|c|c|c|c|c|c|c|c|}
    \hline\hline
  {\bf Type} & {\bf Data source} &  {\bf Species} &  {\bf \# RNAs} &  {\bf Format} & {\bf API} & {\bf Threshold} & {\bf SI} & {\bf Relation with} &  {\bf TI} & {\bf \# Relation} \\ \hline 
\hline
\multirow{20}{*}{\vspace*{-6cm}{\bf miRNA}} &\href{https://www.mirbase.org}{miRBase} \cite{mirbase} & 271 &  87,474 & rel/CSV & no & & WR & premiRNA & WR & 48,885 \\ \cline{2-11} &
\href{https://mirdb.org/}{miRDB}
\cite{mirdb} & 5 & 7,086  & CSV & no & $\sigma>80$ & WR & mRNA & WR & 3,519,884 \\ \cline{2-11}&
\makecell[l]{\href{https://www.mirnet.ca/miRNet/}{miRNet} \cite{mirnet}} & 10 & 7,928 & rel/CSV & yes & & WR & \makecell{SNP \\ mRNA \\ snoRNA \\ chemical \\ TF \\ epi. mod. \\ lncRNA \\ pseudogene \\ circRNA \\ disease} & \makecell{WR \\ WR \\ WR \\ M \\ M \\ M \\ WR \\ WR \\ WR \\ M} & \makecell{67,532 \\ 3,025,487 \\ 9,738 \\ 4,935 \\ 3,311 \\ 1,955 \\ 31,345 \\ 59,417 \\ 804,086 \\ 32,004} \\ \cline{2-11} &
\href{http://c1.accurascience.com/miRecords/}{miRecords}
\cite{mirecords} & 9 & 384 & CSV & no & validated & WR &  mRNA & M & 1,529 \\ \cline{2-11} &
\href{https://www.cuilab.cn/hmdd}{HMDD}
\cite{hmdd} & HS & 1,206 & CSV & no & & WR & disease & M & 35,547 \\ \cline{2-11} & 
\href{http://www.jianglab.cn/EpimiR/}{EpimiR}
\cite{epimir} & 7 & 617 & CSV & no & & WR & epi. mod. & M & 1,974 \\ \cline{2-11} &
\href{http://watson.compbio.iupui.edu:8080/miR2Disease/}{miR2Disease}
\cite{mir2disease} & HS & 349 & CSV & no & & WR & disease & O & 3,273 \\ \cline{2-11} &
\href{https://www.targetscan.org/}{TargetScan}
\cite{Targetscan} & 5 & 5,168 & CSV & no & validated & WR & mRNA & WR & 2,850,014 \\ \cline{2-11} &
\makecell[l]{\href{https://compbio.uthsc.edu/SomamiR/}{SomamiR DB}
\cite{somamir}} & HS & 1,078 & CSV & no & validated & WR & \makecell{mRNA \\ circRNA \\ lncRNA \\ disease} & \makecell{WR \\ WR \\ WR \\ M} & \makecell{2,313,416 \\ 428,237 \\ 127,025 \\ 2,424} \\ \cline{2-11} & 
\href{https://dianalab.e-ce.uth.gr/html/diana/web/index.php?r=tarbasev8}{TarBase}
\cite{tarbase} & 18 & 2,156 & rel/CSV & no & & WR & mRNA & WR & 665,843 \\ \cline{2-11} &
\href{https://mirtarbase.cuhk.edu.cn/~miRTarBase/miRTarBase_2022/php/index.php}{miRTarBase}
\cite{miRTarBase} &	28 & 4,630 & CSV & no & & WR & mRNA & WR & 2,200,449 \\ \cline{2-11} &
\href{http://www.jianglab.cn/SM2miR/}{SM2miR}
\cite{SM2miR} &	21 & 1,658 & CSV & no & & WR & chemical & M & 4,989 \\ \cline{2-11} &
\href{https://www.cuilab.cn/transmir}{TransmiR}
\cite{TransmiR} & 19 & 785 & CSV & no & validated & WR & TF & M & 3,730 \\ \cline{2-11} &
\href{https://compbio.uthsc.edu/miRSNP/}{PolymiRTS}
\cite{PolymiRTS} & HS & 11,182 & rel/CSV & no & & WR & disease & M & 83,516 \\ \cline{2-11} &
\href{https://www.biosino.org/dbDEMC/index}{dbDEMC}
\cite{dbDEMC} & HS & 3,268 & CSV & no& $p_{val}$<0.01 & WR & disease & M & 160,800 \\ \cline{2-11} &
\makecell[l]{\href{http://www.lirmed.com/tam2/}{TAM}
\cite{TAM}} & HS & 1,209 & CSV & no & & WR & \makecell{mol. function \\ miRNA \\ TF \\ disease \\ anatomy} & \makecell{M \\ WR \\ M \\ M \\ M} & \makecell{2,538 \\ 1,218 \\ 165 \\ 12,516 \\ 58} \\ \cline{2-11} &
\href{https://www.isical.ac.in/~bioinfo_miu/TF-miRNA1.php}{PuTmiR}
\cite{putmir} & HS & 1,296 & CSV & no & & WR & TF & M & 12,097 \\ \cline{2-11} &
\href{https://mpd.bioinf.uni-sb.de/overview.html}{miRPathDB}
\cite{miRPathDB} & HS, MM & 29,430 & CSV & no & FDR<0.01 & WR & mol. function & M & 3,063 \\ \cline{2-11} &
\href{http://mircancer.ecu.edu/}{miRCancer}
\cite{miRCancer}  & HS & 57,984 & CSV & no & & WR & disease & M & 9,080 \\ \cline{2-11} &
\makecell[l]{\href{http://mirdsnp.ccr.buffalo.edu/}{miRdSNP}
\cite{miRdSNP}}  & HS & 249 & CSV & no & & WR & \makecell{disease\\SNP} & \makecell{WR\\M} & \makecell{786\\758} \\ \cline{2-11} &
\makecell[l]{\href{http://mirandola.iit.cnr.it/}{miRandola}
\cite{mirandola}}  & 14 & 1,002 & CSV & no & & WR & \makecell{extracell. form\\chemical} & \makecell{M \\ M} &  \makecell{3,262\\25} \\
\hline \hline
\makecell{{\bf mRNA vaccine}}&\makecell[l]{\href{https://go.drugbank.com/categories/DBCAT005631}{DrugBank} \cite{drugbank}}
& & 4 & rel/RDF & 
yes\textsuperscript{S} & & P & disease & M & 7 \\
\hline \hline
\multirow{2}{*}{\vspace*{-0.25cm}{\bf s(i/h)RNA}} &\href{http://web.mit.edu/sirna/}{ICBP siRNA}
\cite{sirna}
& HS, MM & 147 & HTML & no & & P & mRNA & WR & 147 \\ \cline{2-11} &
\makecell[l]{\href{https://go.drugbank.com/indications/DBCOND0093099}{DrugBank} \cite{drugbank}}
& & 4 & rel & 
yes\textsuperscript{S} & & P & \makecell{mRNA \\ disease} & \makecell{WR \\ M} & \makecell{3 \\ 3} \\ \hline\hline
\multirow{2}{*}{\vspace*{-0.25cm}{\makecell{{\bf RNA}\\{\bf aptamer}}}} & \makecell[l]{\href{https://www.aptagen.com/apta-index/}{Apta-Index} \cite{Aptamer}}
& & 230 & rel & no & & P & \makecell{chemical \\ protein} & \makecell{M \\ M} & \makecell{77 \\ 153} \\ \cline{2-11} &
\makecell[l]{\href{https://go.drugbank.com/categories/DBCAT001641}{DrugBank} \cite{drugbank}}
& & 2 & rel/RDF & 
yes\textsuperscript{S} & & P & \makecell{protein \\ disease} & \makecell{M \\ M} & \makecell{2 \\ 2} \\
\hline \hline
\multirow{2}{*}{\vspace*{-0.5cm}{{\bf ASO}}} &\href{https://eskip-finder.org/cgi-bin/input.cgi}{eSkip-Finder}
\cite{eskip}  & 4 & 2,196 & rel & no & & P & mRNA & WR & 11,778 \\ \cline{2-11} &
\makecell[l]{\href{https://go.drugbank.com/categories/DBCAT001709}{DrugBank}
\cite{drugbank}} & & 12 & rel/RDF & 
yes\textsuperscript{S} & & P & \makecell{protein \\ mRNA \\ disease } & \makecell{M \\ WR \\ M} & \makecell{12 \\ 7 \\ 11} \\
\hline \hline
{\bf gRNA} &\href{https://www.addgene.org/crispr/reference/grna-sequence/}{Addgene} \cite{addgene}  & 29 & 296 & HTML & no & & P & gene & WR & 321 \\
\hline \hline
    \end{tabular}
     \end{scriptsize}
    \caption{
    Main data sources (Part I). For each type of RNA molecule, the table reports the corresponding data sources. Moreover, for each data source, {\tt Species} and {\tt \#RNAs} columns specify the number of species and distinct sequences ({\tt HS} and {\tt MM} tags refer to specific species  {\em Homo sapiens} and {\em Mus musculus}); {\tt Relation with} and {\tt \#Relation} columns specify the distinct relationships with bio-entities and their number; {\tt Format} column refers to the data format ({\tt CSV} for flatfiles, {\tt rel} for relational tables, {\tt RDF}, or {\tt HTML} for web pages); {\tt API} column reports the availability of API or SPARQL endpoints (the last one denoted with the superscript {\tt s}) for data access; {\tt Threshold} column provides identified quality threshold within the source. {\tt SI} and {\tt TI} columns contain the class of the identification schemes ({\tt WR} -- {\em Well-Reputed}, {\tt O} -- {\em Ontology-based}, {\tt M} -- {\em Mapping}- {\em based}, and {\tt P} -- {\em Proprietary}) adopted respectively by source and target(s) within a specific resource (the source is the RNA molecule specified in the {\tt Type} column, whereas target(s) are specified in the {\tt Relation with} column).}
    \label{tab:database1}
\end{table}

\begin{table}[htbp]
    \centering
     \begin{scriptsize}
    \begin{tabular}{|c|l|c|c|c|c|c|c|c|c|c|}
    \hline\hline
  {\bf Type} & {\bf Data source} &  {\bf Species} &  {\bf \# RNAs} & {\bf Format} & {\bf API} & {\bf Threshold} &  {\bf SI} & {\bf Relation with} &  {\bf TI} & {\bf \# Relation} \\ \hline 
\hline
\multirow{8}{*}{\vspace*{-3cm}{{\bf lncRNA}}} & \makecell[l]{\href{https://ngdc.cncb.ac.cn/lncbook/}{LncBook}
\cite{LncBook}}  & HS & 323,950 & rel/CSV & no & & WR & \makecell{miRNA \\ small protein \\ disease \\ biological context} & \makecell{WR \\ WR \\ M \\ M} & \makecell{146,092,274 \\ 772,745 \\ 34,536 \\ 95,243} \\ \cline{2-11} &
\href{http://www.rnanut.net/lncrnadisease/}{LncRNADisease}
\cite{LncRNADisease}  & 4 & 6,066 & CSV & no & & WR & disease & M & 20,277 \\ \cline{2-11} &
\href{https://ngdc.cncb.ac.cn/lncexpdb/}{LncExpDB}
\cite{LncExpDB}  & HS & 101,293 & rel/CSV & no & & WR & mRNA & WR & 28,443,865 \\ \cline{2-11} &
\makecell[l]{\href{https://esslnc.pufengdu.org/}{dbEssLnc}
\cite{dbEssLnc}}  & HS, MM & 207 & JSON & no & & WR & \makecell{biological role \\ biological process} & \makecell{P \\ O} & \makecell{207 \\ 28} \\ \cline{2-11} &
\href{https://lncatlas.crg.eu/}{lncATLAS}
\cite{lncATLAS}  & HS & 6,768 & CSV & no & & WR & cell. comp. & M & 2,429,368 \\ \cline{2-11} &
\href{http://www.noncode.org/index.php}{NONCODE}
\cite{NONCODE}  & 39 & 644,510 & rel & no & & WR & disease & O & 32,226 \\ \cline{2-11} &
\href{http://bio-bigdata.hrbmu.edu.cn/lnc2cancer/}{Lnc2Cancer}
\cite{lnc2cancer}  & HS & 3,402 & CSV & no & & WR & disease & O & 9,254 \\ \cline{2-11} &
\makecell[l]{\href{https://ngdc.cncb.ac.cn/lncrnawiki/}{LncRNAWiki} \cite{LncRNAWiki}} & HS & 106,063 & rel/CSV & no & & WR & \makecell{small protein\\disease\\biological context\\cell. comp.\\gene\\miRNA\\TF\\biological process\\mol. function\\chemical\\pathway} & \makecell{WR \\M\\M\\M\\WR\\WR\\M\\M\\M\\M\\M} & \makecell{9,387\\7,634\\18,453\\4,969\\509\\210\\232\\10,806\\1,800\\789\\571} \\
\hline \hline
{\bf Ribozyme} & \href{https://www.ribocentre.org/}{Ribocentre}
\cite{Ribocentre} & 1,195 & 21,084 & rel & no & & P & biological process & M & 34 \\
\hline \hline
\makecell{{\bf Viral  RNA}} & \makecell[l]{\href{https://viroids.org/}{ViroidDB}
\cite{viroiddb}}  & 9 & 9,891 & CSV & no & & WR & ribozyme & P & 17,460 \\
\hline \hline
\multirow{2}{*}{{\bf Riboswitch}}  & \href{https://tbdb.io/}{TBDB}
\cite{tbdb}  & 3,621 & 23,497 & CSV & no & & P & protein & M & 23,535 \\ \cline{2-11} &
\href{https://penchovsky.atwebpages.com/applications.php?page=58}{RSwitch}
\cite{rswitch}  & 50 & 215 & rel/CSV & no & & P & bact. strain & WR & 215 \\
\hline \hline
\multirow{3}{*}{\vspace*{-0.5cm}{\makecell{\textbf{tRF}\\\textbf{\& tsRNA}}}}  & \href{http://genome.bioch.virginia.edu/trfdb/}{tRFdb}
\cite{trfdb}  & 7 & 863 & CSV & no & & P & tRNA & P & 792 \\ \cline{2-11} &
\makecell[l]{\href{https://rna.sysu.edu.cn/tsRFun/}{tsRFun}
\cite{tsrfun}} & HS & 3,940 & CSV & no & \makecell{FDR<0.01} & P &\makecell{miRNA \\ tRNA \\ disease} & \makecell{WR \\ P \\ M} & \makecell{45,165 \\ 46,798 \\ 4,620} \\ \cline{2-11} &
\href{https://cm.jefferson.edu/MINTbase/}{MINTbase}
\cite{mintbase}  & HS & 28,824 & CSV & no & & P & tRNA & P & 125,285 \\
\hline \hline
{\bf snoRNA} & \makecell[l]{\href{https://bioinfo-scottgroup.med.usherbrooke.ca/snoDB/}{snoDB}
\cite{snoDB}} & HS & 751 & CSV & no & & WR & \makecell{gene \\ mRNA \\ lncRNA \\ miRNA \\ pseudogene \\ rRNA \\ snoRNA \\ snRNA \\ tRNA \\ scaRNA} & \makecell{WR \\ WR \\ WR \\ WR \\ WR \\ P \\ WR \\ WR \\ P \\ WR} & \makecell{763 \\ 276 \\ 45 \\ 17 \\ 10 \\ 735 \\ 670 \\ 164 \\ 164 \\ 34} \\
\hline \hline
{\bf tRNA} & \href{http://trna.bioinf.uni-leipzig.de/DataOutput/}{tRNAdb}
\cite{tRNAdb}  & 681 & 9,758 & rel & no & & P & amino acid & M & 8,872 \\
\hline \hline
\multirow{9}{*}{\makecell{{\bf Inter}\\{\bf RNA}}}  & \makecell[l]{\href{http://www.rnainter.org/}{RNAInter}
\cite{rnainter}} & 156 & 455,887 & CSV & yes
& $\sigma \geq 0.2886$ & WR & \makecell{chemical \\ histone mod.\\RBP\\TF\\protein\\gene} & \makecell{M \\ P\\M\\M\\M\\WR} & \makecell{10,890 \\ 1,060,685\\5,200,067\\9,323,690\\22,543,829\\119,377} \\ \cline{2-11} &
\href{http://www.rna-society.org/rnalocate/}{RNALocate}
\cite{rnalocate}  & 104 & 123,592 & CSV & yes & & WR & cell. comp. & M & 213,429 \\ \cline{2-11} &
\href{http://www.rnadisease.org/}{RNADisease}
\cite{rnadisease}  & 117 & 91,245 & CSV & yes & $\sigma \geq 0.95$ & WR & disease & O & 343,273 \\ \cline{2-11} &
\href{https://www.rna-society.org/ncrdeathdb/}{ncRDeathDB}
\cite{ncrdeathdb} & 12 & 648 & CSV & yes & & WR & prog. cell death & M & 4,615 \\ \cline{2-11} &
\href{http://www.rna-society.org/cncrnadb/}{cncRNADB}
\cite{cncrnadb}  & 21 & 2,002 & CSV & yes & & WR & anatomy & M & 2,598 \\ \cline{2-11} &
\href{http://www.rna-society.org/virbase/}{ViRBase}
\cite{virbase}  & 152 & 41,718 & CSV & yes & $\sigma \geq 0.7$ & WR & \makecell{viral RNA\\viral protein} & \makecell{WR\\M} & \makecell{719,214\\195} \\ \cline{2-11} &
\href{http://microvesicles.org/}{Vesiclepedia}
\cite{vesiclepedia}  & 41 & 20,490 & CSV & no & & WR & extracell. form & M & 388,154 \\ \cline{2-11} &
\href{http://www.rnamd.org/directRMDB/index.html}{DirectRMDB} \cite{DirectRMDB}  & 25 & 19,702 & CSV & no & & WR & epi. mod. & WR & 904,712 \\ \cline{2-11} &
\href{https://genesilico.pl/modomics/}{Modomics} \cite{modomics}  & 32 & 225 & rel/RDF & yes & & WR & epi.mod. & WR & 276 \\ 
\hline \hline
    \end{tabular}
 \end{scriptsize}
    \caption{Main  data sources (Part II).} 
    \label{tab:database2}
\end{table}

\begin{table}[htbp]
    \centering
    \begin{small}
    \begin{tabular}{|c|l|c|l|}
    \hline\hline
      {\bf Relation ID}  & {\bf Name} & {\bf Inverse Relation ID} & {\bf Inverse Name} \\ \hline\hline 
      \href{http://purl.obolibrary.org/obo/RO_0000056}{RO:0000056}
          & participates in & \href{http://purl.obolibrary.org/obo/RO_0000057}{RO:0000057} & has participant \\
        \hline
        \href{http://purl.obolibrary.org/obo/RO_0000079}{RO:0000079}
          & function of & \href{http://purl.obolibrary.org/obo/RO_0000085}{RO:0000085} & has function \\
        \hline
        \href{http://purl.obolibrary.org/obo/RO_0001015}{RO:0001015}
          & location of & \href{http://purl.obolibrary.org/obo/RO_0001025}{RO:0001025} & located in \\
        \hline
        \href{http://purl.obolibrary.org/obo/RO_0001018}{RO:0001018}
          & contained in & \href{http://purl.obolibrary.org/obo/RO_0001019}{RO:0001019}
          & contains \\
        \hline
        \href{http://purl.obolibrary.org/obo/RO_0002202}{RO:0002202}
        & develops from & \href{http://purl.obolibrary.org/obo/RO_0002203}{RO:0002203}
        & develops into  \\
        \hline
        \href{http://purl.obolibrary.org/obo/RO_0002204}{RO:0002204}
        & gene product of & \href{http://purl.obolibrary.org/obo/RO_0002205}{RO:0002205}
        & has gene product  \\
        \hline
        \href{http://purl.obolibrary.org/obo/RO_0002245}{RO:0002245}
        & over-expressed in &  & \\
        \hline
        \href{http://purl.obolibrary.org/obo/RO_0002260}{RO:0002260}
        & has biological role &  & \\
        \hline
        \href{http://purl.obolibrary.org/obo/RO_0002246}{RO:0002246}
        & under-expressed in & &  \\
        \hline
        \href{http://purl.obolibrary.org/obo/RO_0002291}{RO:0002291}
        & ubiquitously expressed in &  \href{http://purl.obolibrary.org/obo/RO_0002293}{RO:0002293}
        & ubiquitously expresses \\
        \hline
        \href{http://purl.obolibrary.org/obo/RO_0002302}{RO:0002302}
        & is treated by substance & \href{http://purl.obolibrary.org/obo/RO_0002606}{RO:0002606}
        & is substance that treats  \\
        \hline
        \href{http://purl.obolibrary.org/obo/RO_0002430}{RO:0002430}
          & involved in regulation of & &  \\
        \hline
        \href{http://purl.obolibrary.org/obo/RO_0002430}{RO:0002430}
        & involved in negative regulation of & &  \\
        \hline
        \href{http://purl.obolibrary.org/obo/RO_0002434}{RO:0002434}
        & interacts with* & &  \\
        \hline
        \href{http://purl.obolibrary.org/obo/RO_0002436}{RO:0002436}
        & molecularly interacts with* & & \\
        \hline
        \href{http://purl.obolibrary.org/obo/RO_0002526}{RO:0002526}
        & overlaps sequence of* & &  \\
        \hline
        \href{http://purl.obolibrary.org/obo/RO_0002528}{RO:0002528}
        & is upstream of sequence of & \href{http://purl.obolibrary.org/obo/RO_0002529}{RO:0002529}
        & is downstream of sequence of  \\
        \hline
        \href{http://purl.obolibrary.org/obo/RO_0002559}{RO:0002559}
        & causally influenced by &  \href{http://purl.obolibrary.org/obo/RO_0002566}{RO:0002566}
        & causally influences \\
        \hline
        \href{http://purl.obolibrary.org/obo/RO_0003002}{RO:0003002}
        & represses expression of & &  \\
        \hline
        \href{http://purl.obolibrary.org/obo/RO_0003302}{RO:0003302}
        & causes or contributes to condition &  & \\
        \hline
        \href{http://purl.obolibrary.org/obo/RO_0010001}{RO:0010001}
        & generically depends on & \href{http://purl.obolibrary.org/obo/RO_0010002}{RO:0010002}
        & is carrier of \\
        \hline
        \href{http://purl.obolibrary.org/obo/RO_0011002}{RO:0011002}
        & regulates activity of & & \\
        \hline
        \href{http://purl.obolibrary.org/obo/RO_0011007}{RO:0011007}
        & decreases by repression quantity of & & \\
        \hline
        \hline
    \end{tabular} 
    \end{small}
    \caption{Main relations among bio-entities involving RNA with the RO identifier (* symmetric relationship).}
    \label{tab:relations}
\end{table}

\begin{table}[htbp]
    \centering
        \begin{tabular}{ll}
         Graph parameter\\
         \hline
         Number of nodes & 578,384\\
         Number of directed edges & 8,768,582\\
         Max out degree & 27,109\\
         Max in degree & 117,135\\
         \hline
         Number of edges* & 5,583,802\\ 
         Max degree* & 117,142\\
         Min degree* & 1\\
         Mean degree* & 9.65\\ 
         Diameter* & 36\\
         Upper bound Treewidth* & 8,554\\
         Mean closeness centrality* &
         $3.38\times 10^{-7}$\\
         \hline
    \end{tabular}

    \caption{Basic topological properties of \rnakg (* 
    calculated on the undirected version of the KG).
    }
    \label{tab:basic_props}
\end{table}

\clearpage

\setcounter{table}{0}
\renewcommand{\tablename}{Supplementary Table}
\setcounter{figure}{0}
\renewcommand{\figurename}{Supplementary Figure}
\renewcommand{\lstlistingname}{Supplementary Listing}

\section*{Supplementary material}
Supplementary Table \ref{tab:databasesont2} specifies the bio-ontologies that can be exploited for representing concepts in RNA sources. RNA sources are categorized according to the main treated molecules of RNA (whose characteristics are reported in Tables \ref{tab:database1}-\ref{tab:database2}).

\begin{figure}[htbp]
    \centering
\begin{tikzpicture}

\definecolor{darkgray176}{RGB}{176,176,176}
\definecolor{gray}{RGB}{128,128,128}

\begin{axis}[
x=0.35cm,
y=0.35cm,
tick align=outside,
tick pos=left,
x grid style={darkgray176},
xmin=0, xmax=25,
xtick style={color=black},
xtick={0.5,1.5,2.5,3.5,4.5,5.5,6.5,7.5,8.5,9.5,10.5,11.5,12.5,13.5,14.5,15.5,16.5,17.5,18.5,19.5,20.5,21.5,22.5,23.5,24.5},
xticklabel style={rotate=90.0},
xticklabels={
  aptamer,
  ASO,
  chemical,
  circRNA,
  disease,
  gene,
  gRNA,
  lncRNA,
  miRNA,
  mRNA,
  piRNA,
  protein,
  pseudogene,
  riboswitch,
  ribozyme,
  rRNA,
  scRNA,
  shRNA,
  siRNA,
  snoRNA,
  snRNA,
  tRF,
  tRNA,
  tsRNA,
  viral RNA
},
y dir=reverse,
y grid style={darkgray176},
ymin=0, ymax=51,
ytick style={color=black},
ytick={0.5,1.5,2.5,3.5,4.5,5.5,6.5,7.5,8.5,9.5,10.5,11.5,12.5,13.5,14.5,15.5,16.5,17.5,18.5,19.5,20.5,21.5,22.5,23.5,24.5,25.5,26.5,27.5,28.5,29.5,30.5,31.5,32.5,33.5,34.5,35.5,36.5,37.5,38.5,39.5,40.5,41.5,42.5,43.5,44.5,45.5,46.5,47.5,48.5,49.5,50.5},
yticklabels={
  Addgene,
  Apta-Index,
  cncRNADB,
  dbDEMC,
  dbEssLnc,
  DirectRMDB,
  DrugBank,
  EpimiR,
  eSkip-Finder,
  HMDD,
  ICBP siRNA,
  Lnc2Cancer,
  LncATLAS,
  LncBook,
  LncExpDB,
  LncRNADisease,
  LncRNAWiki,
  MINTbase,
  miR2Disease,
  miRandola,
  miRBase,
  miRCancer,
  miRDB,
  miRdSNP,
  miRecords,
  miRNet,
  miRPathDB,
  miRTarBase,
  Modomics,
  ncRDeathDB,
  PolymiRTS,
  PuTmiR,
  Ribocentre,
  RNADisease,
  RNAInter,
  RNALocate,
  RSwitch database,
  SM2miR,
  snoDB,
  SomamiR,
  TAM,
  TarBase,
  TargetScan,
  TBDB,
  TransmiR,
  tRFdb,
  tRNAdb,
  tsRFun,
  Vesiclepedia,
  ViRBase,
  ViroidDB
}
]
\addplot graphics [includegraphics cmd=\pgfimage,xmin=0, xmax=25, ymin=51, ymax=0] {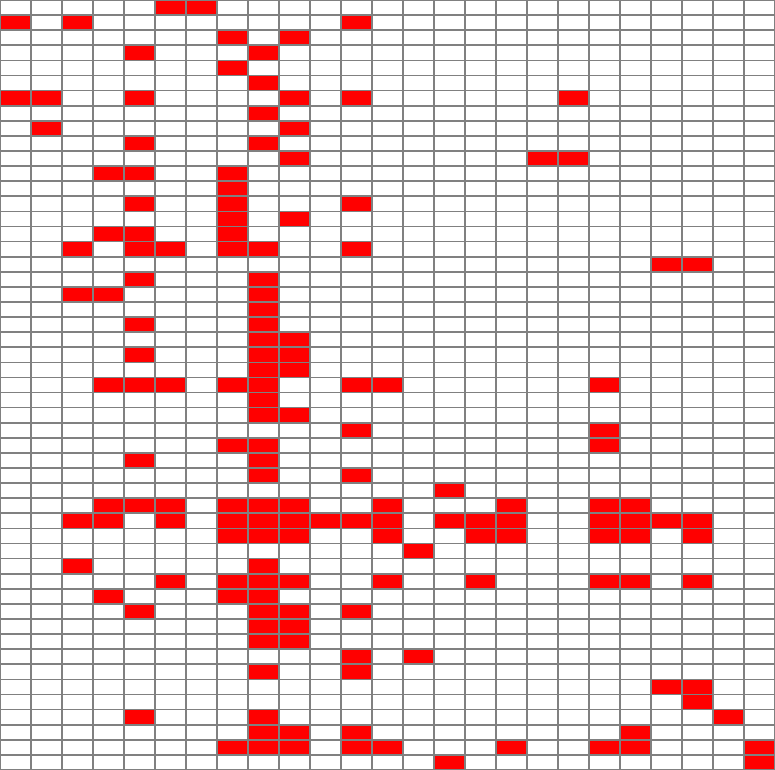};
\addplot [very thick, gray]
table {%
0 51
0 7.105427357601e-15
};
\addplot [very thick, gray]
table {%
25 51
25 7.105427357601e-15
};
\end{axis}

\end{tikzpicture}
    \caption{Relevant bio-entities involved in RNA sources. Columns represent bio-entities involved in the RNA sources (rows) included in \rnakg.}
\label{fig:RNADBxMolecule}
\end{figure}

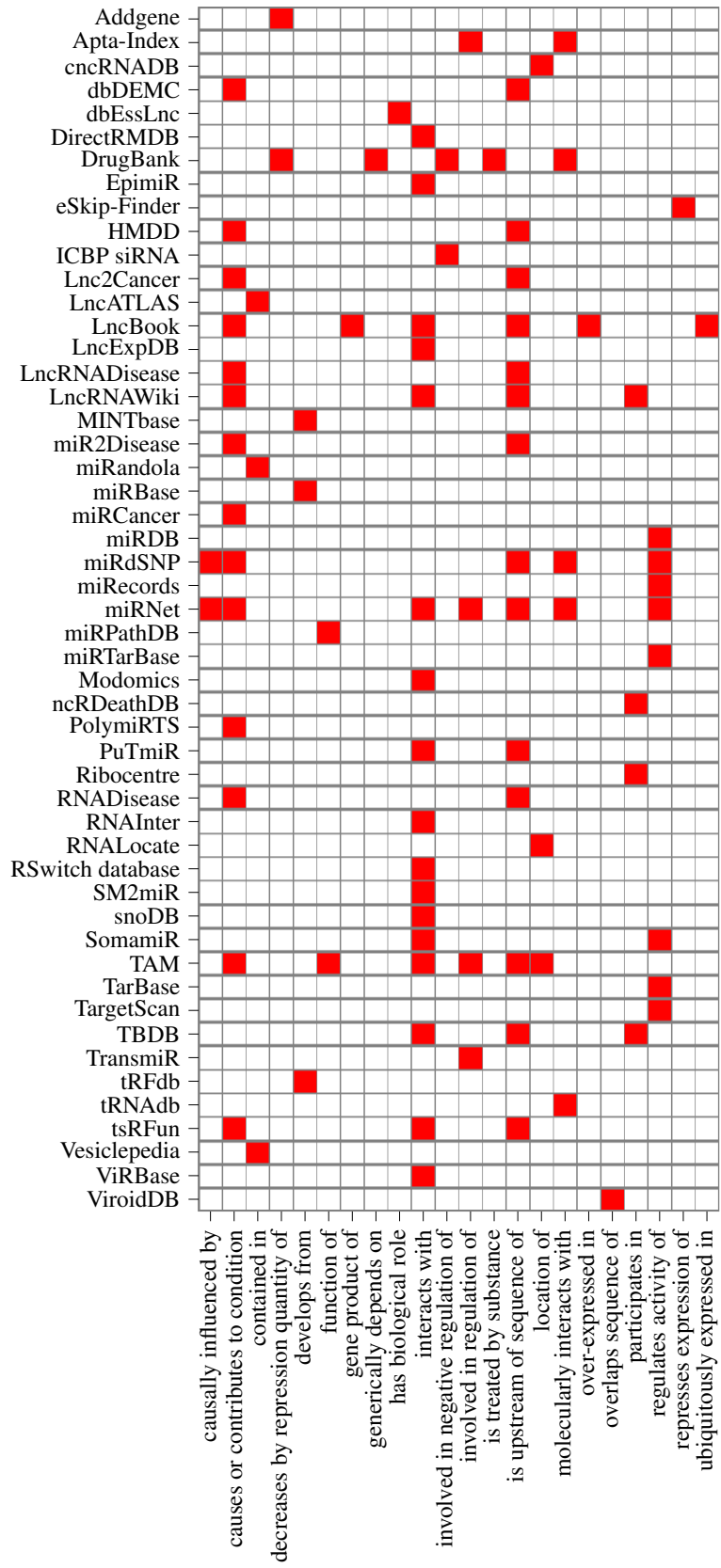
\begin{figure}[htbp]
    \centering
    \adjustbox{width=0.585\textwidth}{
\begin{tikzpicture}

\definecolor{darkgray176}{RGB}{176,176,176}
\definecolor{gray}{RGB}{128,128,128}

\begin{axis}[
x=0.35cm,
y=0.35cm,
tick align=outside,
tick pos=left,
x grid style={darkgray176},
xmin=0, xmax=22,
xtick style={color=black},
xtick={0.5,1.5,2.5,3.5,4.5,5.5,6.5,7.5,8.5,9.5,10.5,11.5,12.5,13.5,14.5,15.5,16.5,17.5,18.5,19.5,20.5,21.5},
xticklabel style={rotate=90.0},
xticklabels={
  causally influenced by,
  causes or contributes to condition,
  contained in,
  decreases by repression quantity of,
  develops from,
  function of,
  gene product of,
  generically depends on,
  has biological role,
  interacts with,
  involved in negative regulation of,
  involved in regulation of,
  is treated by substance,
  is upstream of sequence of,
  location of,
  molecularly interacts with,
  over-expressed in,
  overlaps sequence of,
  participates in,
  regulates activity of,
  represses expression of,
  ubiquitously expressed in,
  under-expressed in
},
y dir=reverse,
y grid style={darkgray176},
ymin=0, ymax=51,
ytick style={color=black},
ytick={0.5,1.5,2.5,3.5,4.5,5.5,6.5,7.5,8.5,9.5,10.5,11.5,12.5,13.5,14.5,15.5,16.5,17.5,18.5,19.5,20.5,21.5,22.5,23.5,24.5,25.5,26.5,27.5,28.5,29.5,30.5,31.5,32.5,33.5,34.5,35.5,36.5,37.5,38.5,39.5,40.5,41.5,42.5,43.5,44.5,45.5,46.5,47.5,48.5,49.5,50.5},
yticklabels={
  Addgene,
  Apta-Index,
  cncRNADB,
  dbDEMC,
  dbEssLnc,
  DirectRMDB,
  DrugBank,
  EpimiR,
  eSkip-Finder,
  HMDD,
  ICBP siRNA,
  Lnc2Cancer,
  LncATLAS,
  LncBook,
  LncExpDB,
  LncRNADisease,
  LncRNAWiki,
  MINTbase,
  miR2Disease,
  miRandola,
  miRBase,
  miRCancer,
  miRDB,
  miRdSNP,
  miRecords,
  miRNet,
  miRPathDB,
  miRTarBase,
  Modomics,
  ncRDeathDB,
  PolymiRTS,
  PuTmiR,
  Ribocentre,
  RNADisease,
  RNAInter,
  RNALocate,
  RSwitch database,
  SM2miR,
  snoDB,
  SomamiR,
  TAM,
  TarBase,
  TargetScan,
  TBDB,
  TransmiR,
  tRFdb,
  tRNAdb,
  tsRFun,
  Vesiclepedia,
  ViRBase,
  ViroidDB
}
]
\addplot graphics [includegraphics cmd=\pgfimage,xmin=0, xmax=23, ymin=51, ymax=0] {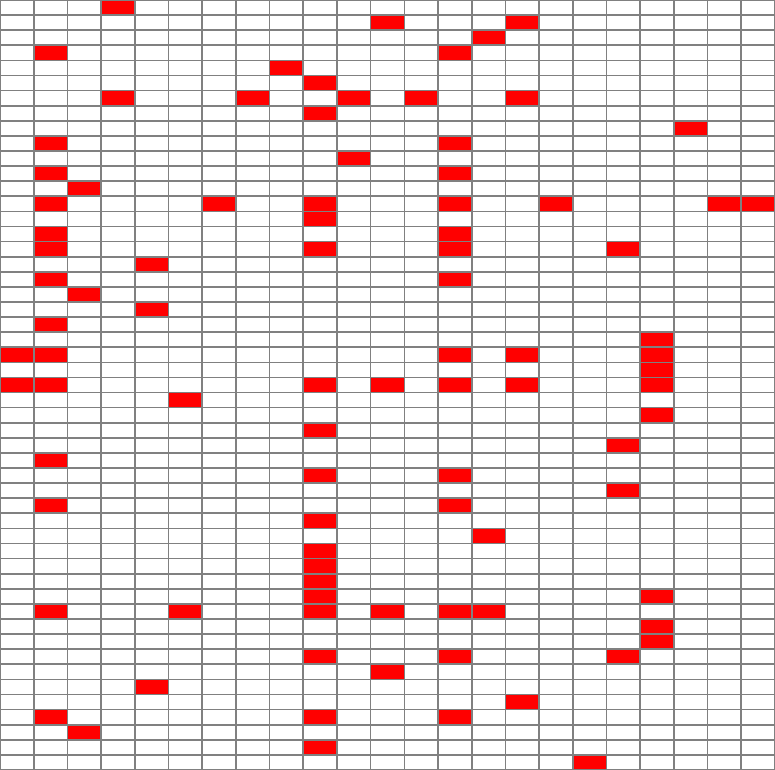};
\addplot [very thick, gray]
table {%
4.44089209850063e-16 51
4.44089209850063e-16 7.105427357601e-15
};
\addplot [very thick, gray]
table {%
23 51
23 7.105427357601e-15
};
\addplot [very thick, gray]
table {%
4.44089209850063e-16 7.105427357601e-15
23 7.105427357601e-15
};
\addplot [very thick, gray]
table {%
4.44089209850063e-16 51
23 51
};
\end{axis}

\end{tikzpicture}}
    \caption{Overlap of of RO terms in RNA sources.
    Columns represent the RO terms involved in the RNA sources (rows) included in \rnakg.}
\label{fig:RNADBxRO}
\end{figure}

Supplementary Tables \ref{tab:stats1}-\ref{tab:stats5} detail the interactions present in \rnakg by showing the two interactors with their numerosity ({\tt Subject} column refers to the numerosity of the first entity while {\tt Object} column refers to the numerosity of the second entity in the {\tt Edge} column), the RO relation we used to semantically describe the interaction ({\tt Relation} column) with its cardinality ({\tt Direct Relation(s)} column), and the sources that have been processed to include the interaction in \rnakg ({\tt Source(s)} column). We omit inverse relationships because they can be retrieved by means of Table \ref{tab:relations}.
Supplementary Figure~\ref{fig:RNADBxMolecule}-\ref{fig:RNADBxRO} show bio-entities present in RNA sources and the mapping on RO terms we used to represent relationships within the sources. Each source is annotated with a red spot if the bioentity or the RO term is present in the source. Confirming our previous analysis, lncRNA, miRNA, mRNA, and protein are among the most represented molecules, whereas {\tt interacts with}, {\tt regulates activity of}, and {\tt causes or contributes to condition} are among the most represented RO terms.  Supplementary Table~\ref{tab:nodeTypes} shows the primary node types and their corresponding identifiers with an instance sample.
Finally, Supplementary Listings~\ref{lst:sparql1}-~\ref{lst:sparql3} show three examples of queries on \rnakg that can be executed through the SPARQL endpoint's query tab.


\newpage

\begin{table}[htbp]
    \centering
\begin{scriptsize}
\begin{tabular}{|c|l||c|c|c|c|c|c|c|c|c|}
    \hline\hline
{\bf Type}& {\bf Data source} & {\bf GO} & {\bf Mondo/HPO}& {\bf VO} & {\bf ChEBI}& {\bf Uberon} & {\bf CLO} & {\bf PRO}  & {\bf SO} & {\bf PW}\\ \hline \hline

\multirow{19}{*}{{\bf miRNA}} &
\href{https://www.mirbase.org}{miRBase} &&&&&&&&x& \\ \cline{2-11} &
\href{https://mirdb.org/}{miRDB}  &&&&&&&&x& \\ \cline{2-11} &
\href{https://www.mirnet.ca/miRNet/}{miRNet} &x&x&&x&&&x&x& \\ \cline{2-11} &\href{http://c1.accurascience.com/miRecords/}{miRecords} &&&&&&&&x& \\ \cline{2-11} &
\href{http://www.jianglab.cn/EpimiR/}{EpimiR}
 &x&&&&&&&x& \\ \cline{2-11} &
\href{https://www.cuilab.cn/hmdd}{HMDD} &&x&&&&&&x& \\ \cline{2-11} &
\href{http://watson.compbio.iupui.edu:8080/miR2Disease/}{miR2Disease} &&x&&&&&&x& \\ \cline{2-11} &
\href{https://www.targetscan.org/}{TargetScan}
&&&&&&&&x& \\ \cline{2-11} &
\href{https://compbio.uthsc.edu/SomamiR/}{SomamiR DB}
 &&x&&&&&&x& \\ \cline{2-11} &
\href{https://dianalab.e-ce.uth.gr/html/diana/web/index.php?r=tarbasev8}{TarBase}
 &&&&&&&&x& \\ \cline{2-11} &
\href{https://mirtarbase.cuhk.edu.cn/~miRTarBase/miRTarBase_2022/php/index.php}{miRTarBase}
&&&&&&&&x& \\ \cline{2-11}&
\href{http://www.jianglab.cn/SM2miR/}{SM2miR}
&&&&x&&&&x& \\ \cline{2-11} &
\href{https://www.cuilab.cn/transmir}{TransmiR}
 &&&&&&&x&x& \\ \cline{2-11} &
\href{https://compbio.uthsc.edu/miRSNP/}{PolymiRTS}
&&x&&&&&&x& \\ \cline{2-11} &
\href{https://www.biosino.org/dbDEMC/index}{dbDEMC}
 &&x&&&&&&x& \\ \cline{2-11} &
\href{http://www.lirmed.com/tam2/}{TAM}
 &x&x&&&x&&x&x& \\ \cline{2-11} &
\href{https://www.isical.ac.in/~bioinfo_miu/TF-miRNA1.php}{PuTmiR}
&&&&&&&x&x& \\ \cline{2-11} &
\href{https://mpd.bioinf.uni-sb.de/overview.html}{miRPathDB}
&x&&&&&&&x& \\ \cline{2-11} &
\href{http://mircancer.ecu.edu/}{miRCancer}
&&x&&&&&&x& \\ \cline{2-11} &
\href{http://mirdsnp.ccr.buffalo.edu/}{miRdSNP}
&&x&&&&&&x& \\ \cline{2-11} &
\href{http://mirandola.iit.cnr.it/}{miRandola}
 &x&&&x&&&&x& \\ \hline
\hline

\makecell{{\bf mRNA  vaccine}} & \makecell[l]{\href{https://go.drugbank.com/categories/DBCAT005631}{DrugBank} } &&x&x&&&&&& \\ \hline
\hline

\multirow{2}{*}{\makecell{{\bf siRNA}}} & \href{http://web.mit.edu/sirna/}{ICBP siRNA}
 &&&&&&&&x& \\ \cline{2-11} &
\href{https://go.drugbank.com/indications/DBCOND0093099}{DrugBank} &&x&&x&&&&x& \\ \hline
\hline

\multirow{2}{*}{\makecell{{\bf RNA}\\{\bf aptamer}}} & \href{https://www.aptagen.com/apta-index/}{Apta-Index}  &&&&x&&&x&& \\ \cline{2-11} &
\href{https://go.drugbank.com/categories/DBCAT001641}{DrugBank}  &&x&&x&&&x&& \\ \hline
\hline

\multirow{2}{*}{\makecell{{\bf ASO}}} & \href{https://eskip-finder.org/cgi-bin/input.cgi}{eSkip-Finder}
&&&&&&&&x& \\ \cline{2-11} &
\href{https://go.drugbank.com/categories/DBCAT001709}{DrugBank}
&&x&&x&&&x&& \\ \hline
\hline

{\bf gRNA} & \href{https://www.addgene.org/crispr/reference/grna-sequence/}{Addgene} &&&&&&&&x& \\ \hline
\hline

\multirow{8}{*}{\makecell{{\bf lncRNA}}} & \href{https://ngdc.cncb.ac.cn/lncbook/}{LncBook}
&x&x&&&x&x&&x& \\ \cline{2-11} &
\href{http://www.rnanut.net/lncrnadisease/}{LncRNADisease}
&&x&&&&&&x& \\ \cline{2-11} &
\href{https://ngdc.cncb.ac.cn/lncexpdb/}{LncExpDB}
&&&&&&&&x& \\ \cline{2-11} &
\href{https://esslnc.pufengdu.org/}{dbEssLnc}
 &x&&&x&&&&x& \\ \cline{2-11} &
\href{https://lncatlas.crg.eu/}{lncATLAS}
&x&&&&&&&x& \\ \cline{2-11} &
\href{http://www.noncode.org/index.php}{NONCODE}
&&x&&&&&&x& \\ \cline{2-11} &
\href{http://bio-bigdata.hrbmu.edu.cn/lnc2cancer/}{Lnc2Cancer}
&&x&&&&&&x& \\ \cline{2-11} &
\href{https://ngdc.cncb.ac.cn/lncrnawiki/}{LncRNAWiki}  &x&x&&x&&&x&x&x \\ \hline
\hline

{\bf Ribozyme} & \href{https://www.ribocentre.org/}{Ribocentre}
 &x&&&&&&&x& \\ \hline
\hline

\makecell{{\bf Viral  RNA}} & \makecell[l]{\href{https://viroids.org/}{ViroidDB}} &&&&&&&&x& \\ \hline
\hline

\multirow{2}{*}{{\bf Riboswitch}} &
\href{https://tbdb.io/}{TBDB}
 &&&&&&&x&x& \\ \cline{2-11} &
\href{https://penchovsky.atwebpages.com/applications.php?page=58}{RSwitch}
 &&&x&&&&&x& \\ \hline
\hline

\multirow{3}{*}{{\bf tRF}} &
\href{http://genome.bioch.virginia.edu/trfdb/}{tRFdb}
 &&&&&&&&x& \\ \cline{2-11} &
\href{https://rna.sysu.edu.cn/tsRFun/}{tsRFun}
 &&x&&&&&&x& \\ \cline{2-11} &
\href{https://cm.jefferson.edu/MINTbase/}{MINTbase}
 &&&&&&&&x& \\ \hline
\hline

{\bf snoRNA} & \href{https://bioinfo-scottgroup.med.usherbrooke.ca/snoDB/}{snoDB} &&&&&&&&x& \\ \hline
\hline

{\bf tRNA} & \href{http://trna.bioinf.uni-leipzig.de/DataOutput/}{tRNAdb}
&&&&x&&&&x& \\ \hline
\hline

\multirow{10}{*}{\makecell{{\bf Inter}\\{\bf RNA}}} &
\href{http://www.rnainter.org/}{RNAInter}
 &&&&x&&&&x& \\ \cline{2-11} &
\href{http://www.rna-society.org/rnalocate/}{RNALocate}
 &x&&&&&&&x& \\ \cline{2-11} &
\href{http://www.rnadisease.org/}{RNADisease}
 &&x&&&&&&x& \\ \cline{2-11} &
\href{https://www.rna-society.org/ncrdeathdb/}{ncRDeathDB}
 &x&&&&&&&x& \\ \cline{2-11} &
\href{http://www.rna-society.org/cncrnadb/}{cncRNADB}
&&&&&&&x&x& \\ \cline{2-11} &
\href{http://www.rna-society.org/virbase/}{ViRBase}
&&&&&&&x&x& \\ \cline{2-11} &
\href{http://microvesicles.org/}{Vesiclepedia}
 &x&&&&&&&x& \\ \cline{2-11} &
\href{http://www.rnamd.org/directRMDB/index.html}{DirectRMDB}  &&&&&&&&x& \\ \cline{2-11} &
\href{https://genesilico.pl/modomics/}{Modomics} &&&&&&&x&x& \\ \hline
\hline
    \end{tabular}
\end{scriptsize}
    \caption{Bio-ontologies that can be exploited for the characterization of data sources.}
    \label{tab:databasesont2}
\end{table}

\begin{table}[htbp]
    \centering
    \begin{scriptsize}
    \begin{tabular}{|l|l|l|c|c|c|}
    \hline\hline
      {\bf Edge}  & {\bf Relation} & {\bf Source(s)} & {\bf Subjects} & {\bf Objects}& {\bf Direct Relation(s)}\\ \hline\hline
premiRNA-miRNA & develops into & miRBase & 1,917 & 2,656 & 2,879 \\ \hline
premiRNA-pseudogene & interacts with & RNAInter & 1 & 1 & 1 \\ \hline
miRNA-protein & interacts with & RNAInter & 10 & 98 & 105 \\ \hline
premiRNA-protein & interacts with & RNAInter & 20 & 22 & 26 \\ \hline
premiRNA-mRNA & regulates activity of & \begin{tabular}[c]{@{}l@{}}miRDB\\ miRecords\\ TarBase\\ miRTarBase\\ TargetScan\\ SomamiR\\ miRdSNP\end{tabular} & 11 & 107 & 117 \\ \hline
miRNA-mRNA & regulates activity of & \begin{tabular}[c]{@{}l@{}}miRDB\\ miRecords\\ TarBase\\ miRTarBase\\ TargetScan\\ SomamiR\\ miRdSNP\end{tabular} & 2,761 & 19,125 & 1,373,343 \\ \hline
miRNA-mRNA & interacts with & RNAInter & 2,599 & 16,718 & 580,001 \\ \hline
premiRNA-mRNA & interacts with & RNAInter & 133 & 175 & 222 \\ \hline
miRNA-miRNA & interacts with & RNAInter & 4 & 4 & 4 \\ \hline
miRNA-pseudogene & regulates activity of & miRNet & 642 & 3,743 & 59,377 \\ \hline
miRNA-pseudogene & interacts with & RNAInter & 466 & 720 & 1,894 \\ \hline
mRNA-pseudogene & interacts with & RNAInter & 856 & 1,043 & 1,173 \\ \hline
pseudogene-pseudogene & interacts with & RNAInter & 79 & 77 & 85 \\ \hline
rRNA-pseudogene & interacts with & RNAInter & 5 & 3 & 5 \\ \hline
miRNA-circRNA & interacts with & RNAInter & 42 & 139 & 322 \\ \hline
miRNA-othersRNA & interacts with & RNAInter & 6 & 12 & 13 \\ \hline
miRNA-epigenetic modification & interacts with & \begin{tabular}[c]{@{}l@{}l@{}}EpimiR\\ RNAInter\\DirectRMDB\end{tabular} & 499 & 22 & 1,911 \\ \hline
premiRNA-epigenetic modification & interacts with & \begin{tabular}[c]{@{}l@{}l@{}}EpimiR\\ RNAInter\\DirectRMDB\end{tabular} & 421 & 27 & 773 \\ \hline
protein-epigenetic modification & interacts with & Modomics & 32 & 7 & 33 \\ \hline
scaRNA-epigenetic modification & interacts with & Modomics & 21 & 4 & 21 \\ \hline
snoRNA-epigenetic modification & interacts with & Modomics & 6 & 3 & 6 \\ \hline
miRNA-disease & causes or contributes to condition & \begin{tabular}[c]{@{}l@{}}miR2Disease\\ HMDD\\ miRNet\\ dbDEMC\\ miRdSNP\\ TAM\\ RNADisease\\ PolymiRTS\\miRCancer\\\end{tabular} & 2,627 & 611 & 38,258 \\ \hline
premiRNA-disease & causes or contributes to condition & \begin{tabular}[c]{@{}l@{}}miR2Disease\\ HMDD\\ miRNet\\ dbDEMC\\ miRdSNP\\ TAM\\ RNADisease \\ PolymiRTS\\miRCancer\\\end{tabular} & 608 & 193 & 5,034 \\ \hline
miRNA-lncRNA & interacts with & \begin{tabular}[c]{@{}l@{}}miRNet\\ LncRNAWiki\\ SomamiR\\ RNAInter\end{tabular} & 2,534 & 777 & 13,893 \\ \hline
premiRNA-lncRNA & interacts with & \begin{tabular}[c]{@{}l@{}}miRNet\\ LncRNAWiki\\ SomamiR\\ RNAInter\end{tabular} & 111 & 535 & 3,450 \\ \hline
\hline
\end{tabular}
    \end{scriptsize}
    \caption{\rnakg  Descriptive statistics by primary edge type (Part I).}
    \label{tab:stats1}
\end{table}

\begin{table}[htbp]
    \centering
    \begin{scriptsize}
    \begin{tabular}{|l|l|l|c|c|c|}
    \hline\hline
      {\bf Edge}  & {\bf Relation} & {\bf Source(s)} & {\bf Subjects} & {\bf Objects}& {\bf Direct Relation(s)}\\ \hline\hline
variant-miRNA & causally influences & \begin{tabular}[c]{@{}l@{}}miRNet\\ miRdSNP\end{tabular} & 300 & 164 & 343 \\ \hline
variant-premiRNA & causally influences & \begin{tabular}[c]{@{}l@{}}miRNet\\ miRdSNP\end{tabular} & 10,756 & 251 & 11,281 \\ \hline
variant-gene & causally influences & miRNet & 357,872 & 16,755 & 357,872 \\ \hline
variant-disease & causally influences & miRdSNP & 321 & 26 & 346 \\ \hline
variant-TF & causally influences & miRNet & 2,080,335 & 113 & 2,082,657 \\ \hline
tsRNA-miRNA & interacts with & tsRFun & 8,044 & 103 & 121,236 \\ \hline
tsRNA-disease & causes or contributes to condition & tsRFun & 449 & 31 & 26,602 \\ \hline
tRF-tRNA & develops from & \begin{tabular}[c]{@{}l@{}}tRFdb\\ MINTbase\end{tabular} & 20,628 & 882 & 114,851 \\ \hline
      tRF-mRNA & interacts with & RNAInter & 269 & 11,150 & 26,314 \\ \hline
tRF-lncRNA & interacts with & RNAInter & 18 & 6 & 18 \\ \hline
tRF-pseudogene & interacts with & RNAInter & 20 & 13 & 26 \\ \hline
      piRNA-lncRNA & interacts with & RNAInter & 1 & 1 & 1 \\ \hline
piRNA-mRNA & interacts with & RNAInter & 1 & 1 & 1 \\ \hline
snoRNA-gene & interacts with & snoDB & 373 & 249 & 379 \\ \hline
snoRNA-premiRNA & interacts with & \begin{tabular}[c]{@{}l@{}}snoDB\\ miRNet\end{tabular} & 1 & 1 & 1 \\ \hline
snoRNA-miRNA & interacts with & \begin{tabular}[c]{@{}l@{}}snoDB\\ miRNet\\ RNAInter\end{tabular} & 249 & 509 & 1,385 \\ \hline
snoRNA-snoRNA & interacts with & snoDB & 74 & 86 & 344 \\ \hline
snoRNA-lncRNA & interacts with & \begin{tabular}[c]{@{}l@{}}snoDB\\ RNAInter\end{tabular} & 146 & 31 & 196 \\ \hline
snoRNA-snRNA & interacts with & \begin{tabular}[c]{@{}l@{}}snoDB\\ RNAInter\end{tabular} & 432 & 329 & 717 \\ \hline
snoRNA-rRNA & interacts with & snoDB & 427 & 318 & 672 \\ \hline
snoRNA-mRNA & interacts with & \begin{tabular}[c]{@{}l@{}}snoDB\\ RNAInter\end{tabular} & 99 & 142 & 180 \\ \hline
snoRNA-tRNA & interacts with & snoDB & 4 & 4 & 4 \\ \hline
snoRNA-retained intron & interacts with & snoDB & 25 & 8 & 33 \\ \hline
snoRNA-scaRNA & interacts with & \begin{tabular}[c]{@{}l@{}}snoDB\\ RNAInter\end{tabular} & 24 & 9 & 34 \\ \hline
snoRNA-pseudogene & interacts with & \begin{tabular}[c]{@{}l@{}}snoDB\\ RNAInter\end{tabular} & 8 & 8 & 9 \\ \hline
lncRNA-gene & interacts with & \begin{tabular}[c]{@{}l@{}}LncRNAWiki\\ RNAInter\end{tabular} & 185 & 620 & 1,138 \\ \hline
othersRNA-gene & interacts with & RNAInter & 1 & 2 & 2 \\ \hline
mRNA-gene & interacts with & RNAInter & 3 & 10 & 10 \\ \hline
lncRNA-disease & causes or contributes to condition & \begin{tabular}[c]{@{}l@{}}LncRNADisease\\ Lnc2Cancer\\ LncRNAWiki\\ LncBook\\ RNADisease\end{tabular} & 1,331 & 347 & 6,557 \\ \hline
circRNA-disease & causes or contributes to condition & \begin{tabular}[c]{@{}l@{}}LncRNADisease\\ Lnc2Cancer\\ RNADisease\end{tabular} & 14 & 8 & 15 \\ \hline
snRNA-disease & causes or contributes to condition & RNADisease & 7 & 13 & 14 \\ \hline
pseudogene-disease & causes or contributes to condition & RNADisease & 101 & 68 & 397 \\ \hline
snoRNA-disease & causes or contributes to condition & RNADisease & 14 & 6 & 14 \\ \hline
scRNA-disease & causes or contributes to condition & RNADisease & 1 & 3 & 3 \\ \hline
mRNA-disease & causes or contributes to condition & RNADisease & 412 & 93 & 502 \\ \hline
lncRNA-chemical & interacts with & \begin{tabular}[c]{@{}l@{}}LncRNAWiki\\ RNAInter\end{tabular} & 822 & 80 & 1,173 \\ \hline
small protein-lncRNA & gene product of & LncBook & 8,178 & 810 & 8,417 \\ \hline
lncRNA-protein & interacts with & \begin{tabular}[c]{@{}l@{}}LncBook\\ LncRNAWiki\\ RNAInter\end{tabular} & 8,412 & 294 & 16,212 \\ \hline
lncRNA-biological context & over-expressed in & LncBook & 822 & 9 & 1,395 \\ \hline
lncRNA-biological context & under-expressed in & LncBook & 1,742 & 9 & 3,128 \\ \hline
\hline
    \end{tabular}
    \end{scriptsize}
    \caption{\rnakg  Descriptive statistics by primary edge type (Part II).}
    \label{tab:stats2}
\end{table}

\begin{table}[htbp]
    \centering
    \begin{scriptsize}
    \begin{tabular}{|l|l|l|c|c|c|}
    \hline\hline
      {\bf Edge}  & {\bf Relation} & {\bf Source(s)} & {\bf Subjects} & {\bf Objects}& {\bf Direct Relation(s)}\\ \hline\hline
      lncRNA-biological role & has biological role & dbEssLnc & 173 & 3 & 173 \\ \hline
      lncRNA-biological context & ubiquitously expressed in & LncBook & 2,221 & 9 & 6,610 \\ \hline
lncRNA-cellular component & contained in & LncATLAS & 10,207 & 6 & 27,287 \\ \hline
lncRNA-pathway & participates in & LncRNAWiki & 30 & 11 & 43 \\ \hline
lncRNA-biological process & participates in & LncRNAWiki & 21 & 2 & 23 \\ \hline
miRNA-TF & is upstream of sequence of & PuTmiR & 4 & 17 & 19 \\ \hline
miRNA-TF & is downstream of sequence of & PuTmiR & 4 & 10 & 10 \\ \hline
      premiRNA-TF & is upstream of sequence of & PuTmiR & 484 & 66 & 1,839 \\ \hline
premiRNA-TF & is downstream of sequence of & PuTmiR & 488 & 65 & 1,949 \\ \hline
premiRNA-TF & involved in regulation of & \begin{tabular}[c]{@{}l@{}}miRNet\\ TransmiR\\ TAM\\\end{tabular} & 330 & 320 & 1,869 \\ \hline
premiRNA-molecular function & has function & \begin{tabular}[c]{@{}l@{}}TAM\\miRPathDB\\\end{tabular} & 26 & 2 & 29 \\ \hline
premiRNA-premiRNA & interacts with & TAM & 320 & 646 & 820 \\ \hline
premiRNA-anatomy & located in & TAM & 57 & 7 & 70 \\ \hline
lncRNA-anatomy & located in & cncRNADB & 4 & 5 & 6 \\ \hline
mRNA-anatomy & located in & cncRNADB & 3 & 3 & 3 \\ \hline
lncRNA-cell & located in & cncRNADB & 49 & 7 & 52 \\ \hline
mRNA-cell & located in & cncRNADB & 41 & 23 & 70 \\ \hline
mRNA-chemical & interacts with & RNAInter & 92 & 12 & 97 \\ \hline
TEC-chemical & interacts with & RNAInter & 2 & 1 & 2 \\ \hline
snoRNA-chemical & interacts with & RNAInter & 15 & 2 & 15 \\ \hline
      pseudogene-chemical & interacts with & RNAInter & 82 & 7 & 84 \\ \hline
premiRNA-chemical & interacts with & \begin{tabular}[c]{@{}l@{}@{}l@{}}SM2miR\\ RNAInter\\ miRNet\\miRandola\\\end{tabular} & 17 & 11 & 22 \\ \hline
miRNA-chemical & interacts with & \begin{tabular}[c]{@{}l@{}@{}l@{}}SM2miR\\ RNAInter\\ miRNet\\miRandola\\\end{tabular} & 724 & 147 & 3,067 \\ \hline
ASO drug-disease & is substance that treats & DrugBank & 10 & 8 & 12 \\ \hline
ASO drug-protein & decreases by repression quantity of & DrugBank & 5 & 11 & 11 \\ \hline
ASO drug-protein & is carrier of & DrugBank & 1 & 1 & 1 \\ \hline
siRNA drug-mRNA & involved in negative regulation of & DrugBank & 3 & 3 & 3 \\ \hline
siRNA-mRNA & involved in negative regulation of & ICBP siRNA & 77 & 54 & 79 \\ \hline
shRNA-mRNA & involved in negative regulation of & ICBP siRNA & 40 & 42 & 42 \\ \hline
siRNA drug-disease & is substance that treats & DrugBank & 4 & 4 & 6 \\ \hline
aptamer-protein & molecularly interacts with & Apta-Index & 33 & 76 & 85 \\ \hline
aptamer-chemical & molecularly interacts with & Apta-Index & 114 & 67 & 114 \\ \hline
aptamer drug-protein & molecularly interacts with & DrugBank & 2 & 4 & 4 \\ \hline
aptamer drug-disease & is substance that treats & DrugBank & 1 & 4 & 4 \\ \hline
mRNA vaccine-disease & is substance that treats & DrugBank & 8 & 1 & 8 \\ \hline
lncRNA-mRNA & interacts with & \begin{tabular}[c]{@{}l@{}}RNAInter\\ LncExpDB\end{tabular} & 2,214 & 18,228 & 2,429,654 \\ \hline
lncRNA-lncRNA & interacts with & RNAInter & 1,431 & 1,431 & 3,187 \\ \hline
lncRNA-rRNA & interacts with & RNAInter & 10 & 8 & 17 \\ \hline
lncRNA-pseudogene & interacts with & RNAInter & 136 & 2,239 & 2,577 \\ \hline
lncRNA-ncRNA & interacts with & RNAInter & 4 & 6 & 6 \\ \hline
lncRNA-scaRNA & interacts with & RNAInter & 7 & 8 & 11 \\ \hline
lncRNA-TF & interacts with & RNAInter & 4,833 & 370 & 173,842 \\ \hline
lncRNA-ribozyme & interacts with & RNAInter & 1 & 1 & 1 \\ \hline
circRNA-protein & interacts with & RNAInter & 1 & 2 & 2 \\ \hline
ncRNA-protein & interacts with & RNAInter & 12 & 2 & 12 \\ \hline
othersRNA-protein & interacts with & RNAInter & 12 & 2 & 12 \\ \hline
pseudogene-protein & interacts with & RNAInter & 44 & 17 & 54 \\ \hline
ribozyme-protein & interacts with & RNAInter & 1 & 2 & 2 \\ \hline
\hline
    \end{tabular}
    \end{scriptsize}
    \caption{\rnakg  Descriptive statistics by primary edge type (Part III).}
    \label{tab:stats3}
\end{table}

\begin{table}[htbp]
    \centering
    \begin{scriptsize}
    \begin{tabular}{|l|c|c|c|c|c|}
    \hline\hline
      {\bf Edge}  & {\bf Relation} & {\bf Source(s)} & {\bf Subjects} & {\bf Objects}& {\bf Direct Relation(s)}\\ \hline\hline
scRNA-protein & interacts with & RNAInter & 1 & 5 & 5 \\ \hline
snRNA-protein & interacts with & RNAInter & 3 & 11 & 11 \\ \hline
      snoRNA-protein & interacts with & RNAInter & 2 & 3 & 3 \\ \hline
snoRNA-miscRNA & interacts with & snoDB & 2 & 2 & 2 \\ \hline
unknownRNA-protein & interacts with & RNAInter & 1 & 1 & 1 \\ \hline
vRNA-protein & interacts with & RNAInter & 1 & 1 & 1 \\ \hline
riboswitch-protein & is downstream of sequence of & TBDB & 9,384 & 1,405 & 148,262 \\ \hline
riboswitch-bacterial strain & interacts with & RSwitch database & 215 & 77 & 215 \\ \hline
riboswitch-biological process & participates in & TBDB & 13,072 & 111 & 13,072 \\ \hline
viralRNA-ribozyme & overlaps sequence of & ViroidDB & 5,654 & 6 & 5,896 \\ \hline
ribozyme-biological process & participates in & Ribocentre & 2 & 6 & 11 \\ \hline
      miRNA-viralmRNA & interacts with & ViRBase & 80 & 38 & 250 \\ \hline
snRNA-viralmRNA & interacts with & ViRBase & 1 & 1 & 1 \\ \hline
lncRNA-viralmRNA & interacts with & ViRBase & 3 & 8 & 8 \\ \hline
pseudogene-viralmRNA & interacts with & ViRBase & 1 & 3 & 3 \\ \hline
snRNA-viralmiRNA & interacts with & ViRBase & 2 & 1 & 2 \\ \hline
pseudogene-viralmiRNA & interacts with & ViRBase & 351 & 4 & 352 \\ \hline
snoRNA-viralmiRNA & interacts with & ViRBase & 8 & 1 & 8 \\ \hline
premiRNA-viralmiRNA & interacts with & ViRBase & 1 & 1 & 1 \\ \hline
unknownRNA-viralmiRNA & interacts with & ViRBase & 3 & 1 & 3 \\ \hline
lncRNA-viralmiRNA & interacts with & ViRBase & 263 & 1 & 263 \\ \hline
othersRNA-viralmiRNA & interacts with & ViRBase & 1 & 1 & 1 \\ \hline
scRNA-viralmiRNA & interacts with & ViRBase & 1 & 1 & 1 \\ \hline
protein-viralmiRNA & interacts with & ViRBase & 11 & 18 & 21 \\ \hline
mRNA-viralmiRNA & interacts with & ViRBase & 17,564 & 63 & 17,923 \\ \hline
miRNA-viralmiRNA & interacts with & ViRBase & 1 & 1 & 1 \\ \hline
lncRNA-viral protein & interacts with & ViRBase & 1 & 1 & 1 \\ \hline
miRNA-viral protein & interacts with & ViRBase & 51 & 15 & 59 \\ \hline
miRNA-ribozyme & interacts with & RNAInter & 1 & 1 & 1 \\ \hline
miRNA-unknownRNA & interacts with & RNAInter & 14 & 4 & 14 \\ \hline
miRNA-scRNA & interacts with & RNAInter & 1 & 1 & 1 \\ \hline
othersRNA-mRNA & interacts with & RNAInter & 58 & 68 & 70 \\ \hline
othersRNA-lncRNA & interacts with & RNAInter & 41 & 16 & 49 \\ \hline
othersRNA-pseudogene & interacts with & RNAInter & 8 & 8 & 8 \\ \hline
othersRNA-rRNA & interacts with & RNAInter & 1 & 1 & 1 \\ \hline
snRNA-snRNA & interacts with & RNAInter & 21 & 21 & 33 \\ \hline
snRNA-lncRNA & interacts with & RNAInter & 34 & 14 & 52 \\ \hline
      snRNA-mRNA & interacts with & RNAInter & 39 & 96 & 100 \\ \hline
snRNA-pseudogene & interacts with & RNAInter & 9 & 11 & 15 \\ \hline
      eRNA-mRNA & interacts with & RNAInter & 2 & 2 & 4 \\ \hline
scRNA-mRNA & interacts with & RNAInter & 1 & 23 & 23 \\ \hline
      mRNA-mRNA & interacts with & RNAInter & 4,459 & 4,459 & 7,035 \\ \hline
mRNA-rRNA & interacts with & RNAInter & 120 & 4 & 133 \\ \hline
mRNA-ncRNA & interacts with & RNAInter & 7 & 5 & 7 \\ \hline
      mRNA-scaRNA & interacts with & RNAInter & 4 & 2 & 4 \\ \hline
mRNA-protein & interacts with & RNAInter & 179 & 130 & 227 \\ \hline
rRNA-rRNA & interacts with & RNAInter & 2 & 3 & 3 \\ \hline
circRNA-RBP & interacts with & RNAInter & 1,027 & 31 & 1,105 \\ \hline
mRNA-RBP & interacts with & RNAInter & 8,856 & 130 & 17,754 \\ \hline
ribozyme-RBP & interacts with & RNAInter & 1 & 13 & 13 \\ \hline
scaRNA-RBP & interacts with & RNAInter & 11 & 17 & 35 \\ \hline
scRNA-RBP & interacts with & RNAInter & 1 & 5 & 5 \\ \hline
othersRNA-RBP & interacts with & RNAInter & 19 & 26 & 54 \\ \hline
snRNA-RBP & interacts with & RNAInter & 44 & 42 & 158 \\ \hline
lncRNA-RBP & interacts with & RNAInter & 1,011 & 140 & 6,136 \\ \hline
snoRNA-RBP & interacts with & RNAInter & 200 & 70 & 753 \\ \hline
ncRNA-RBP & interacts with & RNAInter & 5 & 45 & 48 \\ \hline
rRNA-RBP & interacts with & RNAInter & 19 & 9 & 30 \\ \hline
pseudogene-RBP & interacts with & RNAInter & 1,014 & 92 & 1,907 \\ \hline
        \hline
    \end{tabular}
    \end{scriptsize}
    \caption{\rnakg  Descriptive statistics by primary edge type (Part IV).}
    \label{tab:stats4}
\end{table}

\begin{table}[htbp]
    \centering
    \begin{scriptsize}
    \begin{tabular}{|l|c|c|c|c|c|}
    \hline\hline
      {\bf Edge}  & {\bf Relation} & {\bf Source(s)} & {\bf Subjects} & {\bf Objects}& {\bf Direct Relation(s)}\\ \hline\hline
premiRNA-RBP & interacts with & RNAInter & 32 & 18 & 37 \\ \hline
      unknownRNA-TF & interacts with & RNAInter  & 1 & 74 & 74 \\ \hline
circRNA-TF & interacts with & RNAInter & 33 & 6 & 33 \\ \hline
ribozyme-TF & interacts with & RNAInter  & 3 & 64 & 94 \\ \hline
scaRNA-TF & interacts with & RNAInter & 17 & 83 & 429 \\ \hline
othersRNA-TF & interacts with & RNAInter & 541 & 152 & 11,851 \\ \hline
snRNA-TF & interacts with & RNAInter & 1,124 & 109 & 24,489 \\ \hline
snoRNA-TF & interacts with & RNAInter & 419 & 108 & 11,932 \\ \hline
scRNA-TF & interacts with & RNAInter & 1 & 74 & 74 \\ \hline
tRNA-TF & interacts with & RNAInter & 8 & 10 & 77 \\ \hline
ncRNA-TF & interacts with & RNAInter & 22 & 110 & 716 \\ \hline
rRNA-TF & interacts with & RNAInter & 26 & 55 & 713 \\ \hline
pseudogene-TF & interacts with & RNAInter & 9,524 & 245 & 188,443 \\ \hline
mRNA-TF & interacts with & RNAInter & 5,019 & 222 & 8,629 \\ \hline
premiRNA-TF & interacts with & RNAInter & 1,839 & 140 & 71,662 \\ \hline
miRNA-TF & interacts with & RNAInter & 1,522 & 272 & 4,038 \\ \hline
      circRNA-extracellular vesicle & contained in & miRandola & 6 & 1 & 6 \\ \hline
lipid-extracellular vesicle & contained in & Vesiclepedia & 35 & 3 & 47 \\ \hline
mRNA-extracellular vesicle & contained in & Vesiclepedia & 13,907 & 2 & 14,116 \\ \hline
snRNA-extracellular vesicle & contained in & Vesiclepedia & 1 & 1 & 1 \\ \hline
protein-extracellular vesicle & contained in & Vesiclepedia & 11,924 & 5 & 20,081 \\ \hline
miRNA-extracellular vesicle & contained in & Vesiclepedia & 435 & 2 & 554 \\ \hline
premiRNA-extracellular vesicle & contained in & Vesiclepedia & 26 & 2 & 29 \\ \hline
circRNA-miRNA & interacts with & \begin{tabular}[c]{@{}l@{}}SomamiR\\ miRNet\end{tabular} & 4,309 & 2,590 & 302,197 \\ \hline
circRNA-premiRNA & interacts with & \begin{tabular}[c]{@{}l@{}}SomamiR\\ miRNet\end{tabular} & 8,755 & 110 & 129,990 \\ \hline
unknownRNA-histone modification & interacts with & RNAInter & 1 & 18 & 18 \\ \hline
mRNA-histone modification & interacts with & RNAInter & 42 & 20 & 486 \\ \hline
othersRNA-histone modification & interacts with & RNAInter & 20 & 20 & 263 \\ \hline
lncRNA-histone modification & interacts with & RNAInter & 4,816 & 20 & 51,375 \\ \hline
ncRNA-histone modification & interacts with & RNAInter & 12 & 20 & 130 \\ \hline
pseudogene-histone modification & interacts with & RNAInter & 43 & 20 & 460 \\ \hline
premiRNA-histone modification & interacts with & RNAInter & 1,842 & 20 & 20,754 \\ \hline
premiRNA-subcellular location & located in & RNALocate & 578 & 7 & 926 \\ \hline
miRNA-subcellular location & located in & RNALocate & 2,572 & 15 & 3,496 \\ \hline
snRNA-subcellular location & located in & RNALocate & 1,719 & 4 & 1,719 \\ \hline
ncRNA-subcellular location & located in & RNALocate & 85 & 9 & 107 \\ \hline
pseudogene-subcellular location & located in & RNALocate & 218 & 6 & 230 \\ \hline
snoRNA-subcellular location & located in & RNALocate & 570 & 7 & 763 \\ \hline
scaRNA-subcellular location & located in & RNALocate & 1 & 1 & 1 \\ \hline
rRNA-subcellular location & located in & RNALocate & 2 & 2 & 3 \\ \hline
lncRNA-subcellular location & located in & RNALocate & 1,047 & 11 & 2,313 \\ \hline
lincRNA-subcellular location & located in & RNALocate & 22 & 5 & 28 \\ \hline
othersRNA-subcellular location & located in & RNALocate & 3 & 1 & 3 \\ \hline
tRNA-subcellular location & located in & RNALocate & 1 & 1 & 1 \\ \hline
circRNA-subcellular location & located in & RNALocate & 1 & 1 & 1 \\ \hline
Y RNA-subcellular location & located in & RNALocate & 4 & 1 & 4 \\ \hline
scRNA-subcellular location & located in & RNALocate & 2 & 1 & 2 \\ \hline
mRNA-subcellular location & located in & RNALocate & 13,646 & 19 & 50,903 \\ \hline
vRNA-subcellular location & located in & RNALocate & 3 & 2 & 6 \\ \hline
mtRNA-subcellular location & located in & RNALocate & 10 & 2 & 19 \\ \hline
miRNA-programmed death & participates in & ncRDeathDB & 581 & 3 & 792 \\ \hline
lncRNA-programmed death & participates in & ncRDeathDB & 99 & 2 & 99 \\ \hline
snoRNA-programmed death & participates in & ncRDeathDB & 1 & 1 & 1 \\ \hline
gRNA-gene & decreases by repression quantity of & Addgene & 77 & 43 & 77 \\ \hline
ASO-mRNA & represses expression of & eSkip-Finder & 2,633 & 16 & 2,678 \\ \hline
ASO drug-mRNA & involved in negative regulation of & DrugBank & 6 & 4 & 6 \\ \hline
tRNA-mRNA & interacts with & RNAInter & 4 & 4 & 4 \\ \hline
tRNA-amino acid & molecularly interacts with & tRNAdb & 549 & 19 & 549 \\ \hline
tRNA-lncRNA & interacts with & RNAInter & 8 & 4 & 9 \\ \hline 
        \hline
    \end{tabular}
    \end{scriptsize}
    \caption{\rnakg  Descriptive statistics by primary edge type (Part V).}
    \label{tab:stats5}
\end{table}

\begin{table}[htbp]
    \centering
    \begin{scriptsize}
    \begin{tabular}{|l|l|l|}
\hline \hline
\textbf{Node type} & \textbf{Full name} & \textbf{Identifier(s)} \\ \hline
\hline
Anatomy & - & Uberon (\href{http://purl.obolibrary.org/obo/UBERON_0005253}{http://purl.obolibrary.org/obo/\textbf{UBERON\_0005253}}) \\ \hline
Amino acid & - & ChEBI (\href{http://purl.obolibrary.org/obo/CHEBI_25017}{http://purl.obolibrary.org/obo/\textbf{CHEBI\_25017}}) \\ \hline
ASO & AntiSense Oligonucleotide & Oligo name in literature (\href{https://eskip-finder.org?H45_1-13_18-30}{https://eskip-finder.org?\textbf{H45\_1-13\_18-30}}) \\ \hline
ASO drug & AntiSense Oligonucleotide drug & DrugBank (\href{https://go.drugbank.com/drugs/DB05528}{https://go.drugbank.com/drugs/\textbf{DB05528}}) \\ \hline
Aptamer & RNA aptamer & Apta-Index (\href{https://www.aptagen.com/aptamer-details/?id=608}{https://www.aptagen.com/\textbf{aptamer-details/?id=608}}) \\ \hline
Aptamer drug & RNA aptamer drug & DrugBank (\href{https://go.drugbank.com/drugs/DB04932}{https://go.drugbank.com/drugs/\textbf{DB04932}}) \\ \hline
Bacterial strain & - & NCBI Taxonomy Browser (\href{https://www.ncbi.nlm.nih.gov/Taxonomy/Browser/wwwtax.cgi?id=485}{https://www.ncbi.nlm.nih.gov/Taxonomy/Browser/\textbf{wwwtax.cgi?id=485}}) \\ \hline
Biological context & - & \begin{tabular}[c]{@{}l@{}}Uberon (\href{http://purl.obolibrary.org/obo/UBERON_0000479}{http://purl.obolibrary.org/obo/\textbf{UBERON\_0000479}})\\ GO (\href{http://purl.obolibrary.org/obo/_0070062}{http://purl.obolibrary.org/obo/\textbf{GO\_0070062}})\\ CLO (\href{http://purl.obolibrary.org/obo/CLO_0009828}{http://purl.obolibrary.org/obo/\textbf{CLO\_0009828}})\\ Mondo (\href{http://purl.obolibrary.org/obo/MONDO_0005108}{http://purl.obolibrary.org/obo/\textbf{MONDO\_0005108}})\end{tabular} \\ \hline
Biological process & - & GO (\href{http://purl.obolibrary.org/obo/0044848}{http://purl.obolibrary.org/obo/\textbf{GO\_0044848}}) \\ \hline
Biological role & - & NIH Talking Glossary of Genetic Terms (\href{https://www.genome.gov/genetics-glossary/Oncogene}{https://www.genome.gov/genetics-glossary/\textbf{Oncogene}}) \\ \hline
Cell & - & CLO (\href{http://purl.obolibrary.org/obo/CLO_0009828}{http://purl.obolibrary.org/obo/\textbf{CLO\_0009828}}) \\ \hline
Cellular component & - & GO (\href{http://purl.obolibrary.org/obo/GO_0110165}{http://purl.obolibrary.org/obo/\textbf{GO\_0110165}}) \\ \hline
Chemical & - & ChEBI (\href{http://purl.obolibrary.org/obo/CHEBI_77062}{http://purl.obolibrary.org/obo/\textbf{CHEBI\_77062}}) \\ \hline
circRNA & Circular RNA & NCBI Entrez gene (\href{http://www.ncbi.nlm.nih.gov/gene/6575?circRNA}{http://www.ncbi.nlm.nih.gov/gene/\textbf{6575?circRNA}}) \\ \hline
Disease & - & Mondo (\href{http://purl.obolibrary.org/obo/MONDO_0004971}{http://purl.obolibrary.org/obo/\textbf{MONDO\_0004971}}) \\ \hline
Epigenetic modification & - & \begin{tabular}[c]{@{}l@{}}GO (\href{http://purl.obolibrary.org/obo/GO_0006306}{http://purl.obolibrary.org/obo/\textbf{GO\_0006306}})\\ ENCODE (\href{https://www.encodeproject.org/targets/H3K4me2}{https://www.encodeproject.org/targets/\textbf{H3K4me2}})\end{tabular} \\ \hline
eRNA & RNA enhancer & Human enhancer RNA Atlas (HeRA) (\href{https://hanlab.uth.edu/HeRA?IL1beta-RBT46}{https://hanlab.uth.edu/HeRA?\textbf{IL1beta-RBT46}}) \\ \hline
Extracellular vesicle & - & GO (\href{http://purl.obolibrary.org/obo/GO_1990742}{http://purl.obolibrary.org/obo/\textbf{GO\_1990742}}) \\ \hline
Gene & - & NCBI Entrez gene (\href{http://www.ncbi.nlm.nih.gov/gene/1954}{http://www.ncbi.nlm.nih.gov/gene/\textbf{1954}}) \\ \hline
Genomic sequence & - & SO (\href{http://purl.obolibrary.org/obo/SO_0000704}{http://purl.obolibrary.org/obo/\textbf{SO\_0000704}}) \\ \hline
gRNA & Guide RNA & Addgene (\href{https://www.addgene.org/41818}{https://\textbf{www.addgene.org/41818}}) \\ \hline
Histone modification & - & dbEM (\href{http://crdd.osdd.net/raghava/dbem?H3K9me2}{http://crdd.osdd.net/raghava/dbem?\textbf{H3K9me2}}) \\ \hline
lincRNA & Long intergenic RNA & NCBI Entrez gene (\href{http://www.ncbi.nlm.nih.gov/gene/100287569?lincRNA}{http://www.ncbi.nlm.nih.gov/gene/\textbf{100287569?lincRNA}}) \\ \hline
Lipid & - & ChEBI (\href{http://purl.obolibrary.org/obo/CHEBI_136143}{http://purl.obolibrary.org/obo/\textbf{CHEBI\_136143}}) \\ \hline
lncRNA & Long non-coding RNA & NCBI Entrez gene (\href{http://www.ncbi.nlm.nih.gov/gene/100506207?lncRNA}{http://www.ncbi.nlm.nih.gov/gene/\textbf{100506207?lncRNA}}) \\ \hline
miscRNA & Miscellaneous RNA & NCBI Entrez gene (\href{www.ncbi.nlm.nih.gov/gene/6029?misc_RNA}{www.ncbi.nlm.nih.gov/gene/\textbf{6029?misc\_RNA}}) \\ \hline
mRNA & Messenger RNA & NCBI Entrez gene (\href{http://www.ncbi.nlm.nih.gov/gene/1756?mRNA}{http://www.ncbi.nlm.nih.gov/gene/\textbf{1756?mRNA}}) \\ \hline
mRNA vaccine & - & DrugBank (\href{https://go.drugbank.com/drugs/DB15654}{https://go.drugbank.com/drugs/\textbf{DB15654}}) \\ \hline
miRNA & Mature microRNA & miRBase (\href{https://www.mirbase.org/cgi-bin/mature.pl?mature\_acc=MIMAT0022711}{https://www.mirbase.org/cgi-bin/mature.pl?mature\_acc=\textbf{MIMAT0022711}}) \\ \hline
mtRNA & Mitochondrial RNA & NCBI Entrez gene (\href{http://www.ncbi.nlm.nih.gov/gene/4549?mtRNA}{http://www.ncbi.nlm.nih.gov/gene/\textbf{4549?mtRNA}}) \\ \hline
ncRNA & Non-coding RNA & NCBI Entrez gene (\href{http://www.ncbi.nlm.nih.gov/gene/102723629?ncRNA}{http://www.ncbi.nlm.nih.gov/gene/\textbf{102723629?ncRNA}}) \\ \hline
Other (not classified yet) RNA & - & NCBI Entrez gene (\href{http://www.ncbi.nlm.nih.gov/gene/3537?other}{http://www.ncbi.nlm.nih.gov/gene/\textbf{3537?other}}) \\ \hline
Pathway & - & PW (\href{http://purl.obolibrary.org/obo/PW_0000632}{http://purl.obolibrary.org/obo/\textbf{PW\_0000632}}) \\ \hline
Phenotype & - & HPO (\href{http://purl.obolibrary.org/obo/HP_0005506}{http://purl.obolibrary.org/obo/\textbf{HP\_0005506}}) \\ \hline
piRNA & Piwi-interacting RNA & piRBase (\href{http://bigdata.ibp.ac.cn/piRBase?piR-39980}{http://bigdata.ibp.ac.cn/piRBase?\textbf{piR-39980}}) \\ \hline
premiRNA & Hairpin microRNA & miRBase (\href{https://www.mirbase.org/cgi-bin/mirna\_entry.pl?acc=MI0000067}{https://www.mirbase.org/cgi-bin/mirna\_entry.pl?acc=\textbf{MI0000067}}) \\ \hline
Programmed cell death & - & GO (\href{http://purl.obolibrary.org/obo/GO_0097300}{http://purl.obolibrary.org/obo/\textbf{GO\_0097300}}) \\ \hline
Protein & - & PRO (\href{http://purl.obolibrary.org/obo/PR_Q92506}{http://purl.obolibrary.org/obo/\textbf{PR\_Q92506}}) \\ \hline
Pseudogene & - & NCBI Entrez gene (\href{http://www.ncbi.nlm.nih.gov/gene/442240?pseudo}{http://www.ncbi.nlm.nih.gov/gene/\textbf{442240?pseudo}}) \\ \hline
Retained intron & - & NCBI Entrez gene (\href{http://www.ncbi.nlm.nih.gov/gene/23642?retained_intron}{http://www.ncbi.nlm.nih.gov/gene/\textbf{23642?retained\_intron}}) \\ \hline
RBP & RNA-Binding Protein & PRO (\href{http://purl.obolibrary.org/obo/PR_Q92506}{http://purl.obolibrary.org/obo/\textbf{PR\_Q92506}}) \\ \hline
Riboswitch & - & TBDB (\href{https://tbdb.io/tboxes/UQCD7JYG.html}{https://tbdb.io/tboxes/\textbf{UQCD7JYG.html}}) \\ \hline
Ribozyme & RIBOnucleic acid enZYME & Rfam (\href{http://rfamlive.xfam.org/family/RF02682}{http://rfamlive.xfam.org/family/\textbf{RF02682}}) \\ \hline
rRNA & Ribosomal RNA & snoDB (\href{http://scottgroup.med.usherbrooke.ca/snoDB/28S-3616?snoDBrRNA}{http://scottgroup.med.usherbrooke.ca/snoDB/\textbf{28S-3616?snoDBrRNA}}) \\ \hline
scRNA & Small conditional RNA & NCBI Entrez gene (\href{http://www.ncbi.nlm.nih.gov/gene/618?scRNA}{http://www.ncbi.nlm.nih.gov/gene/\textbf{618?scRNA}}) \\ \hline
scaRNA & Small Cajal body-specific RNA & NCBI Entrez gene (\href{http://www.ncbi.nlm.nih.gov/gene/677767?scaRNA}{http://www.ncbi.nlm.nih.gov/gene/\textbf{677767?scaRNA}}) \\ \hline
shRNA & Short/small hairpin RNA & ICBP siRNA (\href{http://web.mit.edu/sirna/sequences/results-1107.html}{http://web.mit.edu/sirna/sequences/results-\textbf{1107.html}}) \\ \hline
siRNA & Short interfering RNA & ICBP siRNA (\href{http://web.mit.edu/sirna/sequences/results-1053.html}{http://web.mit.edu/sirna/sequences/results-\textbf{1053.html}}) \\ \hline
siRNA drug & Short interfering RNA drug & DrugBank (\href{https://go.drugbank.com/drugs/DB15935}{https://go.drugbank.com/drugs/\textbf{DB15935}}) \\ \hline
Small protein & - & SmProt (\href{http://bigdata.ibp.ac.cn/SmProt/SmProt.php?ID=SPROHSA53815}{http://bigdata.ibp.ac.cn/SmProt/SmProt.php?ID=\textbf{SPROHSA53815}}) \\ \hline
snRNA & Small nuclear RNA & NCBI Entrez gene (\href{http://www.ncbi.nlm.nih.gov/gene/26824?snRNA}{http://www.ncbi.nlm.nih.gov/gene/\textbf{26824?snRNA}}) \\ \hline
snoRNA & Small nucleolar RNA & NCBI Entrez gene (\href{http://www.ncbi.nlm.nih.gov/gene/727676?snoRNA}{http://www.ncbi.nlm.nih.gov/gene/\textbf{727676?snoRNA}}) \\ \hline
Subcellular location & - & GO (\href{http://purl.obolibrary.org/obo/GO_0005840}{http://purl.obolibrary.org/obo/\textbf{GO\_0005840}}) \\ \hline
TEC & To be Experimentally Confirmed RNA & NCBI Entrez gene (\href{https://www.ncbi.nlm.nih.gov/gene/8123?TEC}{https://www.ncbi.nlm.nih.gov/gene/\textbf{8123?TEC}}) \\ \hline
TF & Transcription Factor & PRO (\href{http://purl.obolibrary.org/obo/PR_000007035}{http://purl.obolibrary.org/obo/\textbf{PR\_000007035}}) \\ \hline
tRF & tRNA-derived fragment & \begin{tabular}[c]{@{}l@{}}tRFdb (\href{http://genome.bioch.virginia.edu/trfdb?tRF-3018b}{http://genome.bioch.virginia.edu/trfdb?\textbf{tRF-3018b}})\\ MINTbase (\href{https://cm.jefferson.edu/MINTbase/InputController?v=f&g=GRCh37&e=1&search=submit&t=All&am=All&an=All&da=&tn=&fs=&fn=tRF-16-4YDRFUE}{https://cm.jefferson.edu/MINTbase/InputController?v=f\&g=GRCh37\&e=1\&} \\ \href{https://cm.jefferson.edu/MINTbase/InputController?v=f&g=GRCh37&e=1&search=submit&t=All&am=All&an=All&da=&tn=&fs=&fn=tRF-16-4YDRFUE}{search=submit\&t=All\&am=All\&an=All\&da=\&tn=\&fs=\&fn=\textbf{tRF-16-4YDRFUE}})
\end{tabular} \\ \hline
tRNA & Transfer RNA & \begin{tabular}[c]{@{}l@{}}GtRNAdb (\href{http://gtrnadb.ucsc.edu/genomes/eukaryota/Hsapi19/genes/tRNA-Gln-CTG-1-3.html}{http://gtrnadb.ucsc.edu/genomes/eukaryota/Hsapi19/genes/\textbf{tRNA-Gln-CTG-1-3.html}})\\ tRFdb (\href{http://genome.bioch.virginia.edu/trfdb?chr8.trna4-TyrGTA}{http://genome.bioch.virginia.edu/trfdb?\textbf{chr8.trna4-TyrGTA}})\\ NCBI Entrez gene (\href{https://www.ncbi.nlm.nih.gov/gene/4567?tRNA}{https://www.ncbi.nlm.nih.gov/gene/\textbf{4567?tRNA}})\end{tabular} \\ \hline
tsRNA & tRNA-derived small RNA & tsRFun (\href{https://rna.sysu.edu.cn/tsRFun/searchDetail-tsRNA.php?tsRNAid=tsRNA-Gly-i-0605}{https://rna.sysu.edu.cn/tsRFun/searchDetail-tsRNA.php?tsRNAid=\textbf{tsRNA-Gly-i-0605}}) \\ \hline
Unknown RNA & - & NCBI Entrez gene (\href{https://www.ncbi.nlm.nih.gov/gene/100128998?unknown}{https://www.ncbi.nlm.nih.gov/gene/\textbf{100128998?unknown}}) \\ \hline
Vaccine & - & VO (\href{http://purl.obolibrary.org/obo/VO_0000186}{http://purl.obolibrary.org/obo/\textbf{VO\_0000186}}) \\ \hline
Variant (SNP) & - & NCBI dbSNP (\href{https://www.ncbi.nlm.nih.gov/snp/rs71354105}{https://www.ncbi.nlm.nih.gov/snp/\textbf{rs71354105}}) \\ \hline
Viral miRNA & - & miRBase (\href{https://www.mirbase.org/cgi-bin/mature.pl?mature\_acc=MIMAT0001581}{https://www.mirbase.org/cgi-bin/mature.pl?mature\_acc=\textbf{MIMAT0001581}}) \\ \hline
Viral mRNA & - & NCBI Entrez gene (\href{https://www.ncbi.nlm.nih.gov/gene/43740578?viral\_mRNA}{https://www.ncbi.nlm.nih.gov/gene/\textbf{43740578?viral\_mRNA}}) \\ \hline
Viral protein & - & PRO (\href{http://purl.obolibrary.org/obo/PR_000036828}{http://purl.obolibrary.org/obo/\textbf{PR\_000036828}}) \\ \hline
Viral RNA & - & NCBI Nucleotide database (\href{https://www.ncbi.nlm.nih.gov/nuccore/KF869252.1}{https://www.ncbi.nlm.nih.gov/nuccore/\textbf{KF869252.1}}) \\ \hline
vRNA & Vault RNA & NCBI Entrez gene (\href{https://www.ncbi.nlm.nih.gov/gene/56664?vRNA}{https://www.ncbi.nlm.nih.gov/gene/\textbf{56664?vRNA}}) \\ \hline
Y RNA & - & NCBI Entrez gene (\href{https://www.ncbi.nlm.nih.gov/gene/6090?Y_RNA}{https://www.ncbi.nlm.nih.gov/gene/\textbf{6090?Y\_RNA}}) \\ \hline \hline
    \end{tabular}
    \end{scriptsize}
    \caption{\rnakg primary node types and their corresponding identifiers with an instance sample.}\label{tab:nodeTypes}
\end{table}

\clearpage
\begin{lstlisting}[language=sparql,label={lst:sparql1},caption={SPARQL query to retrieve all miRNA molecules that %are known to 
causes or contributes to the development of leukemia.}]
PREFIX rdfs: <http://www.w3.org/2000/01/rdf-schema#>
PREFIX obo: <http://purl.obolibrary.org/obo/>

SELECT ?miRNA
WHERE {
  ?miRNA rdfs:subClassOf obo:SO_0000276. # SO_0000276 ->label-> miRNA
  ?miRNA obo:RO_0003302 obo:MONDO_0005059.
    # RO_0003302 ->label-> causes or contributes to condition;
    # MONDO_0005059 ->label-> leukemia
}
\end{lstlisting}  

\begin{lstlisting}[language=sparql,label={lst:sparql2},caption={SPARQL query to retrieve all premiRNA molecules that develop into mature miRNAs known to be located in an apoptotic body and cause or contribute to the development of a cancer.}]
PREFIX rdfs: <http://www.w3.org/2000/01/rdf-schema#>
PREFIX obo: <http://purl.obolibrary.org/obo/>

SELECT ?premiRNA ?disease
WHERE {
  ?premiRNA rdfs:subClassOf obo:SO_0000647; # SO_0000647 ->label-> premiRNA
            obo:RO_0002203 ?miRNA. # RO_0002203 ->label-> develops into
  ?miRNA rdfs:subClassOf obo:SO_0000276; 
         obo:RO_0003302 ?disease.
  ?miRNA obo:RO_0001025 obo:GO_0097189.
    # RO_0001025 ->label-> located in; GO_0097189 ->label-> apoptotic body
  ?disease rdfs:subClassOf obo:MONDO_0004992. # MONDO_0004992 ->label-> cancer
}
\end{lstlisting}  

\begin{lstlisting}[language=sparql,label={lst:sparql3},caption={SPARQL query to retrieve all lncRNA molecules that are over-expressed in a viral infectious disease and are known to cause or contribute to a disease treated by at least one RNA drug.}]
PREFIX rdfs: <http://www.w3.org/2000/01/rdf-schema#>
PREFIX obo: <http://purl.obolibrary.org/obo/>
PREFIX disease: <http://purl.obolibrary.org/obo/MONDO_>
PREFIX RNAdrug: <https://go.drugbank.com/drugs/>

SELECT ?lncRNA ?disease ?RNAdrug (COUNT(DISTINCT ?RNAdrug) as ?numRNAdrugs)
WHERE {
    ?lncRNA rdfs:subClassOf obo:SO_0001877; # SO_0001877 ->label-> lncRNA
        obo:RO_0002245 obo:MONDO_0005108;
          # MONDO_0005108 ->label-> viral infectious disease
        obo:RO_0003302 ?disease.
    ?disease obo:RO_0002302 ?RNAdrug.
      # RO_0002302 ->label-> is treated by substance

    FILTER(STRSTARTS(STR(?disease), STR(disease:)))
    FILTER(STRSTARTS(STR(?RNAdrug), STR(RNAdrug:)))
}
GROUP BY ?lncRNA ?disease ?RNAdrug
HAVING (COUNT(DISTINCT ?RNAdrug) >= 1)
\end{lstlisting}  

\nascondi{
\begin{table}[htbp]
    \centering
    \begin{scriptsize}
    \begin{tabular}{|c|l|}
\hline \hline
\textbf{Concept} & \textbf{Definition} \\
\hline
\hline
API & Application Programming Interfaces\\
\hline
ASO & AntiSense Oligonucleotide\\
\hline
bact. strain & Bacterial strain\\
\hline
CCDF & Complementary Cumulative Distribution Function\\
\hline
cell. comp. & Cellular component\\
\hline
ChEBI & Chemical Entities of Biological Interest ontology\\
\hline
circRNA & circular RNA\\
\hline
CLO & Cell Line Ontology\\
\hline
CSV & Comma-Separated Values\\
\hline
DNA & DeoxyriboNucleic Acid\\
\hline
EFO & Experimental Factor Ontology\\
\hline
epi. mod. & Epigenetic modification\\
\hline
extracell. form & Extracellular form\\
\hline
eRNA & RNA enhancer\\
\hline
FAIR & Findable, Accessible, Interoperable, and Reproducible\\
\hline
gaf & GO annotation file format\\
\hline
GO & Gene Ontology\\
\hline
gRNA & guide RNA\\
\hline
histone mod. & Histone modification\\
\hline
HPO & Human Phenotype Ontology\\
\hline
hpoa & HPO annotation file format\\
\hline
HTML & HyperText Markup Language\\
\hline
JSON & JavaScript Object Notation\\
\hline
KG & Knowledge Graph\\
\hline
lincRNA & long intergenic RNA\\
\hline
lncRNA & long non-coding RNA\\
\hline
miscRNA & miscellaneous RNA\\
\hline
mRNA & messenger RNA\\
\hline
miRNA & mature microRNA\\
\hline
mtRNA & mitochondrial RNA\\
\hline
mol. function & Molecular function\\
\hline
Mondo & Mondo Disease Ontology\\
\hline
ncRNA & non-coding RNA\\
\hline
OMA & Orthologous MAtrix\\
\hline
Unclassified RNA & other (not classified yet) RNA\\
\hline
OWL & Web Ontology Language\\
\hline
PheKnowLator & Phenotype Knowledge TransLator\\
\hline
piRNA & Piwi-interacting RNA\\
\hline
premiRNA & hairpin microRNA\\
\hline
prog. cell death & Programmed cell death\\
\hline
PRO & Protein Ontology\\
\hline
PW & Pathway Ontology\\
\hline
RBP & RNA-Binding Protein\\
\hline
RDF & Resource Description Framework\\
\hline
R2RML & RDB to RDF Mapping Language\\
\hline
RDFS & Resource Description Framework Schema\\
\hline
Ribozyme & Ribonucleic acid enzyme\\
\hline
RML & Relational Meta-Language\\
\hline
RNA & RiboNucleic Acid\\
\hline
RO & Relation Ontology\\
\hline
rRNA & ribosomal RNA\\
\hline
scRNA & Small conditional RNA\\
\hline
ShExML & Shape Expressions Mapping Language\\
\hline
shRNA & Short/small hairpin RNA\\
\hline
siRNA & Short interfering RNA\\
\hline
sncRNA & Small non-coding RNA\\
\hline
snRNA & Small nuclear RNA\\
\hline
snoRNA & Small nucleolar RNA\\
\hline
SNP & Single Nucleotide Polymorphism\\
\hline
SO & Sequence Ontology\\
\hline
SPARQL & SPARQL Protocol and RDF Query Language\\
\hline
t-SNE & t-Distributed Stochastic Neighbor Embedding\\
\hline
TEC & To be Experimentally Confirmed RNA\\
\hline
TF & Transcription Factor\\
\hline
tRF & tRNA-derived fragment\\
\hline
tRNA & Transfer RNA\\
\hline
tsRNA & tRNA-derived small RNA\\
\hline
TSV & Tab-Separated Values\\
\hline
Uberon & Uber-Anatomy Ontology\\
\hline
UniProtKB & UniProt KnowledgeBase\\
\hline
VO & Vaccine Ontology\\
\hline
vRNA & vault RNA\\
\hline
xlsx & Microsoft Excel Spreadsheet\\
\hline
XML & eXtensible Markup Language\\
\hline \hline
    \end{tabular}
    \end{scriptsize}
    \caption{Acronyms used in the Manuscript.}
\end{table}
}

\end{document}